\documentclass{aa}  
\usepackage{graphicx}
\usepackage{txfonts}
\usepackage[]{hyperref}
\usepackage{xcolor}
\usepackage{soul}
\usepackage{booktabs}
\usepackage{tabularx}
\usepackage{breqn}
\usepackage{ulem}
\begin{document} 

\title{Accelerating exoplanet climate modelling:%\\
}
\subtitle{A machine learning approach to complement 3D GCM grid simulations}
\titlerunning{}
\author{
Alexander Plaschzug\inst{1,2,3} $^{\href{https://orcid.org/0009-0002-0113-3449}{\includegraphics[scale=0.5]{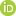}}}$,
Amit Reza\inst{1} $^{\href{https://orcid.org/0000-0001-7934-0259}{\includegraphics[scale=0.5]{FIGURES/orcid.jpg}}}$,
Ludmila Carone\inst{1}$^{\href{https://orcid.org/0000-0001-9355-3752}{\includegraphics[scale=0.5]{FIGURES/orcid.jpg}}}$,
Sebastian Gernjak\inst{1,2,3} $^{\href{https://orcid.org/0009-0003-1951-7384}{\includegraphics[scale=0.5]{FIGURES/orcid.jpg}}}$,
Christiane Helling\inst{1,2} $^{\href{https://orcid.org/0000-0002-8275-1371}{\includegraphics[scale=0.5]{FIGURES/orcid.jpg}}}$
}
\authorrunning{Plaschzug et al.}
\institute{
$^1$Space Research Institute, Austrian Academy of Sciences, Schmiedlstrasse 6, A-8042 Graz, Austria\\
$^2$Institute for Theoretical Physics and Computational Physics, Graz University of Technology, Petersgasse 16 8010 Graz\\
$^3$Institute of Physics, University of Graz, Universitätsplatz 3 8010 Graz\\
\email{amit.reza@oeaw.ac.at}    
}
   
\date{Received 22 May 2025; Accepted 03 August 2025}
%-
\abstract
{With the development of ever-improving telescopes capable of observing exoplanet atmospheres in greater detail and number, there is a growing demand for enhanced 3D climate models to support and help interpret observational data from space missions like CHEOPS, TESS, JWST, PLATO, and Ariel. However, the computationally intensive and time-consuming nature of general circulation models (GCMs) poses significant challenges in simulating a wide range of exoplanetary atmospheres.}
%Aims (mandatory)
{
The aim of this study is to determine whether machine learning (ML) algorithms can be used to predict the 3D temperature and wind structure of arbitrary tidally-locked gaseous exoplanets in a range of planetary parameters.
}
{
A new 3D GCM grid has been introduced with 60 inflated hot Jupiters orbiting A, F, G, K, and M-type host stars modelled with \texttt{Exorad}. Four new climate characteristics are introduced to characterise these planets: Day-Nightside temperature difference, Evening-Morning temperature difference, Maximum zonal Wind speed, and Wind jet width. A dense neural network (DNN) and a decision tree algorithm (XGBoost) are trained on this grid to predict local gas temperatures along with horizontal and vertical winds. To ensure the reliability and quality of the ML model predictions, WASP-121 b, HATS-42 b, NGTS-17 b, WASP-23 b, and NGTS-1 b -like planets, which are all targets for PLATO observation, are selected and modelled with \texttt{ExoRad} as well as the two ML methods as test cases. For these planets, the equilibrium gas-phase composition and transmission spectra are calculated to test whether differences in local gas temperature between the GCM and ML models would significantly affect the predicted chemical composition and transmission spectra.
}
{The DNN predictions for the gas temperatures are to such a degree that the calculated spectra agree within 32 ppm for all but one planet, for which only one single HCN feature reaches a 100 ppm difference. The XGBoost predictions are somewhat worse but never exceed 380 ppm differences. Generally, the resulting deviations are too small to be detectable with the observational capabilities of modern space telescopes, including JWST. For the DNN, only the WASP-121 b-like planet, which is the hottest investigated planet, shows a general offset that is smaller than 16~ppm. Horizontal wind predictions are less accurate but can capture the most general trends. As with temperature, the DNN also outperforms XGBoost in this respect. Predicting vertical wind remains challenging for all ML methods explored in this study.
}
{
The developed ML emulators can, within one second, reliably predict the complete 3D temperature field of an inflated warm to ultra-hot tidally locked Jupiter around A to M-type host stars. It provides a fast and computationally inexpensive tool to complement and extend traditional GCM grids for exoplanet ensemble studies. The quality of the predictions is such that no or minimal effects on the gas phase chemistry, hence on the cloud formation and transmission spectra, are to be expected. 
} 
%-
\keywords{}
   
\maketitle
%-
\nolinenumbers
%-

\section{Introduction}
\label{sec:intro}
Exoplanet research is driven by a combination of discovering exoplanets to study their demographics and evolution (e.g.,\citealt{2023ASPC..534..839L,2024arXiv240605447R,2025arXiv250201510R}), and by a detailed investigation of individual extrasolar planets. This is done to understand, for example, their atmosphere chemistry and physics \citep[e.g.][]{Carone2023,Bell2024,Deline2025,Coulombe2023,Demangeon2024},  or to prove the presence of an atmosphere (e.g., \citealt{2025ApJ...979L..19R}).  
The gas giant extrasolar planets WASP-39 b (\citealt{2024Natur.632.1017E},  WASP-43 b (\citealt{2024AJ....168....4H}), HD~189733 b (\citealt{2025AJ....169...38Z}), WASP-107 b (\citealt{2024NatAs...8.1562M}) are among the first planets observed with JWST. JWST demonstrated that such gas giants exhibit day/night differences that may be traced by observing terminator asymmetries as suggested by early models in, e.g., \cite{2020SSRv..216..139S, 2015A&A...580A..12L, 2016MNRAS.460..855H,VonParis2016,Line2016}.  The logic follow-up will be to utilise the exquisite data quality from JWST's instruments to conduct comparative studies of an ensemble of gas giant exoplanets. The interpretation of the steadily increasing number of exoplanets to be characterised by JWST, in combination with smaller telescopes like TESS and CHEOPS (e.g. \citealt{2024A&A...692A.129S}), requires simulations that model the relevant chemical and physical processes (gas dynamics, radiative transfer, gas phase chemistry, cloud formation). A fully consistent 3D model atmosphere solution including clouds has so far been presented only for two gas giants: HD~189733b (\citealt{2016A&A...594A..48L}) and HD~209458 b (\citealt{2018A&A...615A..97L}. 1D atmosphere simulations allow for a wider variety of global parameters, hence a larger ensemble of objects, but they average out all horizontal atmosphere characteristics (e.g., \citealt{2020MNRAS.498.4680G,2024ApJ...975...59M,2024A&A...690A.127J,2025arXiv250105521C}). The study of 3D atmosphere and chemistry structures was confined to the use of cloud parametrisation (\citealt{2016ApJ...828...22P,2022ApJ...934...79K}) or by a hierarchical approach (\citealt{Helling2023}), both having a pre-defined set of model planets with set global parameters (planetary global temperature, host star, orbital distances, mass, element abundances). 
%-

The diversity of known exoplanets, even within the class of gas giants, is substantial, ranging from warm to ultra-hot. Classical grid studies are useful to reveal the underlying general principles of chemistry and climate dynamics as in \cite{2015MNRAS.453.2412C, 2016ApJ...821....9K, 2016ApJ...828...22P, 2020SSRv..216..139S,Helling2023,2025ApJ...978...82K}. By necessity, these studies only cover a small subset of global parameters like planetary and stellar radii and stellar effective temperature, which makes them difficult to apply for detailed atmosphere characterisation of specific planets.
In order to answer fundamental questions about the formation and evolution of exoplanet atmospheres in different galactic environments, the number of targets that undergo characterisation will increase  (TESS: ca. 500, PLATO: ca. 200 (from \citealt{2024arXiv240605447R})). Yield studies for the PLATO mission \citealt{2023A&A...677A.133M} suggest that a minimum of 800 new gas giants will be discovered with unprecedented precision in planetary radii (3\%), which then require characterisation. The LOPS2 (long-pointing field southern hemisphere) field that will be first observed by PLATO, contains a fair number of planets with vastly diverse properties (Sec. 3.4.1. in \citealt{2025arXiv250107687N}).
%-

Interpolating model atmosphere grids is a long-standing challenge for observers who wish to use these modelling data to help physically interpret their data (e.g., Sect. 4.1. in \citealt{2024ApJ...966L..11P}, \citealt{2012A&A...540A..85P,2016A&A...593A..75S}). An alternative is to apply highly parametrised model atmospheres and tune the parameters until observations fit (e.g., \citealt{2024A&A...690A.336S}), or train neural networks on spectral grids derived from forward models for fast data interpretation (e.g., \citealt{2024ApJ...961...30Y}). \cite{2023ApJ...954...22L} presents a study on 1D profiles of brown dwarfs, comparing consistently calculated model atmospheres (though with parametrised clouds) with those derived from simplifying parametrisations used in their retrieval approach to validate what information can be deduced from such a fast parametrisation that reproduces an observed spectrum.
%-

The present paper acknowledges the limits inherent to classical grid studies in general and in particular for 3D atmosphere simulations, which may hamper rapid data interpretation for ensemble studies, as well as for in-depth studies of specific planets, since a precalculated atmosphere grid seldom contains the observed planet specifically (e.g. \citealt{2012A&A...540A..85P,2024ApJ...966L..11P}).
\begin{figure}
\centering
\includegraphics[width=1\linewidth]{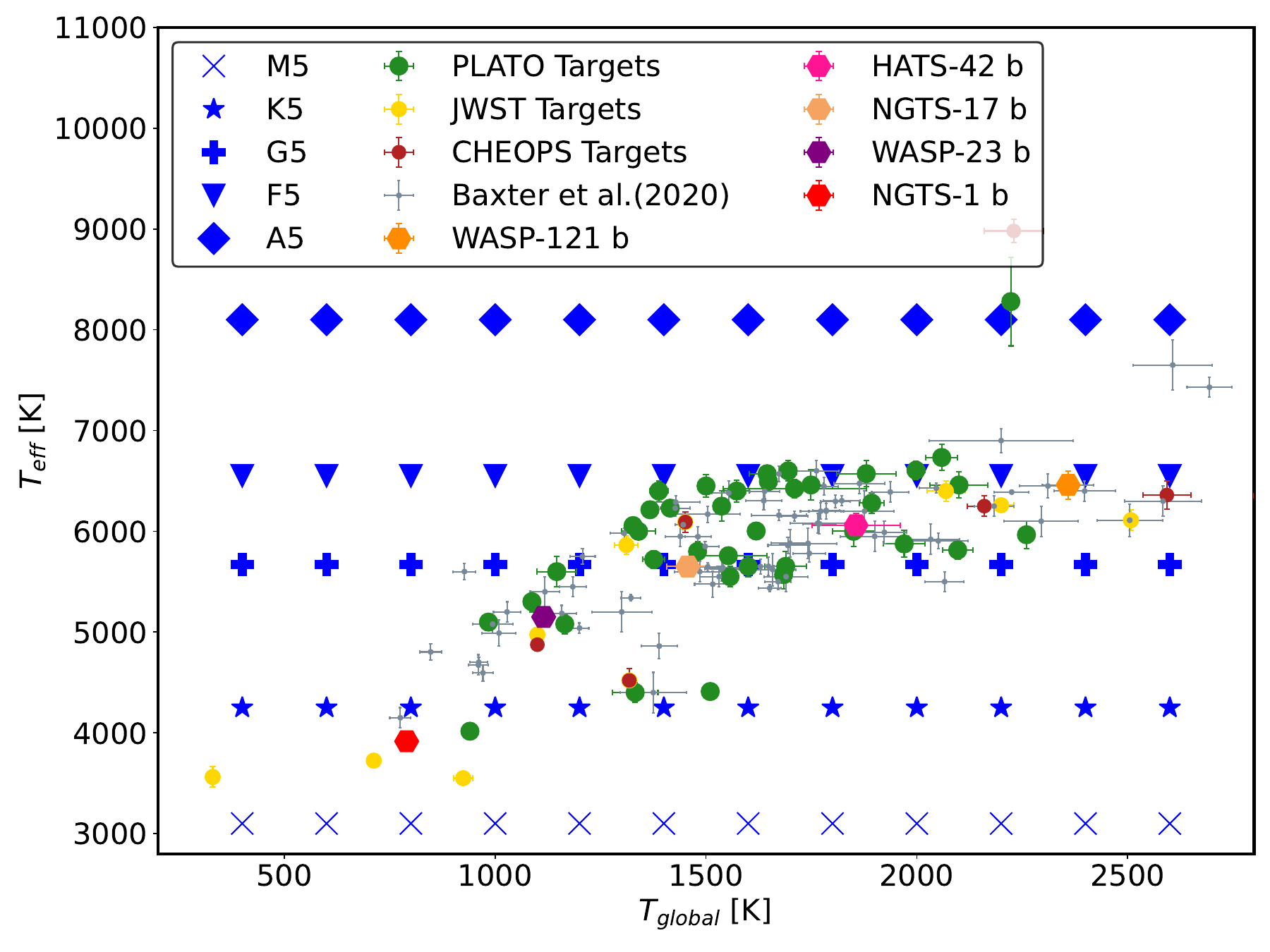}
\caption{3D \texttt{ExoRad} GCM grid of 60 simulated planets (blue) for different host stars from M to A ($T_{\rm eff}$ [K], see Tab.\ref{tab: stellar_params} ) and global planetary temperatures $T_{\rm global}$ [K]. Observation targets for JWST (yellow), PLATO (green), and CHEOPS (red) are overlayed. Highlighted in different colours are the hot Jupiters WASP-121 b, HATS-42 b, NGTS-17 b, WASP-23 b, and NGTS-1 b, which are used to test the prediction quality of our designed ML models. The grey points are objects from \citep{Baxter2020}.}
\label{fig:motivation}
\end{figure}
%-
Fig.~\ref{fig:motivation} shows the planets that are current targets for the PLATO (green) and CHEOPS (dark red) missions, planets that JWST observed (gold), and a list of ideally suitable targets for exoplanet atmosphere studies collected by \citep{Baxter2020} (grey). The coverage of the global planetary ($T_{\rm global}=400\,\ldots 2600$ K) and stellar effective temperature ($T_{\rm eff}$ [K] for A, F, G, K and M host stars) of our new 3D AFGKM \texttt{ExoRad} (blue; introduced in Sec. \ref{s:ExoRadII}) is shown for comparison: It covers well the overall global parameter range of the observational ensemble of gas giant planets but not necessarily specific planets like, for example, the PLATO targets WASP-121 b, HATS-42 b, NGTS-17 b, WASP-23 b or NGTS-1 b. These planets are highlighted specifically with different colours as they are used to test our ML predictions.
%-

The study presented here is therefore dedicated to exploring the potential of Machine Learning (ML) as a reliable and time-efficient alternative to complex 3D GCM models for completing existing model grids (filling the gaps) and interpreting observational data. ML models are designed to predict the thermodynamic and hydrodynamic properties of exoplanetary atmospheres, which are critical to determine the chemical composition and, eventually, the observable transmission spectrum. This leads us to explore the following research questions.
%-
\begin{enumerate}
\item  Can ML models predict the 3D temperature and wind structure of un-modelled (unknown) exoplanet atmospheres? 
\item  Which ML models can efficiently fill the gaps in the grid? Do different ML models and methods provide consistent results? Can computational costs be reduced by using ML methods?
\item  Do ML-predicted atmospheric structures differ from those obtained with traditional 3D GCM modelling ($\texttt{ExoRad}$)? How do potential differences affect the chemical composition of the atmospheric gas as the necessary precursor for cloud formation? Would these differences be observable in a transmission spectrum?
\end{enumerate}
%-
We apply several ML models using the $\texttt{ExoRad}$ 3D GCM data to address the research questions posed. A newly simulated $\texttt{ExoRad}$ grid, introduced in Sec.~\ref{s:ExoRadII}, provides the training data that represent the diversity of exoplanet atmospheres. A set of four parameters, that is, day-to-night side and evening-to-morning terminator temperature differences, wind jet speed and width (Fig. \ref{fig:Grid_Temp_Windjet_Diag}) are introduced to characterise gas giant climate states that complement the global system parameters ($\text{T}_{\rm eff}$, $\text{T}_{\rm global}$, $\text{P}_{\rm orb}$, [M/H]=0). Sec.~\ref{s:approach} outlines our ML and validation methodology, including the training and testing strategy applied to assess the ML model's performance, while Sec.~\ref{s:method} details the ML models employed in this work. The results are presented in Sec.~\ref{s:results}, where predictions from the ML models for unseen planets are compared against $\texttt{ExoRad}$ simulation results. It includes case studies for specific PLATO target gas giant exoplanets to demonstrate that ML models offer an efficient solution to fill the gaps of 3D GCM atmosphere grids, making them a viable addition to integrate into data interpretation pipelines. Sec.~\ref{s:results2} explores how prediction errors from ML models propagate into the chemical equilibrium composition and transmission spectra. Finally, Sec.~\ref{sec:time-comp} discusses and compares the computational cost of ML models and $\texttt{ExoRad}$ simulations. Sec.~\ref{s:concl} presents the conclusions, including that a light-weight dense neural network can predict the local gas temperature for unseen gas giant planets to a precision indistinguishable from the forward 3D GCM solution by the present observational facilities. The predictions are much more accurate than the differences between different GCM frameworks.
%-
\begin{figure*}[htbp]
\centering
\begin{minipage}[t]{0.49\linewidth}
\includegraphics[width=\linewidth]{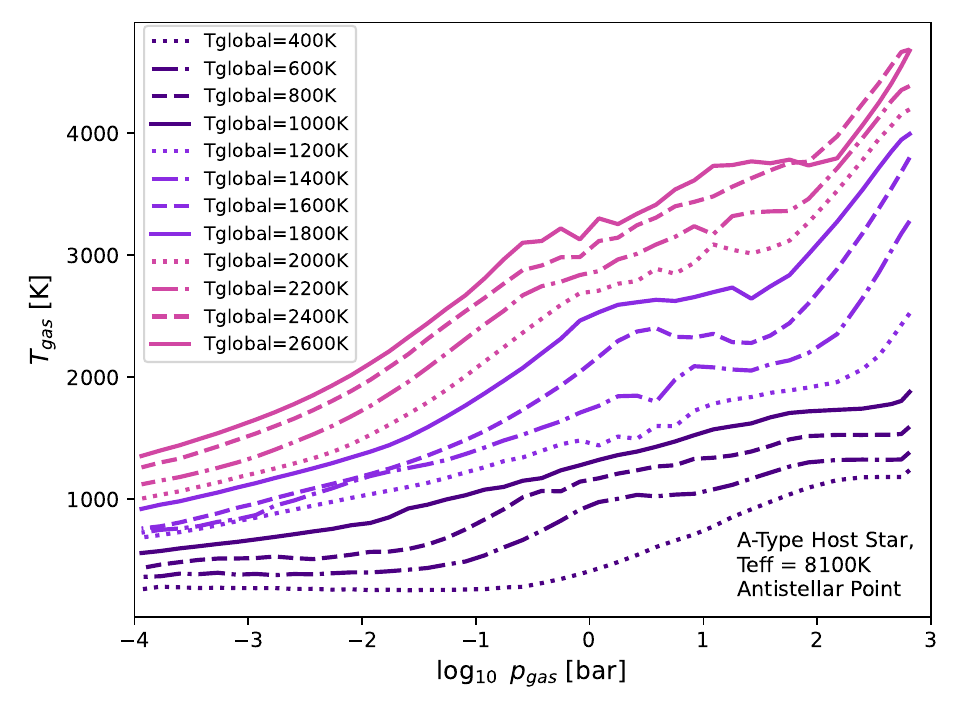}
\end{minipage}
\begin{minipage}[t]{0.49\linewidth}
\includegraphics[width=\linewidth]{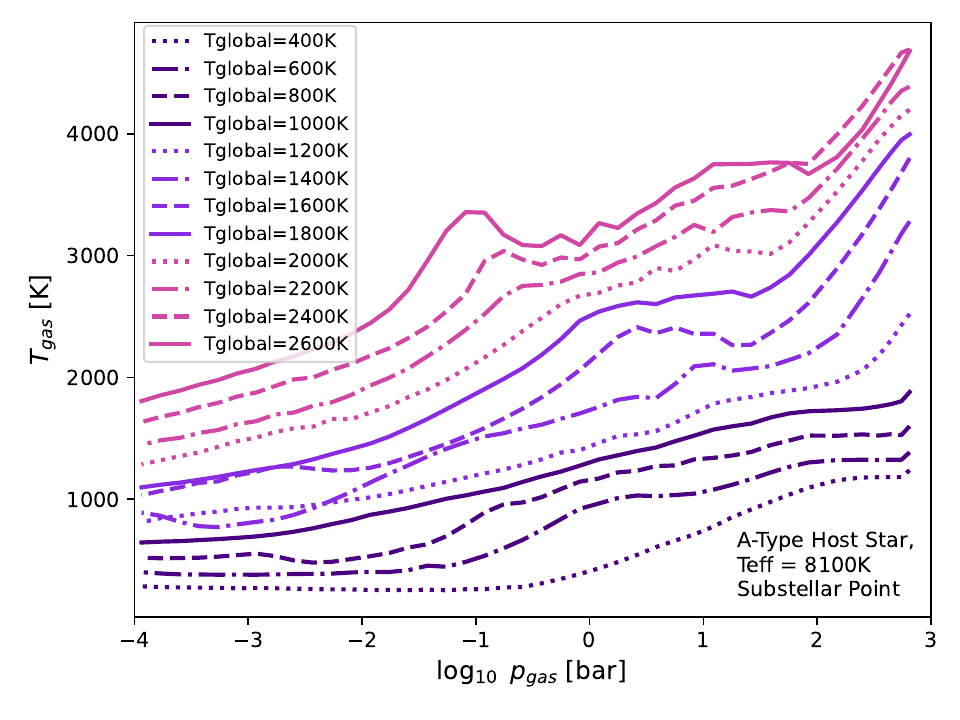}
\end{minipage}
%-
\begin{minipage}[t]{0.49\linewidth}
\includegraphics[width=\linewidth]{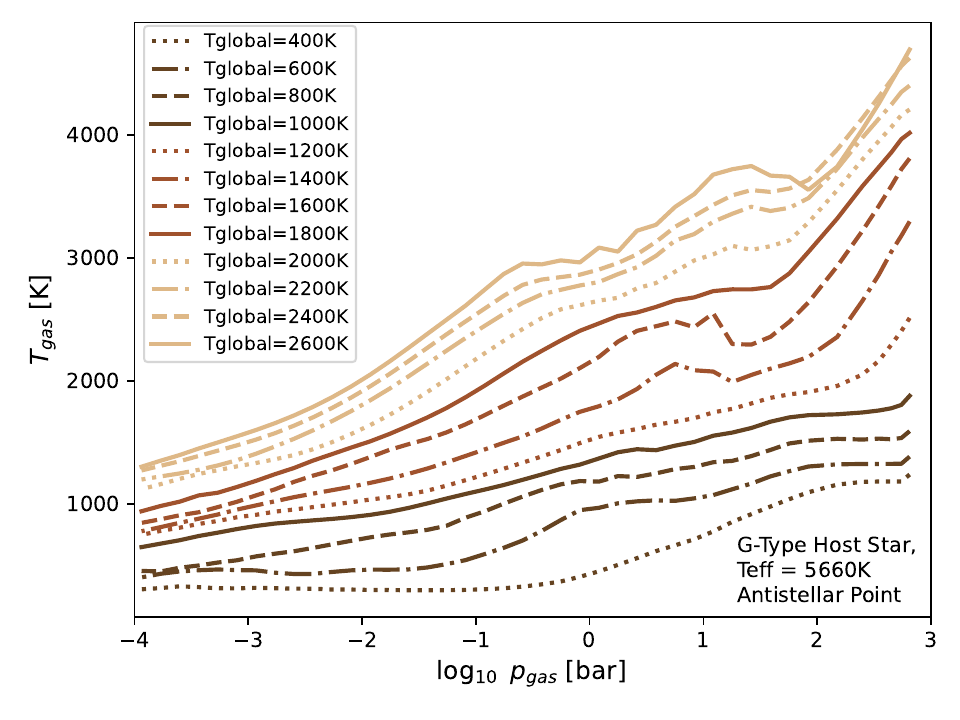}
\end{minipage}
\begin{minipage}[t]{0.49\linewidth}
\includegraphics[width=\linewidth]{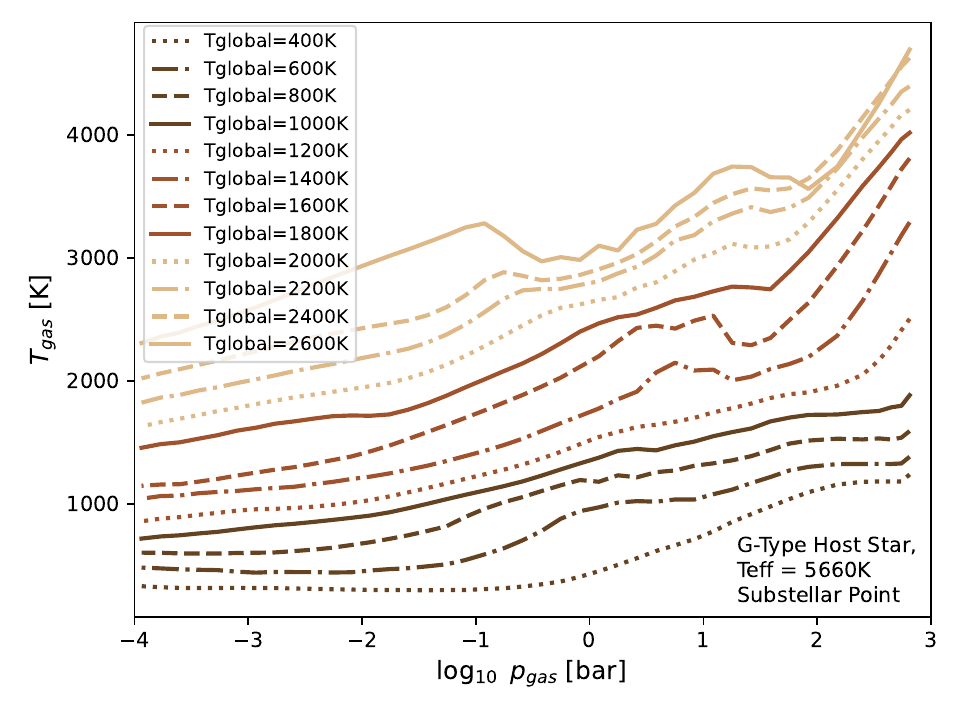}
\end{minipage}
%-
\begin{minipage}[t]{0.49\linewidth}
\includegraphics[width=\linewidth]{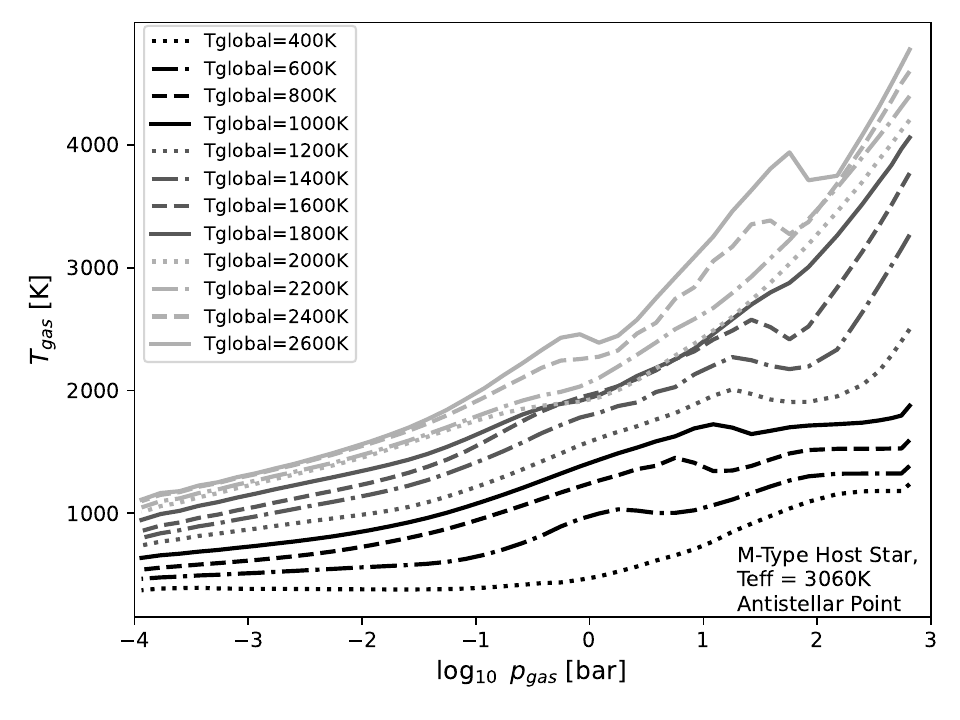}
\end{minipage}
\begin{minipage}[t]{0.49\linewidth}
\includegraphics[width=\linewidth]{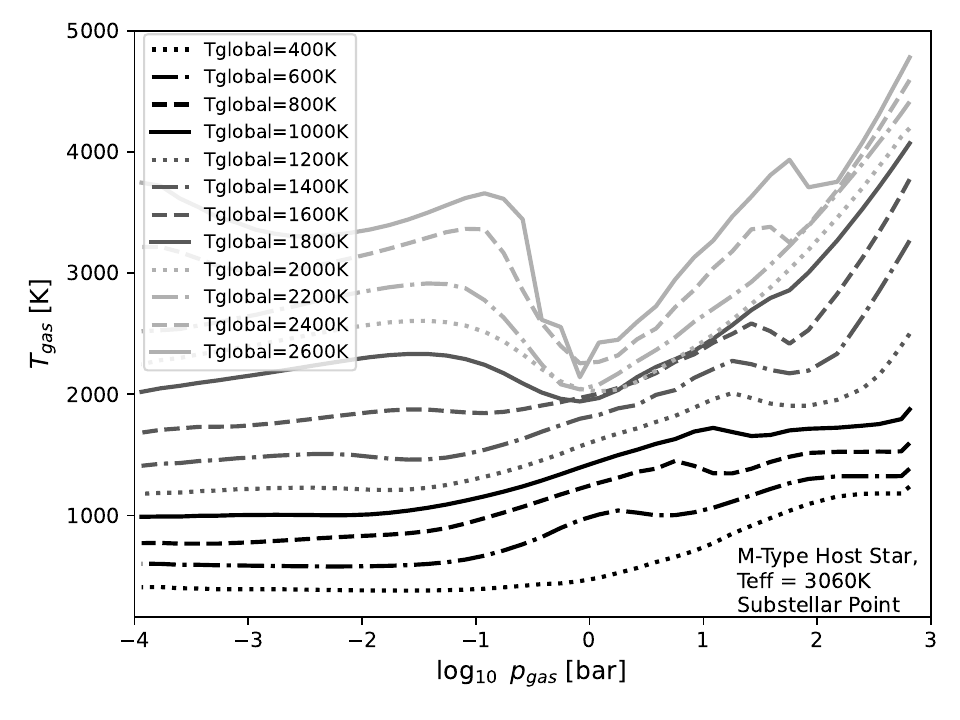}
\end{minipage}
\caption{The A (top), G (middle), and M (bottom) host star results for the 3D AFGKM \texttt{ExoRad}  GCM grid for gas giant exoplanets: 1D (T$_{\rm gas}$, p$_{\rm gas}$)- profiles extracted from the  3D GCM models for the anti-stellar point (nightside, left) and the substellar point (dayside, right).}
\label{fig:1D_AS_SS_PT}
\end{figure*}
\begin{figure*}
\includegraphics[width=1\linewidth]{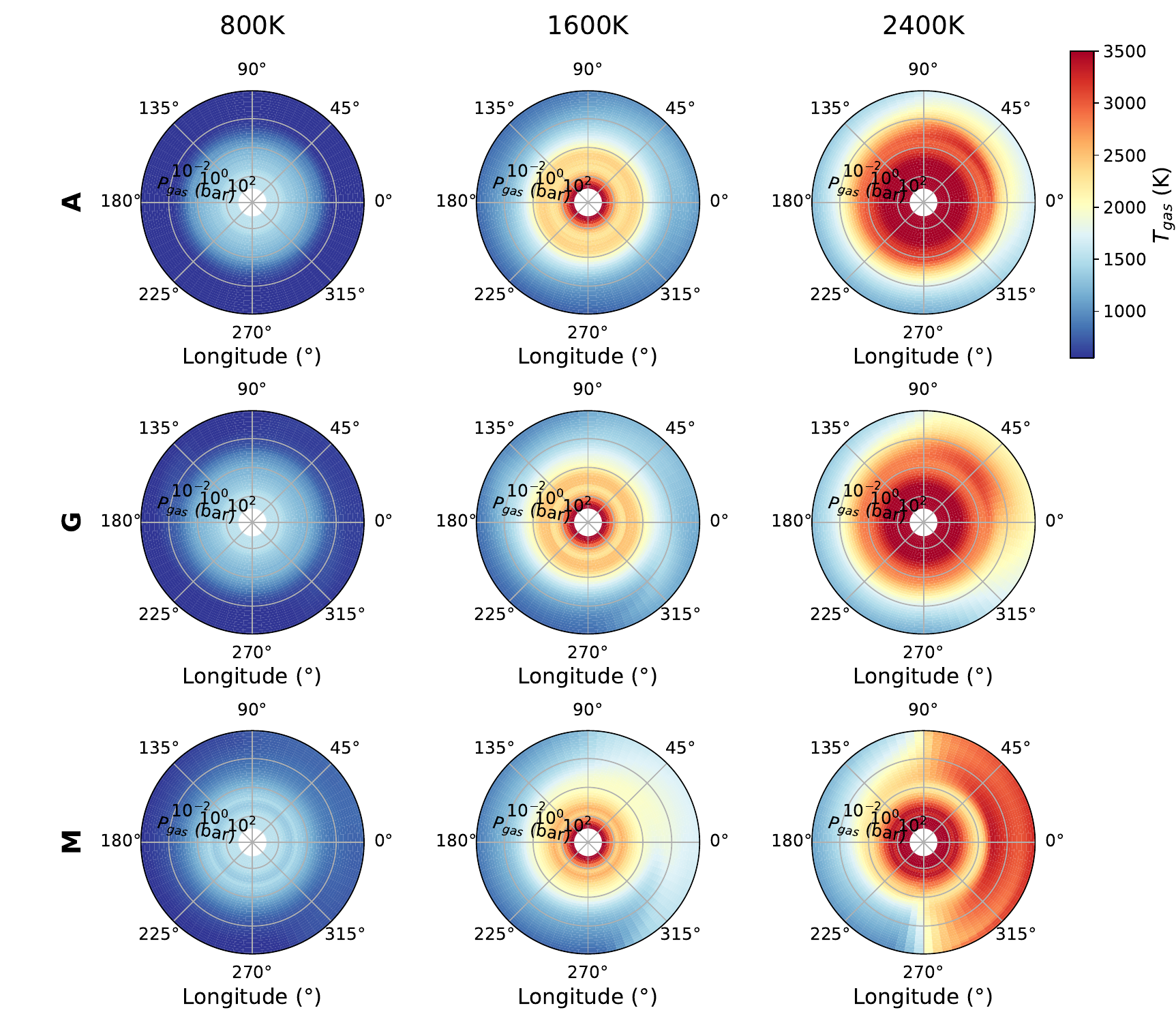}
\caption{The radial gas temperature maps of planets from the 3D AFGKM \texttt{ExoRad} with a global temperature, T$_{\rm global}$ of 800K (left), 1600K (center) and 2400K (right) that orbit different host stars (top to bottom: A, G, M). The (T$_{\rm gas}, p_{\rm gas}$)-maps gas are shown as equatorial slice plots. 
}
\label{fig:EQ_slice}
\end{figure*}
%-
\begin{figure*}
\includegraphics[width=1\linewidth]{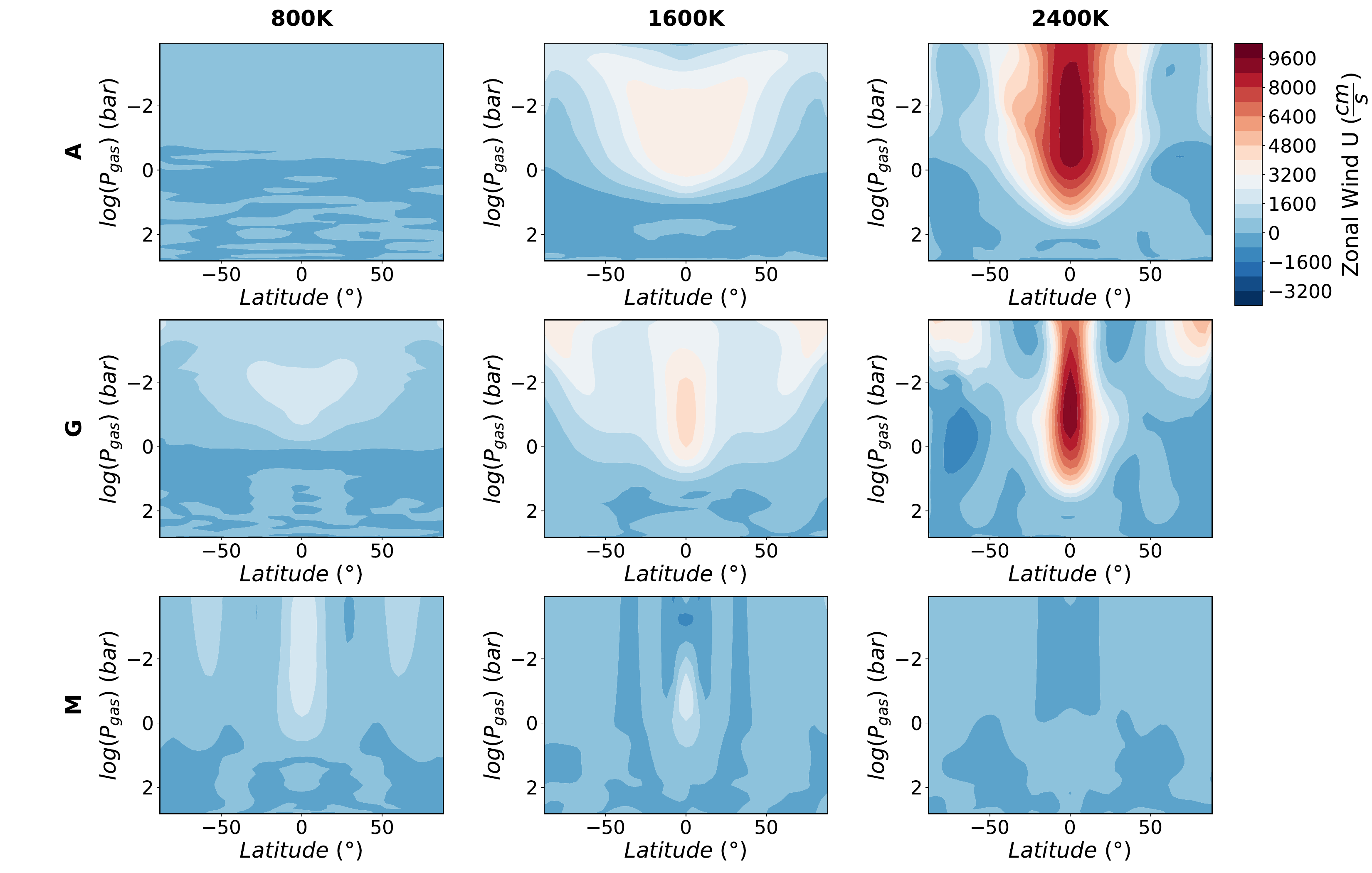}
\caption{Zonal wind speeds, $U$ [cm/s] of planets from the 3D AFGKM \texttt{ExoRad}  with a global average temperature of 800K (left), 1600K (center) and 2400K (right) around different host stars (top to bottom: A, G, M). The values in these latitude-pressure maps are taken at the fixed longitude of the substellar point. Fig.~\ref{fig:EQ_slice} shows the equatorial (T$_{\rm gas}, p_{\rm gas}$)-maps for the same models. }
\label{fig:zonal3x3}
\end{figure*}
%-
\begin{figure*}[htbp]
\centering
\begin{minipage}[t]{0.49\linewidth}
\includegraphics[width=\linewidth]{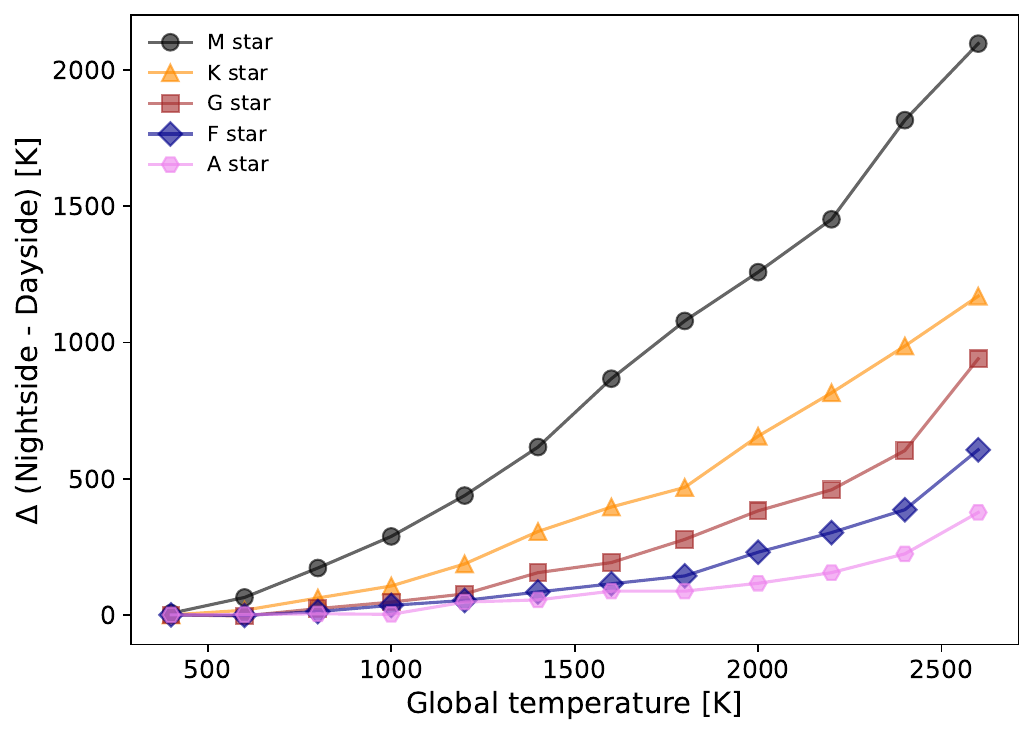}
\end{minipage}
\begin{minipage}[t]{0.49\linewidth}
\includegraphics[width=\linewidth]{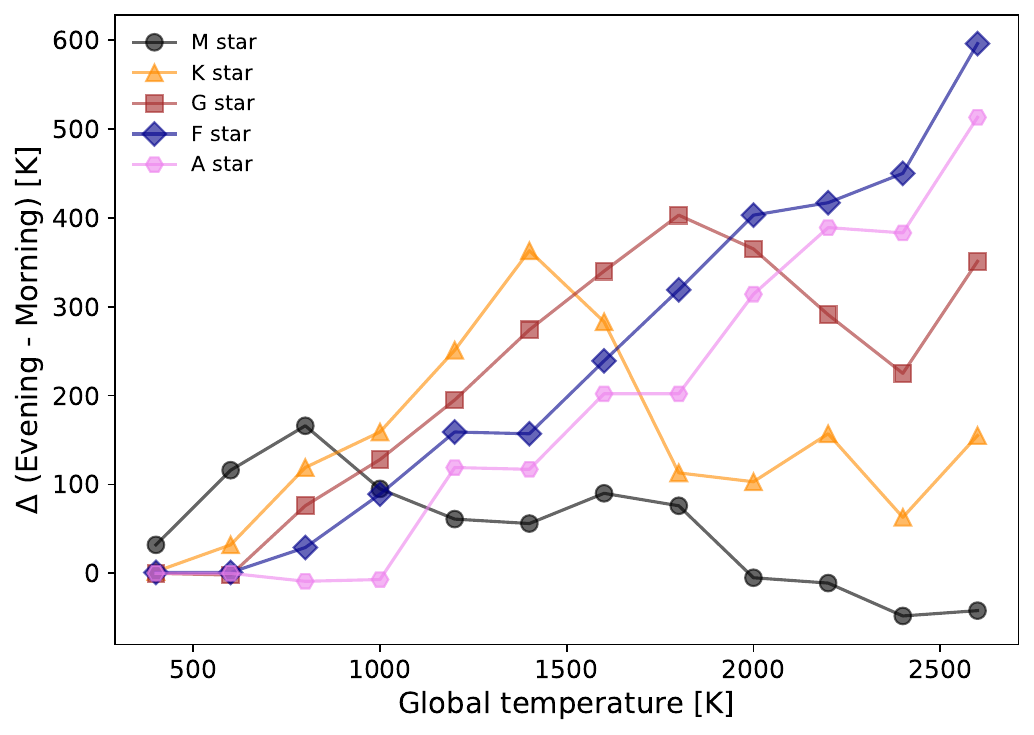}
\end{minipage}
%-
\begin{minipage}[t]{0.49\linewidth}
\includegraphics[width=\linewidth]{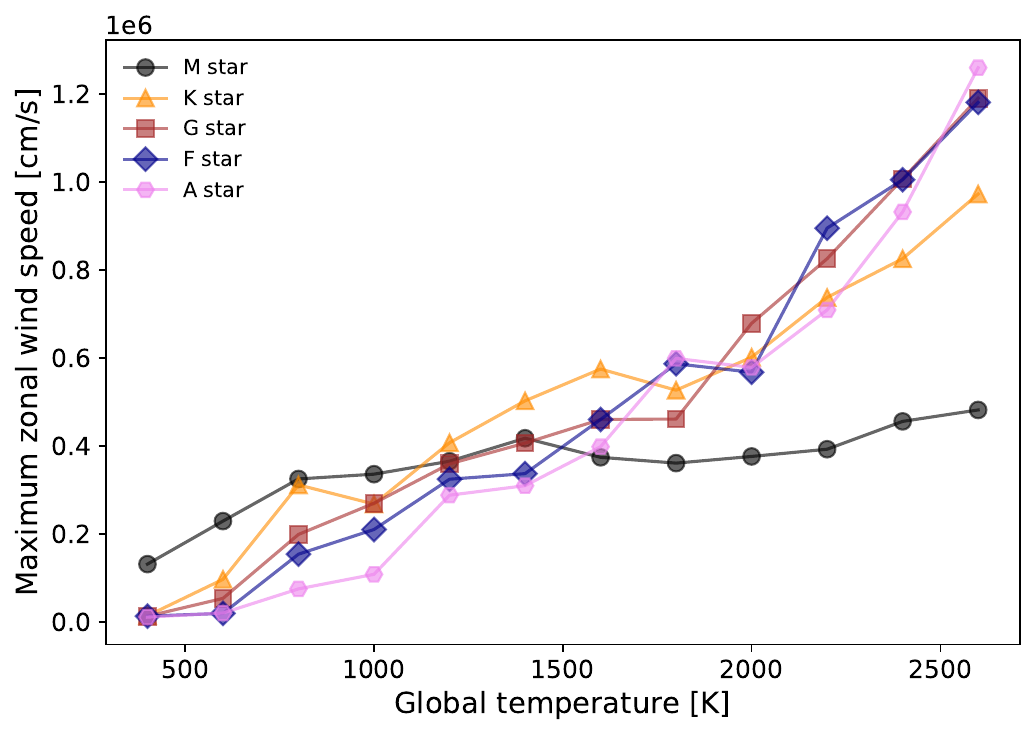}
\end{minipage}
\begin{minipage}[t]{0.49\linewidth}
\includegraphics[width=\linewidth]{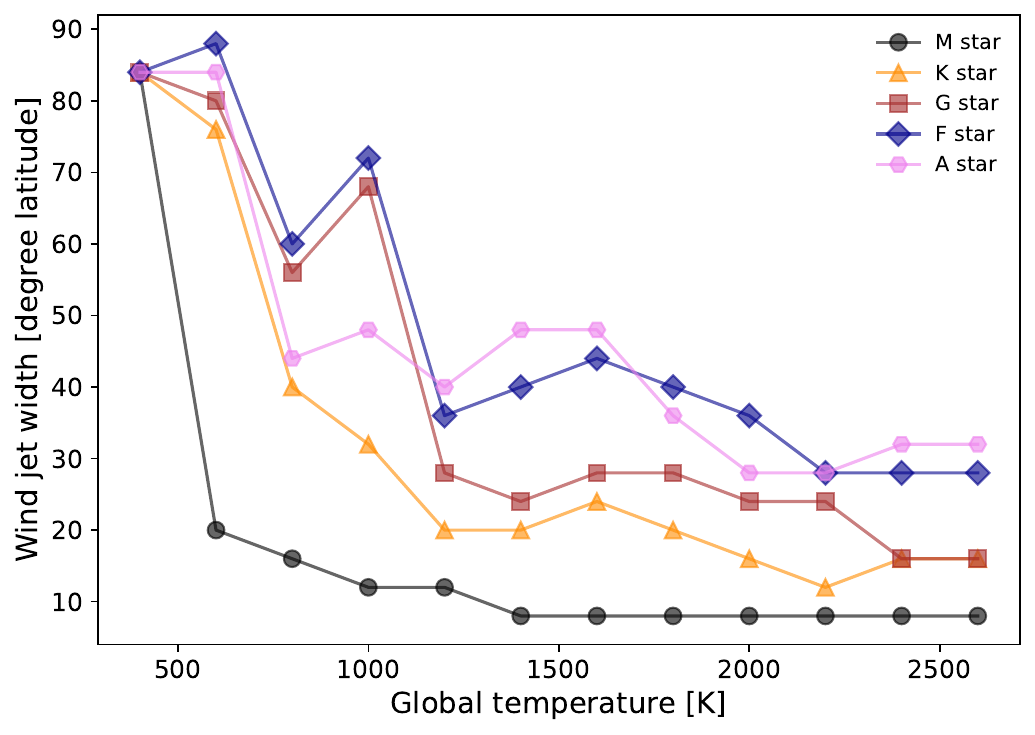}
\end{minipage}
\caption{Basic climate state diagnostics across the 3D AFGKM \texttt{ExoRad}  for the A, F, G, K, and M host stars. {Top left:} Differences between the day and night side averaged temperatures at $p_{\rm gas}=10^{-3}$~bar. {Top right:} Differences between the morning and evening terminator temperatures, averaged over all latitudes and covering $\pm 7.5^{\circ}$ along the morning ($-90^{\circ}$) and evening terminator (($+90^{\circ}$) longitudes at $p_{\rm gas}=10^{-3}$~bar. {  Bottom left:} Maximum zonal mean of the zonal wind speed.  {  Bottom right:} Equatorial wind jet width in degrees latitude.}
\label{fig:Grid_Temp_Windjet_Diag}
\end{figure*}
%-
\section{3D GCM grid \texttt{ExoRad} for gas giants\\ orbiting M, G, K, F, and A stars}
\label{s:ExoRadII}
%-

3D climate grid studies have been extremely useful to explore the impact of diverse physical and chemical processes (e.g., kinetic chemistry \citep{Komacek2019,Baeyens2021}, photochemistry \citep{Baeyens2022} and cloud distribution and ionisation states \citep{HellingCloud2021}) for the warm to hot Jupiter population. \citet{Baeyens2021,Baeyens2022,HellingCloud2021} used the \texttt{ExoRad} 3D climate framework for hot Jupiters \citep{Carone2020} with the dynamical core of \texttt{MITgcm}, employing Newtonian cooling to describe the irradiation of the tidally locked planets. In the last years, \texttt{ExoRad} was updated with \texttt{expeRT/MITgcm} to include full radiative transfer and equilibrium gas phase chemistry \citep{Schneider2022a,Schneider2022b}. This new grid of 3D climate models provides the training data for the ML frameworks presented in this work.
%-
\subsection{Computational set-up for 3D GCM grid}
%-
\paragraph{Star and planet setup:} For comparability with previous work, the new grid assumes a moderately inflated Jupiter sized planet with $\text{R}_{\rm P}=1$ $\text{R}_{\rm Jup}$ and $\text{g} = 10$~m/s$^{2}$ similar to \citet{Baeyens2021,Baeyens2022,Helling2023}, which yields a mass of 0.39 $\text{M}_{\rm Jup}$. For the host stars, stellar parameters for M5V, KV5, GV5, F5V and A5V stars are adopted according to \citet{Pecaut2013} (See Tab.\ref{tab: stellar_params}). We thus extend the grid towards planets around higher mass host stars. The setup is such that the tidally locked planets are placed at a semi major axis so that the global temperatures, $\text{T}_{\rm global}$ [K], averaged over the planet\footnote{$\text{T}_{\rm global}^4= \text{T}_{\text{eq}}^4 + \text{T}_{\rm int}^4$, where $\text{T}_{\rm eq} = \text{T}_{\rm star} \sqrt{\text{R}_{\rm star}/\text{2a}}$} assumes values between 400~K and 2600~K in 200~K steps. The interior temperature, $\text{T}_{\rm int}$ [K], is determined via the parametric fit of \citep{Thorngren2019}. For the planetary parameters ($\text{T}_{\rm global}$ [K], $\text{T}_{\rm int}$ [K], semi major axis $a$ [au], and orbital period $\text{P}_{\rm orb}$ [d]), please see Tables~\ref{tab: A_params} - ~\ref{tab: M_params}.
%-

\paragraph{Radiative transfer and chemistry set up:}
%-
The gas opacities are implemented by correlated-k tabulated opacities combined in 11 spectral bins, corresponding to the S1 resolution in \citet{Schneider2022a} for the following species: H$_2$O (from ExoMol\footnote{\url{https://www.exomol.com}} -- \citealt{Tennyson2016_ExoMol, TennysonEtal2020jqsrtExomol2020}), Na \citep{Allard19_Na_K}, K \citep{Allard19_Na_K}\footnote{We note that the Na and K opacities also include pressure broadening.}, CO$_2$, CH$_4$, NH$_3$, CO, H$_2$S, HCN, SiO, PH$_3$ and FeH, as well as H$^-$ scattering suitable for an ionised atmosphere as listed in \citet[][Tab. 1]{Schneider2022a} excluding TiO and VO. Including them would create an upper atmosphere inversion for a sub-set of planets ($\text{T}_{\text{global}}$ $\geq 1800$~K). The grid is set-up such to represent the impact of advection and irradiation with a similar chemical make-up, including major heating sources across a large range of global temperatures 400~K - 2600 K. Collision-induced absorption for H$_2$--H$_2$ \citep{BorysowEtal2001jqsrtH2H2highT, Borysow2002jqsrtH2H2lowT, Richard2012} and H$_2$--He \citep{BorysowEtal1988apjH2HeRT, BorysowFrommhold1989apjH2HeOvertones, BorysowEtal1989apjH2HeRVRT}, Rayleigh scattering for H$_2$ \citep{Dalgarno1962}, He \citep{Chan1965}, and H$^-$ free-free and bound-free opacities\citep{Gray2008} are included.  Solar metallicity ([M/H] = 0)  and a solar C/O ratio of 0.55 according to \citet{Asplund2009} are adopted for the planetary atmosphere chemistry calculated in LTE. 
%-
\subsection{The 3D AFGKM \texttt{ExoRad} GCM grid} 
\label{ss:3dAFGKM}
%-
The 3D AFGKM \texttt{ExoRad} grid consists of 60 3D GCM models for gas giant exoplanets orbiting A, F, G, K, and M type host stars (blue symbols in Fig.~\ref{fig:motivation}). The grid spans global planetary temperatures T$_{\rm global}=400\,\ldots\,2600$K.
To introduce the new grid, selected 1D gas pressure-temperature profiles for all 60 grid planets (Fig.\ref{fig:1D_AS_SS_PT}) are explored. The 2D equatorial gas temperature maps for planets orbiting selected host stars (A, G, M; Fig.~\ref{fig:EQ_slice}) demonstrate the change in local gas temperatures and atmospheric asymmetries that emerge from our 3D GCM simulations. The zonal wind speeds (Fig.~\ref{fig:zonal3x3}), as drivers of the atmospheric asymmetries, are shown for the same set of host stars and planets. Lastly, in order to be able to characterise the climate state of the 60 3D AFGKM \texttt{ExoRad} GCM grid planets, four properties are introduced: day/night gas temperature difference, morning/evening gas temperature difference, maximum zonal wind speed and width of the wind jet\footnote{We note that for slow rotations, that is in particular for hotter host stars, wind jets can be broad with little horizontal wind jet gradient (Fig.~\ref{fig:zonal3x3}). Thus, 10\% variations in local wind speed can result in a change of the estimate of the wind jet width by $\pm 10$~degrees in latitude.} (Fig.~\ref{fig:Grid_Temp_Windjet_Diag}). These further allow for a comparison within the ensemble of gas giant planet climate states.
%-

\paragraph{ Atmospheric gas temperature-pressure:} The global temperature asymmetries result from the interplay between advection (fluid dynamics) and host star irradiation, where the planet is subject to a strongly asymmetric irradiation field due to tidal locking, as it always faces its host star with the same side. The local gas temperature structure can also be strongly impacted by differences in planetary rotation. A planet with a given global temperature orbiting an M dwarf star is on a shorter orbital period with a faster tidally locked rotation rate compared to a planet with the same global temperature orbiting an A star (see 
Appendix~\ref{apx:completgris} for the complete set of 3D GCM simulations).
%-

\paragraph{ Equatorial wind jets in gas giant exoplanets:}
Tidally locked warm and hot gas giant exoplanets develop a strong equatorial wind jet that efficiently transports warm dayside gases to the cold nightside. The changing gas temperature can cause observable terminator asymmetries.
The equatorial jet changes depending on global system parameters (host star, orbital period; Fig.~\ref{fig:zonal3x3}), resulting in the changing of local gas temperature distributions as shown in Fig.~\ref{fig:EQ_slice}.
The structure of the equatorial wind jet is diagnosed by displaying the maximum equatorial wind speed (bottom left) and the half width of the equatorial jet (bottom right), that is, the latitudinal location at which the wind speed decreases to half of the maximum wind speed of the equatorial jet (Fig.~\ref{fig:Grid_Temp_Windjet_Diag}). Day and night side temperature differences (top left, Fig.~\ref{fig:Grid_Temp_Windjet_Diag}) further elucidate the efficiency of horizontal heat transport. Morning and evening terminator differences (top right, Fig.~\ref{fig:Grid_Temp_Windjet_Diag}) are indicators for potential terminator asymmetries which will eventually be amplified by cloud formation \citep{Helling2023, HellingCloud2021}.
%-

\paragraph{ Climate state characteristics:}
Fig.~\ref{fig:Grid_Temp_Windjet_Diag} characterises the climate of each of the 60 
AFGKM \texttt{ExoRad}  grid 3D GCM models by four properties: day/night gas temperature difference (top left), morning/evening gas temperature difference (top right), maximum zonal wind speed (bottom left) and width of the wind jet (bottom left). These four characteristic properties show that tidally locked gas giant planets around M dwarfs have the least efficient wind flow compared to all other grid planets. The main reasons for this effect is the comparatively fast rotation of the M dwarf grid planets. The equatorial wind jet tends to narrow down and is accompanied by at least one pair of additional strong wind jets for orbital periods shorter than 1.5~days \citep[e.g.][]{Carone2020}. Such short orbital periods are reached with $\text{T}_{\rm global} \geq 600$~K for M dwarf gas giant planets. 
%-

The fast rotating M dwarf gas giant planets thus never develop wind speeds higher than 5000~m/s that cover more than $\pm 20$ degrees latitude of the planet, whereas all other planets generally exhibit a steady increase of wind speed with global planetary temperature. As a consequence, day-to-night side heat transport is very inefficient for M dwarf planets, as evidenced by the day-to-night side temperature differences that steadily increase with higher global temperatures. Conversely, the day-to-night side temperature differences barely increase with global planetary temperature for the slowly rotating A dwarf planets and never exceed 400~K due to the highly efficient, broad wind jets. 
%-

The morning and evening temperature differences show the clearest picture of the wind jet structure. Generally, the contrast between the hotter evening terminator and the cooler morning terminator tends to increase for higher global temperatures in the presence of a dominant equatorial wind jet of at least $\pm$~25~degrees latitudinal extent. In addition, as long as an efficient enough broad wind jet is maintained, planets with faster rotations for a given global temperature tend to have larger terminator contrasts.
%-

The condition for an efficient equatorial wind jet, that is, orbital periods larger than 1.5~days, is no longer fulfilled with global temperatures $\text{T}_{\rm global} \geq$~1400~K for K, $\text{T}_{\rm global} \geq$ ~1800~K for G, $\text{T}_{\rm global} \geq$ 2400~K for F and never reached for A star planets. This is also evident from the wind jet width (bottom right), which narrows down to less than 25~degrees at these global average temperatures.
%-
Thus, F and K dwarf planets show the strongest contrast between the hotter evening terminator and the cooler morning terminator for global temperatures $\text{T}_{\rm global} = 2000 \,\ldots\,2600$~K. These planets still maintain an efficient equatorial wind jet with a latitudinal extent of at least 30~degrees, also for such high temperatures. Once the equatorial wind jet is no longer the dominant wind jet structure, the terminator contrast decreases with increasing global temperature, as evidenced by the M, K and G grid host stars. Thus, the observation of terminator temperature asymmetries is better suited to diagnose the complexity of wind jet structures on tidally locked gas giants than the canonical day-to-night side temperature differences\citep{Cowan2011}.
%-

The zonal wind speed cross-section for selected grid planets further confirms the differences in wind jet structure across the grid (Fig.~\ref{fig:zonal3x3}). The M dwarf planets are clearly an outlier in the \texttt{ExoRad} grid, due to the very inefficient equatorial wind jet accompanied by additional latitudinal jets. The G grid planets generally display an efficient equatorial wind jet, that narrows down to less than 25~degrees for very high global temperatures (\rm $\text{T}_{\rm global} = 2400$~K). Conversely, the A grid planets display a strong and broad equatorial wind jet for intermediate to hot global temperatures. For colder temperatures ($\text{T}_{\rm global}\leq 800$~K) no efficient equatorial wind jet can form on a grid planet, because the equatorial Rossby wave number becomes larger than the planetary radius for slower rotations, that is, $\text{P}_{\rm orb} \gtrsim 20$~days \citep[][Fig.~2]{Baeyens2021}. In this case, no standing Rossby wave can form, which is a necessary condition for the formation of a super-rotating equatorial wind jet \citep{Carone2020,Showman2011,Showman2015}. Thus, for slow-rotating A star planets, an equatorial wind jet with wind speeds larger than 1000~m/s only develops for relatively large global temperatures ($\text{T}_{\rm global} \geq 1200~\text{K}$).
%-

\paragraph{Summary:} The AFGKM \texttt{ExoRad} grid describes the 3D atmosphere structures of 60 gas giant exoplanets on a regularly spaced grid in T$_{\rm global}$. Fig.~\ref{fig:motivation} demonstrates that the ensemble of known planets is reasonably well covered by these 60 models, and that known extrasolar planets seldom fall onto the grid points. Hence, in order to support data interpretation with our models as in \citep{Carone2023,Demangeon2024,Deline2025}, new 3D atmosphere simulations were required. This classical target-focused simulation approach has been successful for individual planets but will reach its limit if ensemble studies need to be conducted, for example, as part of a data interpretation pipeline or a retrieval approach.
%-

The AFGKM \texttt{ExoRad} grid covers a rich variety of atmosphere dynamics that shape the 3D temperature structure of gas giant planets that will be diagnosed in the coming years with multi-messenger observations from PLATO, CHEOPS, HST, JWST, ARIEL, ELT, NewAthena across different wavelengths and with different techniques. It displays diverse equatorial wind jet dynamics with very different shapes and strengths, including cases where a superrotating jet either does not develop or is extremely narrow and thus provides inefficient horizontal heat transport. Capturing the diversity of these climates and their impact on the 3D temperature structure is an ambitious task that, for now, requires time-consuming GCM simulations. The grid simulations thus provide by focusing on the two major forces that shape 3D hot Jupiters, advection and irradiation around different host stars, a large diversity in 3D temperatures and wind structure. In this work, we therefore investigate whether ML techniques are up to the task of capturing the resulting thermodynamic diversity, aiding with ensemble studies. Based on the results of this work, future ML studies for capturing additional complexity can be informed like kinetic chemistry that impact colder planets ($<1000$~K), TiO/VO that would result for ultra-hot Jupiters in upper atmosphere thermal inversions ($>1500$~K), clouds.
%-
\section{Approach to accelerate exoplanet climate modelling}
\label{s:approach} 
\begin{figure}
\centering
\includegraphics[width=0.9\linewidth]{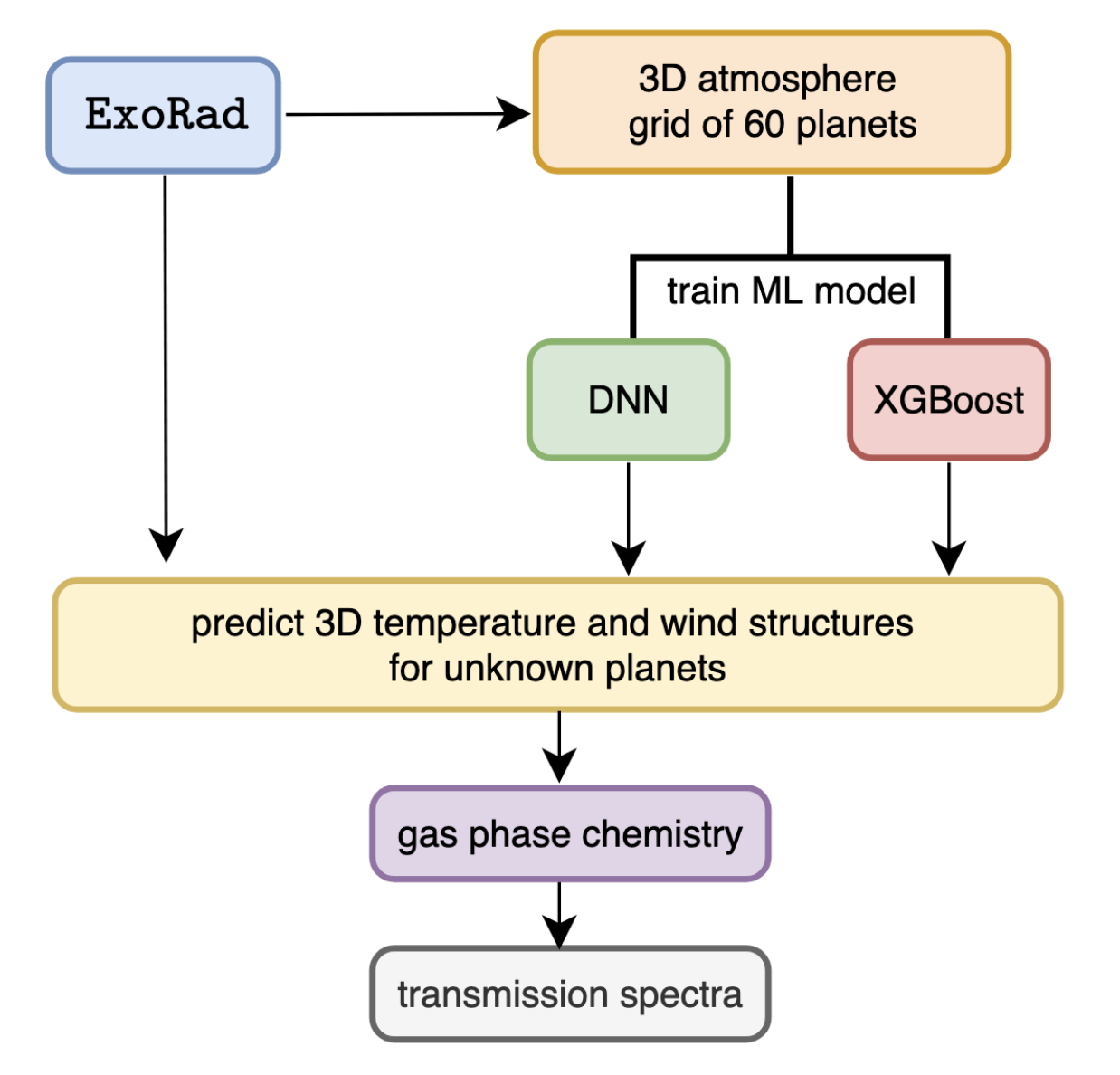}
\caption{
The flowchart to derive ML accelerated 3D atmosphere models: ML models (DNN, XGBoost) are trained to predict the gas temperature and wind structure of exoplanetary atmospheres. Step 1: A grid of pre-calculated 3D planetary atmospheres ($\texttt{ExoRad}$) is used for training. Step 2: The performance of the ML models is evaluated on unseen planets. Their predicted 1D (T$_{\rm gas}$, p$_{\rm gas}$)-profiles are compared to reference ground truth (T$_{\rm gas}$, p$_{\rm gas}$)-profiles ($\texttt{ExoRad}$). Step 3: To assess how prediction errors propagate, equilibrium gas phase chemistry calculations ($\texttt {GGchem}$), performed on selected 1D (T$_{\rm gas}$, p$_{\rm gas}$)-profiles, and simulated transmission spectra ($\texttt{petitRADTRANS}$) are compared.
}
\label{fig:workflow}
\end{figure}

%-
The approach to accelerate exoplanet climate modelling is to develop, train and test suitable ML frameworks, and compare the ML results to ground truths in the form of deterministic GCM results, equilibrium chemistry and transmission spectra (Fig.\ref{fig:workflow}). The whole ML process starts with a dedicated study of the data structure, which utilises the simulation data from the AFGKM \texttt{ExoRad} grid introduced in Sec.\ref{s:ExoRadII}. We explore supervised learning techniques, including decision trees and dense neural networks, to capture the underlying structure of the 3D GCM data. Effective learning of ML models is highly dependent on the quality and structure of training data. Therefore, we start by describing our data representation and necessary pre-processing details before integrating it with the ML models. 
%-
%-
\subsection{Data structure} 
\label{ss:datastruc}
%-
A comprehensive study of data representation is an essential prerequisite for enhancing both training and predictive performance of ML models. The 3D AFGKM \texttt{ExoRad} GCM grid is used for training and testing the ML models. The grid consists of 12 planets orbiting different host stars, each with different global average temperatures $\text{T}_{\rm global} = 400\,\ldots\,2600$~K. These planets orbit 5 stars with different stellar types (Tab.\ref{tab: stellar_params}). Each planetary atmosphere simulation output has 42 pressure layers (p$_{\rm gas} = 650\,\ldots\, 1.16 \times 10^{-4}$ bar), 45 equally spaced points in latitude ($\theta$), and 72 equally spaced points of longitude ($\phi$). This data set-up defines a five-dimensional parameter space, and encompasses a total of $ 8,164,800 = (5 \times 12 \times 42 \times 45 \times 72) $  data points. For each data point, we use the information of the following quantities:
%-
\begin{enumerate}
\item Local gas temperature: $\text{T}_{\text{gas}}$ [K]
\item Zonal wind speed: $\text{U}$ (East-West) [cm/s]
\item Meridional wind speed: $\text{V}$ (North-South) [cm/s]
\item Vertical wind speed: $\text{W}$ (Up-down) [Pa/s]
\end{enumerate} 
%-
For clarity in our notation, the $\textit{grid}$ specifically refers to the 3D AFGKM \texttt{ExoRad} GCM grid, which encompasses the entire ensemble of 60 3D GCMs. A data point, on the other hand, is a singular point identified by five input parameters ($\text{T}_{\rm eff}, \text{T}_{\rm global}, \text{p}_{\rm gas}, \theta, \phi $), each containing detailed information on four corresponding output quantities ($\text{T}_{\rm gas}, \text{U}, \text{V}, \text{W}$).
For easy identification, we will refer to a specific grid planet by stellar type + global average temperature. As an example, the planet orbiting the F5V star with a global average temperature of 2200K will be referred to as: F2200.
%-
\subsection{Choice of ML models}
%-
The aim is to develop an ML model $\mathcal{M}_{\eta}$ that captures a non-linear mapping from five-dimensional input data points $x \in \mathbb{R}^5$ ($x=\{\text{T}_{\rm eff}, \text{T}_{\rm global}, \text{p}_{\rm gas}, \theta, \phi\}$) to four atmospheric output variables $y \in \mathbb{R}^{4}$ ($y=\{\text{T}_{\rm gas}, \text{U}, \text{V}, \text{W} \}$). During training, the ML model $\mathcal{M}_{\eta}$ learns this mapping by minimising the error between the predicted outputs and the reference data, where $\eta$ represents the trainable parameters of a model. Once trained, the model can be queried with any new data point $x^{(q)}$, producing a prediction $\hat{y}^{(q)} = \mathcal{M}_{\eta}(x^{(q)})$. We consider two such models: a dense neural network (DNN) denoted by $\mathcal{M}_{\eta}^{\text{DNN}}$ (see Fig.\ref{fig:NN}) and an Extreme Gradient Boosting (XGBoost) model denoted by $\mathcal{M}_{\eta}^{\text{XGBoost}}$. A key advantage of this mapping is adaptability because all five input parameters $x \in \mathbb{R}^5$ are treated as free variables, and the ML model can predict any arbitrary combination of inputs. That implies predicting the atmospheric gas temperature profile at a specific location or extracting a particular latitudinal slice without predicting every point on a planet is possible. 
%-
Furthermore, this mapping maximises the use of available data to train the ML models, allowing the models to learn something from each data point. However, a challenge arises because the number of required queries for full-planet predictions scales with the resolution of data points. For example, a complete planetary atmosphere output would require $ 42 \times 45 \times 72 = 136,080$ predictions (see Sec.~\ref{ss:datastruc}). ML models are inherently designed to handle such complexity by performing batch inference over large input sets. That means that increasing the number of queries has a negligible impact on computational performance, thus maintaining the practicality of the approach. In fact, for our study $\mathcal{M}_{\eta}^{\text{DNN}}$ and $\mathcal{M}_{\eta}^{\text{XGBoost}}$ generate predictions for all atmospheric output variables across one entire planet in under two seconds on standard CPU hardware (see Sec.\ref{sec:time-comp} for details).
%-

Our ML application strategy is to systematically study and compare different ML models to find the best model for our 3D GCM dataset. We take into account the strengths and limitations of each model, such as overfitting, underfitting, training efficiency, scalability, and generalisation performance. Importantly, we evaluate these factors across a diverse range of planetary conditions, ensuring the broad applicability of the chosen model.
%-

In addition, we explored an alternative strategy that involves four separate ML models, each dedicated to predicting a single atmospheric variable. We also investigated long short-term memory (LSTM) networks ~\citep{hochreiter1997long} as an alternative to dense architecture. However, neither the variable-specific models nor the LSTM approach yielded satisfactory performance. Therefore, this study presents results based on XGBoost and DNN trained to simultaneously predict the four atmospheric variables ($\text{T}_{\rm gas}, \text{U}, \text{V}, \text{W}$).
\subsection{Performance evaluation}
\label{ss:error}
%-
\paragraph{ Training and testing strategies:}
%-
We follow a standard strategy for model development by dividing the data into $80\%$ training and $20\%$ validation. Once we have good prediction accuracy with the trained models, we evaluate model performance on five unseen real planetary-like test planets (e.g., WASP-121 b *), which are not used during training or validation. Additionally, we conduct a sanity check to estimate model performance on grid-aligned test points. For this, we excluded a pair of planets (e.g., [F1800, K800]) from the training set and treated them solely for testing. While the sanity check is not strictly required, it serves as a helpful baseline to verify the predictive ability of the ML models under ideal conditions, i.e., the test data lie within the same structured parameter space as the training data. 
%-

\paragraph{ {Error analysis:}} This study uses the root mean squared error (rMSE) and coefficient of determination ($\text{R}^{2}$) \citep{ozer1985correlation, nagelkerke1991note, di2008coefficient, chicco2021coefficient} as primary statistical metrics to evaluate the performance of the ML models for our regression task. Both these measures assess the differences between actual and predicted values, in our case, for all atmospheric variables corresponding to the 3D data of a planet.
%-
The rMSE metric is a good indicator for comparing different model structures, while $\text{R}^{2}$ reflects how well the prediction fits the ground truth. Mathematically $\text{R}^{2}$ is defined as: $\text{R}^{2} = 1 - \frac{\sum_{i = 1}^{n}(y_{i} - \hat{y}_{i})^{2}}{\sum_{i = 1}^{n}(y_{i} - \bar{y}_{i})^{2}} \, ,$ where $y_{i}$, $\hat{y}_{i}$ represent the actual (observed) value of the target variable and the predicted value of the target variable obtained from the ML model, respectively. $\bar{y}_{i}$ defines the mean of the actual target values for all data points, and $n$ is the total number of data points. 
%-

The $\text{R}^{2}$ score ranges from ($-\infty$, 1], where values closer to one reflect high prediction accuracy, which means high ML model performance, while values near zero imply that the model has limited explanatory power, leading to high prediction error. A negative $\text{R}^{2}$ suggests that the ML model performs worse than a mean prediction.
%-

We initially calculated rMSE and $\text{R}^{2}$ for training and testing atmospheric variables in normalised space to assess the performance of the ML model. A high $\text{R}^{2}$ value indicates that the model effectively captures most of the data variance. However, we would anticipate a high prediction error once the predicted values are transformed into the original physical scale. Even minor prediction errors in normalised space can translate into significant deviations in the original scale, leading to higher rMSE and lower $\text{R}^{2}$. Therefore, we report rMSE and $\text{R}^{2}$ on the original scales to provide a more realistic measure of prediction error.
%-
\subsection{Testing unknown planets}
%-
\paragraph{ {Selection of unknown test planets:}}
%-
\begin{figure} 
\centering
\includegraphics[width=1\linewidth]{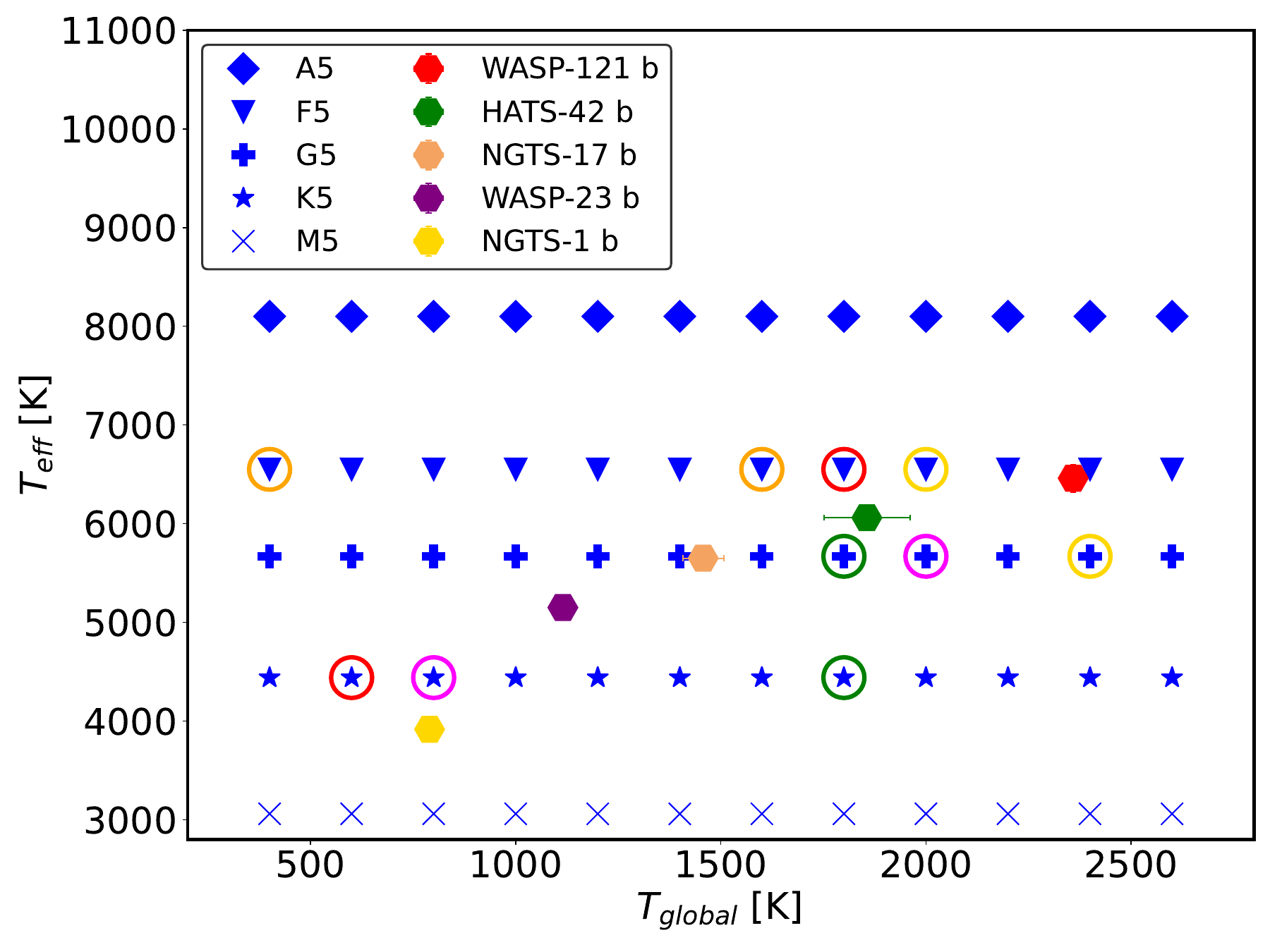}
\caption{3D AFGKM \texttt{ExoRad} GCM grid of all grid planets (blue), all randomly selected pairs of test planets removed from the training set during the sanity check (i.e., test) of ML models are highlighted with differently coloured circles for each pair, and the five target planets are put in the context of the grid.}
\label{fig:TestTargetsComb}
\end{figure}
%-
The key question in working with ML models is how useful the resulting ML predictions are in supporting the interpretation of data from current and upcoming space telescopes. Beyond validating the numerical performance of the ML models, we are particularly interested in the physical implications, specifically, whether the atmospheric chemistry derived from the ML predictions matches that of the original 3D AFGKM \texttt{ExoRad} GCM (Sec.\ref{s:ExoRadII}). To explore this,  five representative planets are selected from the AFGKM \texttt{ExoRad} GCM grid that reflect the properties of many exoplanets expected to be characterised in the coming years (Fig.~\ref{fig:motivation}). Explored are ML/\texttt{ExoRad} differences for the PLATO target planets WASP-121 b, HATS-42 b, NGTS-17 b, WASP-23 b, and NGTS-1 b (see Sec. \ref{s:results2}). We note that at least a subset of planets located in the PLATO's LOPS2 field either have been or will be observed with other current and future telescopes to yield complementary data \citep{Eschen2024}. Among the selected planets, WASP-121~b is the best characterized so far \citep[][e.g.]{Davenport2025,Prinoth2025,Changeat_2023}. NGTS-1 b represents the rare example of a gas giant orbiting an M dwarf star, the atmosphere of which is ideally observable with JWST. Characterising the atmospheres of such planets is an emerging research topic to illuminate the planet formation process around low-mass stars \citep{Kiefer2024,Canas2025}. 
%-
The planets in the context of the grid can be seen in Fig. \ref{fig:TestTargetsComb}. It is important to note that it is expected that planets that lie closer to a grid planet (e.g., WASP-121 b) will perform better than planets that are further away (e.g., WASP-23 b) due to the grid structure of the training data.
%-

\paragraph{Modifications:}
%-
The planets and their respective properties selected for this study are listed in Tab. \ref{tab:TestPlanets}. All grid planets have a fixed mass (0.39~$\text{M}_{\text{j}}$), a fixed radius (1~$\text{R}_{\text{j}}$) 
%-
and therefore a surface gravity of $10 \, \text{ms}^{-2}$ while the actual planets have the surface gravities: WASP-121 b ($8.827 \, \text{ms}^{-2}$),  HATS-42 b ($23.82 \, \text{ms}^{-2}$), NGTS-17 b ($12.82 \, \text{ms}^{-2}$), WASP-23 b ($20.89 \, \text{ms}^{-2}$), NGTS-1 b ($11.9 \, \text{ms}^{-2}$), 
%-
so we also assigned the same values to our five target planets for consistency. Furthermore, as explained in Sec.~ \ref{s:ExoRadII}, the calculated values for $\text{T}_{\rm global}$ are obtained by fixing the semi-major axis in the \texttt{ExoRad} simulation. Upon examining the available numbers, we observed minor inconsistencies between the semi-major axis and the resulting global temperatures reported in the literature. For NGTS-17 b, we choose $\text{T}_{\rm global} = 1515 \text{K}$ as our model input. To highlight that the planets are a modified version of themselves we will refer to them by adding '*'. e.g. "WASP-121 b*"
%-
\begin{table} 
\caption{Test cases - selected target planets and their parameters.}
\label{tab:TestPlanets}
\centering
\begin{tabular}{l c c c c}
\hline\hline
 {Planet} &  {$\text{T}_{\rm global} (\rm K)$}&  {$\text{T}_{\rm star} (\rm K)$}&  {$\text{M}_{\rm P}$ ($\text{M}_{\rm j}$)}& {$\text{R}_{\rm P}$ ($\text{R}_{\rm j}$)}\\
\hline
WASP-121 b & $2359$ & $6460 \pm 140$ & $1.184$ & $1.865$\\
HATS-42 b& $1856 \pm 105$ & $6060 \pm 120$ & $1.8$ & $1.4$ \\
NGTS-17 b& $1457 \pm 50$ & $5650 \pm 100$ & $0.76$ & $1.24$ \\
WASP-23 b& $1115$ & $5150 \pm 100$ & $0.917$ & $1.067$ \\
NGTS-1 b& $790 \pm 20$ & $3916 \pm 71$ & $0.812$ & $1.33$\\
\hline
\end{tabular}
\tablefoot{Target planets selected as test cases with values for stellar temperature ($\text{T}_{\text{eff}}$), global average temperature ($\text{T}_{\text{global}}$), planetary mass ($\text{M}_{\text{P}}$) and planetary radius ($\text{R}_{\text{P}}$) \citep{exoplaneteu}. }
%-
\end{table}
%-

\paragraph{ {Prediction assessment:}}
%-
To evaluate ML model performance, we directly compare predictions of gas temperature and wind patterns against the corresponding outputs from the reference \texttt{ExoRad} GCM results. This comparison is further facilitated by analysing 1D ($\text{T}_{\rm gas}$–$\text{p}_{\rm gas}$) profiles, generating 2D wind maps, and quantifying the level of agreement using $\text{R}^{2}$ as our error metric.
%-

\paragraph{ {Gas-phase chemistry:}}
%-
To understand the impact of the temperature deviations, we first calculate the equilibrium chemistry abundances of essential gas-phase species at the morning and evening terminators with \texttt{GGCHEM} \citep{GGchem}. In Figs. \ref{fig:Wasp121_Morning} and \ref{fig:NGTS1_Morning}, we show the abundances of the following atoms and ions: H$_2$O, CO, CO$_2$, H, H+, CH4, NH$_3$, HCN, SiO, TiO, MgO, FeO, SiO$_2$, and TiO$_2$. Substantial differences for molecules like TiO, MgO, and SiO$_2$ may affect cloud formation, and deviations for major gas-phase opacity sources like CO, H$_2$O, and TiO may impact the observable spectrum.
%-

\paragraph{ {Transmission spectra:}}
%-
To understand how strongly differences in chemical abundances could affect observations, we used \texttt{petitRADTRANS} \citep{p_Mollière_2019} to generate synthetic spectra based on both \texttt{ExoRad} and ML-predicted atmospheric states. The resulting spectral differences are then compared with the expected observational precision of JWST \citep{PandExo_2017}, Hubble \citep{Changeat_2023}, and PLATO \citep{PLATO_filter}, allowing us to evaluate whether the discrepancies may be detectable. The transmission spectra shown in Figs. \ref{fig:Wasp121_Spectra}, \ref{fig:NGTS1_Spectra}, \ref{fig:HATS42_Spectra}, \ref{fig:NGTS17_Spectra}, and \ref{fig:WASP23_Spectra} are calculated at the morning and evening terminator and then averaged over each limb.
The Hubble resolution was taken from \citep{Changeat_2023}, which also observed WASP-121 b in this spectral range. The error for PLATO is an estimate from \citep{PLATO_filter} where the feasibility of the blue and red filters was discussed. The JWST MIRI LRS and NIRSpec PRISM spectral accuracy was calculated by performing Pandexo \citep{PandExo_2017} simulations for two transits and assuming 20 pixels per bin. Just as with the \texttt{ExoRad} runs, the planetary mass and radius were changed to 0.39 $\text{M}_j$ and 1 $\text{R}_j$.
%-
\section{Machine learning models}
\label{s:method} 
%-
This section introduces the ML models to accelerate 3D exoplanet climate modelling and the corresponding training strategies. Our approach involves training, testing, and evaluating various ML models to determine which is most effective for our data purpose, that is, representing the 3D GCM thermodynamic structure.
%-
\subsection{Decision trees (XGBoost)}
%-
Decision trees (DT) are algorithms commonly used for classification and regression, which iteratively divide data into substructures based on feature values to create a tree-like structure \citep{rokach2005decision, kingsford2008decision, de2013decision}. Each tree consists of a node, a branch, and a leaf, where a node represents a decision based on a feature, a branch represents an outcome of that decision, and a leaf node represents a final prediction. The aim is to partition the data to maximise the similarity of the target variable within each subset based on criteria such as the mean squared error for the regression task. This work explores a special kind of DT algorithm, named XGBoost, which is based on the foundation of ensemble learning \citep{chen2015xgboost, chen2016xgboost}. It consists of multiple shallow decision trees (called weak learners) combined sequentially to obtain a strong predictive model. The learning process in XGBoost is recursive, as a new decision tree of the current state fits the residual error obtained from the tree of the previous state. That means that the current decision tree effectively corrects the prediction mistakes of the previous trees, engaging the model in the learning process. Although XGBoost is fast to train and can provide good prediction accuracy for low-dimensional data points, it is prone to overfitting and may not generalise well to unknown test data. Hence, this work thoroughly investigates the performance of the DNN and XGBoost algorithms and compares their performance on training and testing data. 
%-

\paragraph{ Hyperparameter optimisation:}
%-
The performance of the XGBoost algorithm depends on several model parameters, such as the number of estimators, the depth of the tree structure, and others. Instead of fixing these parameters to arbitrary values to train a model, finding the best combination of parameters to obtain the model corresponding to the best choice is more effective. Finding the best set of parameters is known as hyperparameter optimisation. For our data, during training, we also performed hyperparameter optimisation using the ``Random Search" scheme to optimise the hyperparameters for the XGBoost model. For the extensive search of parameters, we define certain boundaries for each parameter (see Tab. \ref{tab:XG}), and the search algorithm provides the best combination of parameters by exploring this parameter space. The search was performed by fitting five folds for each of the 50 candidates and totalling 250 fits. It is essential to note that the optimal parameter set could differ depending on the prior choice of the boundaries. 
%-
\begin{table} 
\caption{Parameter space explored in XGBoost model fine-tuning.}
\label{tab:XG}
\centering
\begin{tabular}{l c}
\hline\hline
 {Parameter} &  {Range}\\
\hline
$n\_{\text{estimators}}$ & 500, 1000, 1500, 3000, 5000 \\
$\text{max}\_{\text{depth}}$ & 3, 4, 5, 6, 9 \\
$\text{learning}\_{\text{rate}}$ & 0.01, 0.05, 0.1, 0.3, 0.5  \\ 
$\gamma$ & 0, 0.1, 0.2 \\ 
subsample& 0.4, 0.6, 0.8, 1.0\\
$\text{reg}\_{\alpha}$ & 0, 0.01, 0.1, 0.5, 1\\
$\text{reg}\_{\lambda}$ & 1, 1.5, 2, 3\\
$\text{min}\_{\text{child}}\_{\text{weight}}$ & 1, 3, 5\\
$\text{colsample}\_\text{bytree}$ & 0.6, 0.8, 1.0\\
\hline
\end{tabular}
\tablefoot{
$n\_{\text{estimators}}$ - the number of trees (i.e., boosting rounds) to build sequentially; $\text{max}\_{\text{depth}}$ - maximum depth of a tree, controlling model complexity; $\text{learning}\_{\text{rate}}$ - controls how much the model adjusts with each new tree. Smaller values make the model learn slowly but more robustly; $\gamma$ - minimum loss reduction required to make a split, acting as a regularization term; $\text{reg}\_{\alpha}$ - Lasso regularization on weights; $\text{reg}\_{\lambda}$ - Ridge regularization on weights (controls model complexity); $\text{subsample}$ - a fraction of training data used per tree, controlling overfitting; $\text{min}\_{\text{child}}\_{\text{weight}}$ - the minimum sum of instance weight needed in a child node; $\text{colsample}\_\text{bytree}$ - a fraction of features used in each tree ~\citep{chen2015xgboost, chen2016xgboost}.}
\end{table}
%-
\subsection{Neural network (DNN)}
%-
With the goal of predicting gas temperature and wind speeds at arbitrary query points, i.e., locations not included in the training data, this study employs a DNN architecture to encapsulate the relationship between input data points and atmospheric variables. The problem is formulated as a regression problem. The choice of neural network model is guided by the fact that DNN is well-suited to solve regression tasks and to capture complex, nonlinear relationships with relatively low-dimensional data structures \citep{bishop2006pattern, bartlett2021deep}. Further, we aim to minimise architecture complexity by limiting the number of neurons per layer and the total number of trainable parameters. With such architectures, defined as ``lightweight,'' reducing training time and facilitating rapid predictions is achievable.
%-
\begin{figure} 
\centering
\includegraphics[width=1\linewidth]{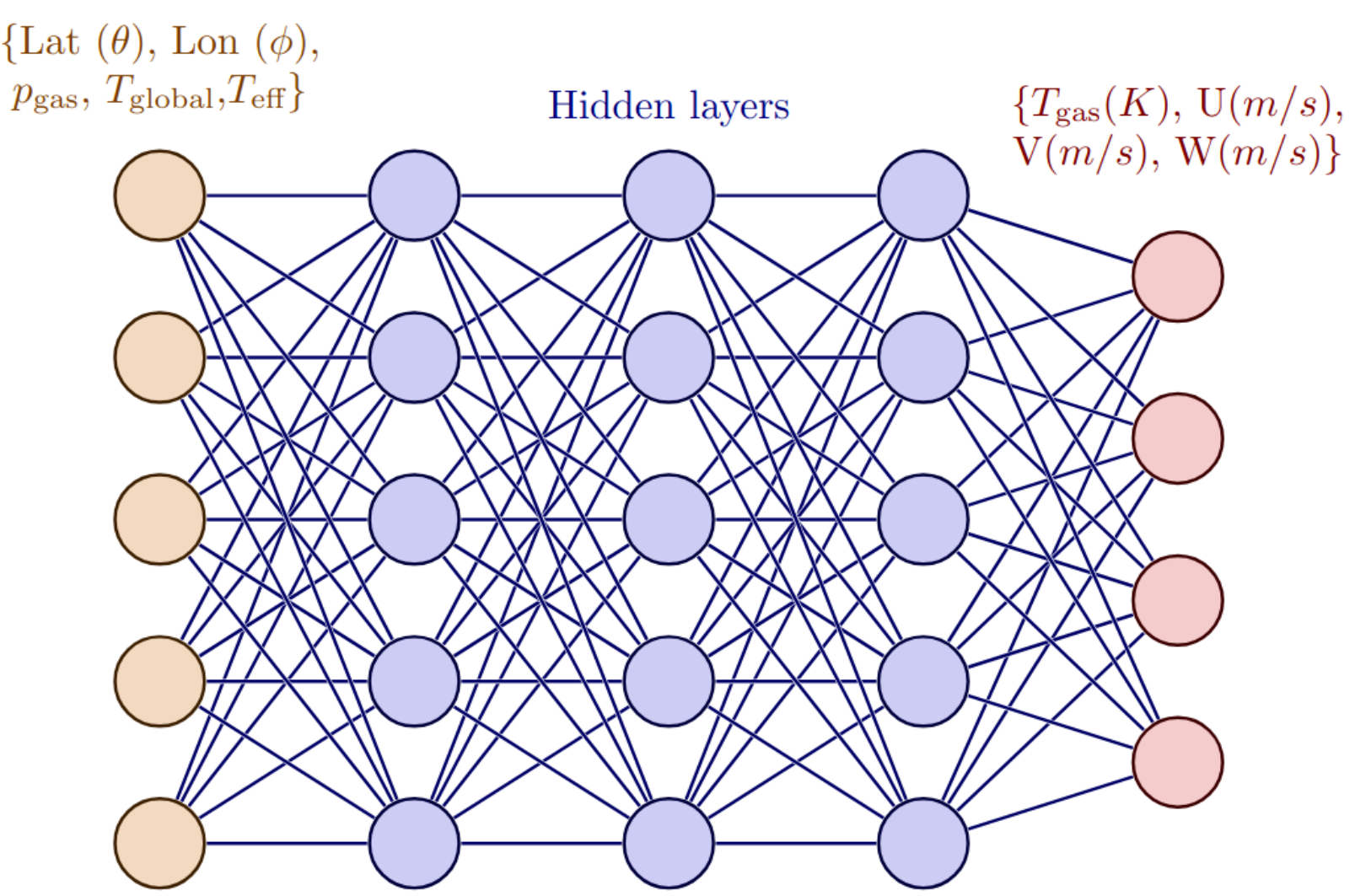}
\caption{
Schematic diagram of a dense neural network that maps the five input parameters ($\text{T}_{\text{eff}}$, $\text{T}_{\text{global}}$, $\text{p}_{\text{gas}}$, $\text{Lat} (\theta)$, $\text{Lon} (\phi)$) to the four output parameters ($\text{T}_{\text{gas}}$, $\text{U}$, $\text{V}$, and $\text{W}$) with fully connected hidden layers in between. We finally opted for six layers and 256 neurons each in this work.
}
\label{fig:NN}
\end{figure}
A DNN, a fundamental type of neural network architecture, comprises dense layers (i.e. fully connected) only. The dense layers connect each neuron to every neuron in the subsequent layer, ensuring immediate and efficient information flow. A standard DNN architecture connects the input and output layers through one or more of these intermediate hidden layers (see Fig.~\ref{fig:NN}). The input layer acquires input data, e.g., data points for our study, which are then transformed by the hidden layers through a sequence of linear and non-linear operations, called forward propagation. Linear transformations are performed using weights and biases, while nonlinear transformations are introduced through activation functions \citep{narayan1997generalized, sibi2013analysis, sharma2017activation, ramachandran2017searching}. The activation function represents a non-linear transformation and can take various functional forms, for example $\texttt{Sigmoid}$, Rectified Linear Unit ($\texttt{ReLU}$), Exponential Linear Unit ($\texttt{ELU}$) or $\texttt{Tanh}$. The output layer, which provides the final predictions, has the number of neurons corresponding to the number of prediction targets (in our case, four atmospheric variables, local gas temperature, zonal, meridional, and vertical wind). The loss function is used to measure the correctness of the predictions. The loss is usually defined as the mean squared error (MSE) between the predicted and actual values for regression tasks. During the initial stages of training, the loss is expected to be high and will gradually decrease with a fine-tuned model. Fine-tuning the model is done by iteratively adjusting its weights and biases to reduce it through backpropagation combined with gradient descent. Although composed only of dense layers, such simple networks can model complex, non-linear relationships, making them highly effective for regression tasks \citep{hornik1989multilayer, bartlett2021deep}.
%-

\paragraph{ {ML model design and training strategy:}} Building on the objective of designing a lightweight DNN architecture, our study focuses on systematically identifying an ML model that balances predictive performance with computational efficiency in both training and inference. Initially, we start with a simple ML model (e.g., with two fully connected layers), gradually increase complexity in terms of the number of layers (up to ten) and neurons (up to 512), and obtain prediction accuracy on the test dataset for each ML model. The training of the ML models has been carried out for a maximum of 1000 training iterations (epochs), with a learning rate of $10^{-4}$, a batch size of 128, the $\texttt{RMSProp}$ optimiser, and using the mean squared error as the loss function. Early stopping has been introduced and is set with a maximum of 1000 epochs, with the constraint that the training process stops if the validation loss does not improve over 25 consecutive epochs. During training, $20 \%$ of the ExoRad GCM data has been used as the validation set. Incorporating early stopping into the training of ML models helps to prevent overfitting and reduces overall training time. Among all the variations, we found that a DNN model with six fully connected (linear) layers,  followed by five $\texttt{tanh}$ activation (nonlinear) layers, each having 256 neurons, provides the best predictive performance while maintaining a reasonable model size, aligning with our goal of creating a lightweight yet accurate model.
%-

\section{ML modelling results for an accelerated\\ 
exoplanet climate modelling}
%-
\label{s:results}
%-
The results from applying optimised ML models (DNN, XGBoost) to predict the 3D exoplanet atmosphere gas temperature and wind structure are presented here. We adopt two complementary test approaches to thoroughly evaluate the performance of ML models. First, we remove pairs of planets from the training dataset and use them as test data. This allows us to examine the ability of the ML model to generalise unseen but on-grid planetary conditions, as presented in Sec.\ref{s:on-grid val}. Second, we evaluate the ML models using five exoplanets chosen as representatives of the LOPS2 PLATO target field in Sec.\ref{s:results2}. Since these planets lie between the training grid points, this approach uniquely probes the models' interpolation capabilities and their relevance to realistic observational scenarios. This section concludes with an analysis of ML-induced differences in the resulting chemical equilibrium gas compositions and the transmission spectra.
%-
\subsection{Sanity check: On-grid test}
\label{s:on-grid val}
\begin{figure} 
\centering
\includegraphics[width=0.9\linewidth]{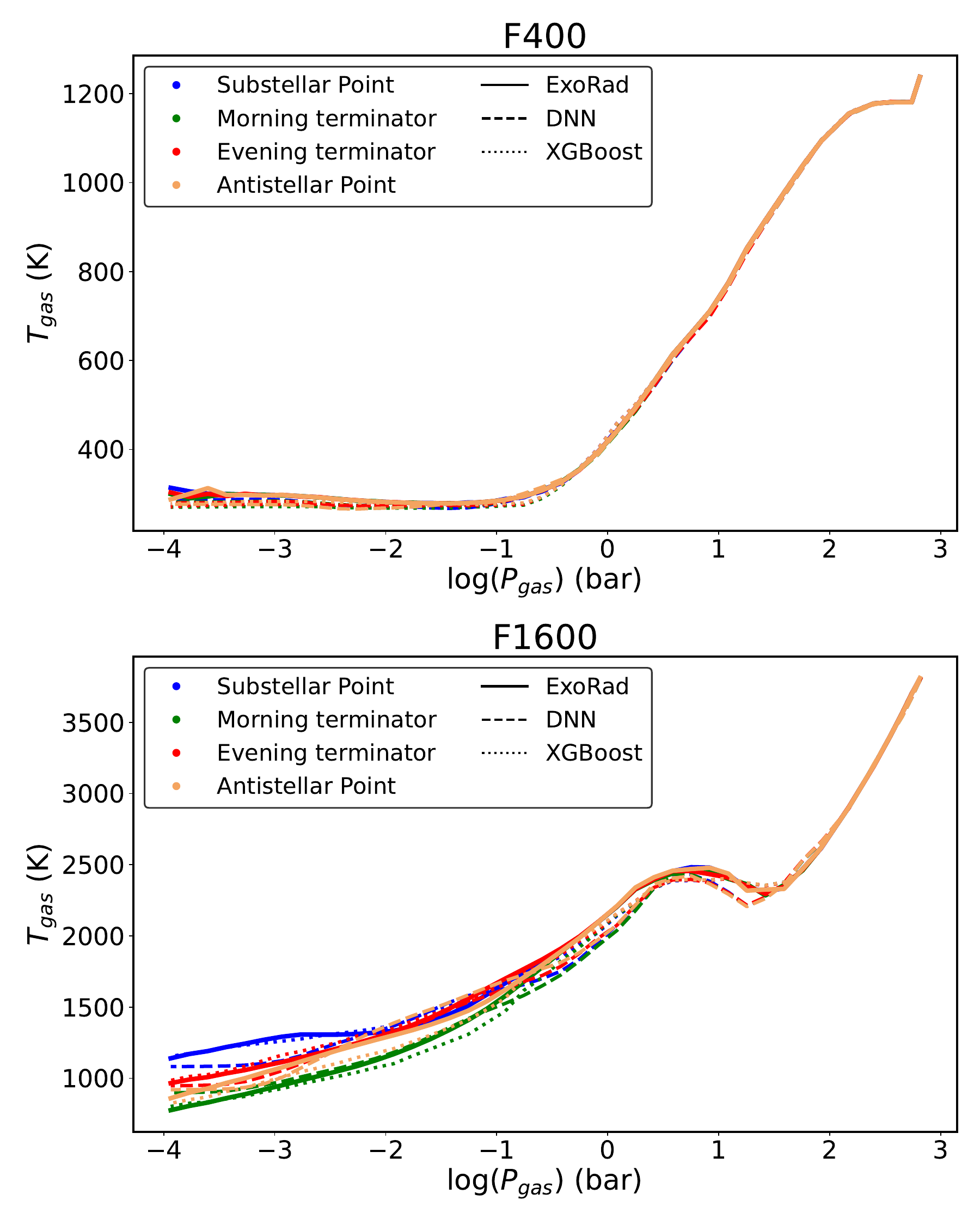}
\caption{1D (T$_{\rm gas}$, p$_{\rm gas}$)-profiles of the two planets chosen for test (F400, F1600). Showing the substellar point (blue), the antistellar point (orange), the morning terminator (green), and the evening terminator (red). The solid line shows the \texttt{ExoRad} simulated values, the dashed line the DNN prediction, and the dotted line the XGBoost prediction.}
\label{fig:1DTeststackF400F1600}
\end{figure}

%-
As outlined in Sec.~\ref{s:approach}, to examine the prediction accuracy of ML models, we withhold a pair of planets (e.g., F1800 and K800) from the training set. Treating this pair as an unseen test sample, we then compare the ML predictions with the corresponding \texttt{ExoRad} outputs. We prepared five pairs of planets (F1800, K800), (G1800, K1800), (F2000, G2200), (G2000, K600), and (F400, F1600) as shown in Fig. \ref{fig:TestTargetsComb} and evaluated the statistical error using rMSE and $\text{R}^{2}$ metrics between the predicted and original values of atmospheric variables. The resulting errors of the local gas temperature for all these five planets are listed in Tab. \ref{tab:ResultsPlanetsMulti}. For DNN and XGBoost, $\text{R}^{2}$ values are nearly one and show relatively consistent model performance across the grid. Fig.~\ref{fig:1DTeststackF400F1600} displays the 1D (T$_{\rm gas}$, p$_{\rm gas}$) profiles of the pair of planets F400 and F1600 at four key locations, i.e., the substellar point, the antistellar point, and the two terminators. The close agreement between the ML-predicted and actual profiles from $\texttt{ExoRad}$ confirms that planet-wide high $\text{R}^{2}$ value translates into equally accurate individual predicted 1D profiles. The same visualisation for the other planet pairs exhibits similar qualitative behaviour (see Figs.~\ref{fig:1DTeststackF1800K800} - \ref{fig:1DTeststackG2000K600}), underscoring the robustness of ML predictions carried out here.
%-
%-
\begin{table} 
\caption{Error metric for five pairs of randomly selected planets.}
\label{tab:ResultsPlanetsMulti}
\centering
\begin{tabular}{l l c c }
\hline\hline
 {Planets} &  {Error Metric} &  {DNN} &  {XG} \\
\hline
F1800   & $\text{R}^{2}$   & 0.9971    & 0.9971    \\
K800    & rMSE      & 48.7536   &  49.1047  \\
\hline  
G1800   & $\text{R}^{2}$   & 0.9817    & 0.9813    \\
K1800   & rMSE      &117.3391   & 118.8262  \\
\hline
F2000   & $\text{R}^{2}$   & 0.9835    & 0.9843    \\
G2200   & rMSE      & 121.7374  & 118.7824  \\
\hline
G2000   & $\text{R}^{2}$   & 0.9931    & 0.9953    \\
K600    & rMSE      & 77.0609   & 63.4323   \\
\hline
F400    & $\text{R}^{2}$   & 0.9975    & 0.9987    \\
F1600   & rMSE      & 44.6912   & 32.3724   \\
\hline
%-
\end{tabular}
\tablefoot{$\text{R}^{2}$ and rMSE for $\rm T_{\rm gas}$ data of the five independent pairs of randomly selected planets removed from training (Fig.~\ref{fig:1DTeststackF400F1600}, Figs.~\ref{fig:1DTeststackF1800K800} - \ref{fig:1DTeststackG2000K600}). All of the $\text{R}^{2}$ values are very close to one, showing good agreement between the prediction and the ground truth. Although the rMSE is high due to large absolute deviations, the ML models successfully capture the underlying trends and variability, as evidenced by the high $\text{R}^{2}$ values.}
\end{table}
%-
\subsection{Application to PLATO exoplanet targets}
\label{s:results2}
%-
This study aims to demonstrate that suitably developed ML models (here: DNN, XGBoost) can reliably enrich the number of classically calculated 3D exoplanet atmosphere structures. ML-accelerated 3D model atmosphere would then enable us to conduct ensemble studies as envisioned for the PLATO targets. Therefore, we selected five inflated exoplanets that are located in PLATO's first long pointing field, LOPS2 \citep{Nascimbeni2025,2024arXiv240605447R}. They are spread between the grid of synthetic planets used for training (Fig. \ref{fig:TestTargetsComb}). We judge the quality of the ML predictions across the training grid by comparing them to additional \texttt{ExoRad} models generated specifically for these planets. Sec.~\ref{ss:dtdm} explores whether the differences between the ML results and the ground truth (\texttt{ExoRad} simulations) matter in terms of chemical equilibrium abundances and observability in transmission spectra for those planets.
%-
\subsubsection{ML performance for the selected planets}
%-
\begin{table*} 
\caption{Error estimation - $\text{R}^{2}$ values for PLATO target test planets.}
\label{tab:ResultsNewPlanets}
\centering
\begin{tabular}{l c c c c c }
\hline\hline
{Predicted quantity DNN} &  {WASP-121 b*}&  {HATS-42 b*}&  {NGTS-17 b*}&  {WASP-23 b*}&  {NGTS-1 b*}\\
\hline
Temperature     & 0.9979 & 0.9975 & 0.9950 & 0.9967 & 0.9912\\
Vertical Wind   & 0.7536 & 0.2036 & 0.0686 & -0.2430 & -6.9838\\
Zonal Wind      & 0.9894 & 0.9525 & 0.8694 & 0.9331 & 0.8598\\
Meridional Wind & 0.9217 & 0.8113 & 0.7734 & 0.9542 & 0.4140\\
\hline
 {Predicted quantity XGBoost} &  {WASP-121 b*}&  {HATS-42 b*}&  {NGTS-17 b*}&  {WASP-23 b*}&  {NGTS-1 b*}\\
\hline
Temperature     & 0.9716 & 0.9934 & 0.9570 & 0.9248 & 0.7059\\
Vertical Wind   & 0.0769 & 0.0384 & 0.0560 & 0.1222 & 0.0383\\
Zonal Wind      & 0.8122 & 0.8195 & 0.5517 & 0.8997 & 0.5996\\
Meridional Wind & 0.4960 & 0.6371 & 0.5835 & 0.7080 & 0.3329\\
\hline
\end{tabular}
\tablefoot{$\text{R}^{2}$ (Sec. \ref{ss:error}) of DNN and XGBoost model predictions compared to the \texttt{ExoRad} GCM values for the 5 target planets, for each prediction quantity. All $\text{R}^{2}$ of the temperature prediction of the DNN are above 0.99, while the wind predictions, especially for the vertical wind, show worse results. The XGBoost predictions follow the same trend but are a bit worse.}
\end{table*}
%-
\begin{figure}
\centering
\includegraphics[width=1\linewidth]{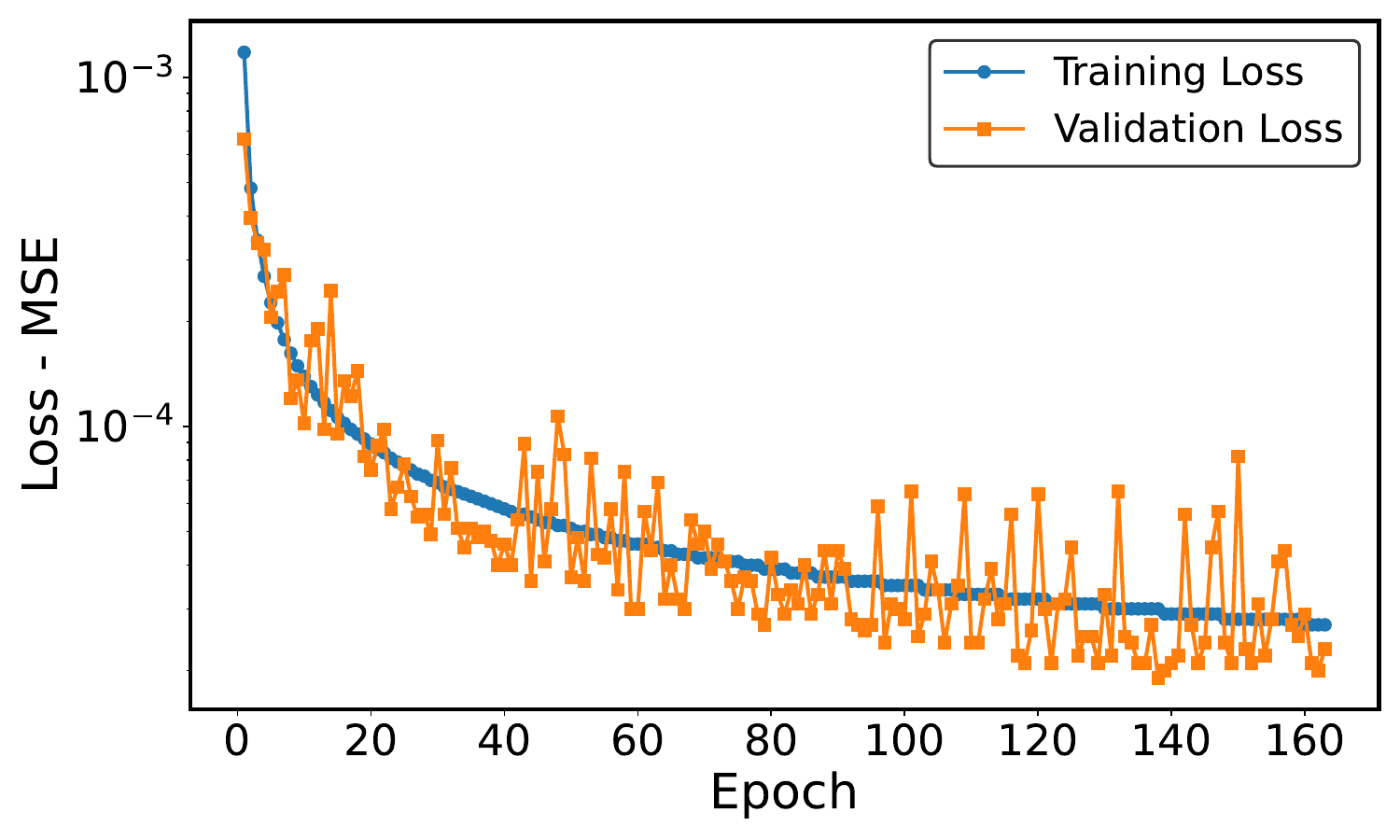}
\caption{Training and validation loss of the DNN used for predicting the test data (5 target exoplanets).}
\label{fig:train_val_loss}
\end{figure}
%-

Fig. \ref{fig:train_val_loss} shows the training and validation loss profiles during the training process of the DNN. The loss, representing the MSE after each epoch, is calculated on 80\% of the data used for training and 20\% for validation. The early stopping is triggered after 163 epochs, with the lowest validation loss of $1.9 \times 10^{-5}$ occurring at epoch 138. This error is computed in normalised space. The training and validation loss profiles align closely after 25 epochs, indicating that the model is learning well and generalizing effectively, with no significant signs of overfitting. The model with minimal validation loss is saved and subsequently used to make predictions on the test data ( i.e., the five target exoplanets). For the XGBoost model, hyperparameter optimization indicates that the best performance is achieved with a subsample ratio of 1. While other values such as 0.4, 0.6, and 0.8 are also tested (Tab.\ref{tab:XG}), using the full dataset (i.e., a subsample ratio of 1) in each boosting round yields the most accurate results.
%-
Tab.~\ref{tab:ResultsNewPlanets} compares the predictions from DNN and XGBoost with the reference values from \texttt{ExoRad}, using the $\text{R}^{2}$ score for each atmospheric variable across the five test planets: The DNN consistently achieves high accuracy in temperature prediction, with $\text{R}^{2}$ values exceeding 0.99 for all test cases. In contrast, predictions for the horizontal wind yield lower scores overall, and as observed in earlier tests, the vertical wind remains the most challenging variable to predict accurately and reliably. For zonal and meridional winds, WASP-121 b* and WASP-23 b* maintain good performance, with $\text{R}^{2}$ values above 0.9. However, NGTS-17 b* and NGTS-1 b* fall below this threshold, particularly NGTS-1 b*, whose meridional wind prediction drops to 0.41. HATS-42 b* shows comparatively better results, achieving 0.95 for zonal wind and 0.81 for meridional wind. While XGBoost exhibits similar overall trends, a key difference from the results in Sec.\ref{s:on-grid val} is that it is consistently outperformed by the DNN. This is an expected outcome as XGBoost generally tends to excel along structured lines of the grid, whereas the DNN, seems better equipped to learn the underlying physical patterns across the grid. The most striking exception occurs for the temperature prediction of NGTS-1 b*, where XGBoost achieves only 0.7 compared to the DNN’s 0.99.
%-
\subsubsection{Do the differences matter?}
\label{ss:dtdm} 
%-
We already saw that the gas temperature predictions across all planets agree quite well numerically.
We now explore if and how these differences matter for the numerically best and worst performing planets (Tab. \ref{tab:ResultsNewPlanets}).
These are the hottest and coolest selected planets: WASP-121 b* and NGTS-1 b*. The results for these planets are therefore representative of the whole set of selection PLATO target test planets (see also Appendix~\ref{sec: Spectra_all}).
%-
\paragraph{WASP-121 b*:}
%-
\begin{figure} 
\centering
\includegraphics[width=1.1\linewidth]{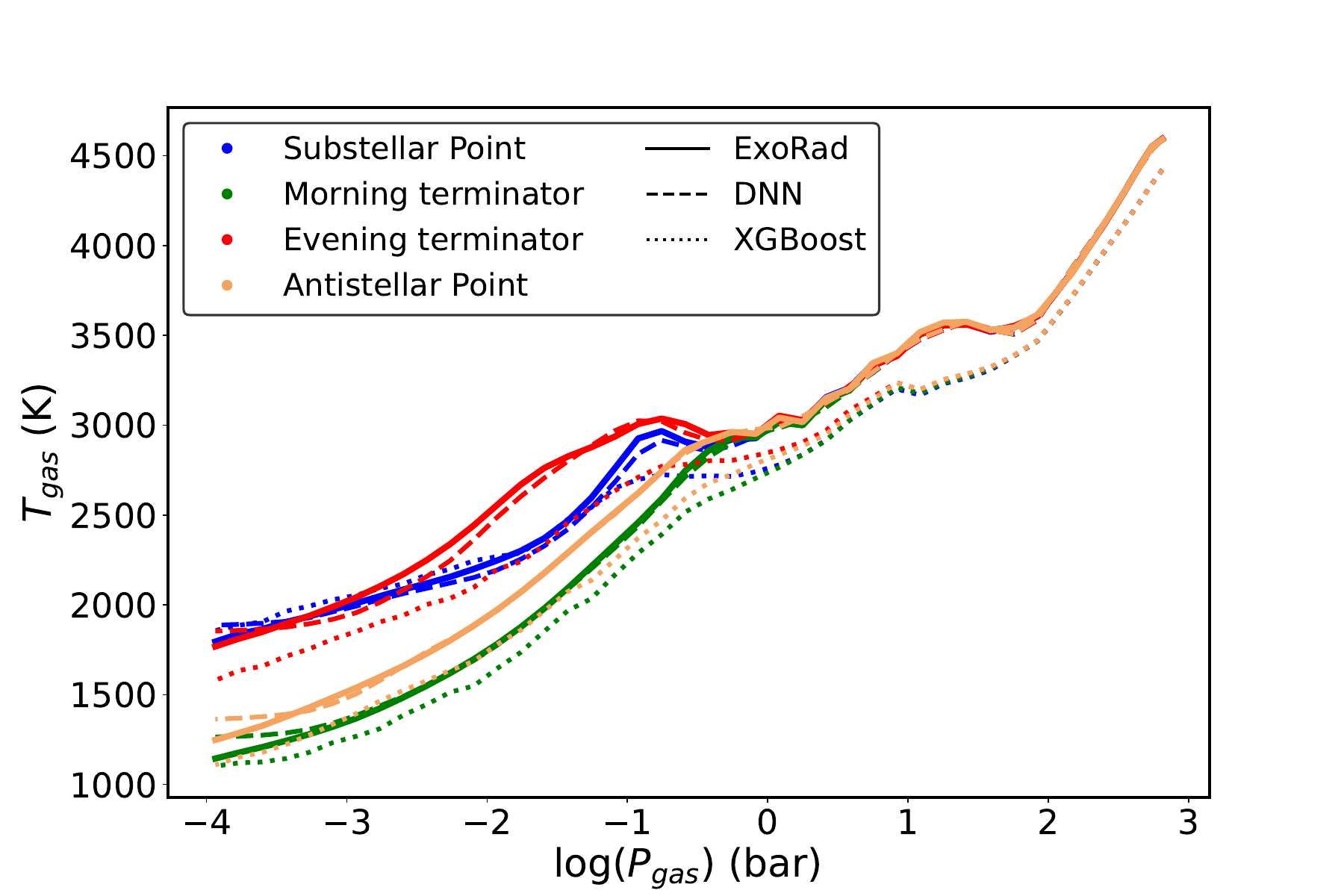}
\caption{WASP-121 b* 1D (T$_{\rm gas}$, p$_{\rm gas}$)-
profiles for the substellar (blue) and antistellar points (orange), the morning (green) and evening terminators (red). The solid line is the \texttt{ExoRad} simulated values, the dashed line the DNN prediction and the dotted line the XGBoost prediction.}
\label{fig:Wasp121_T}
\end{figure}
\begin{figure*} 
\centering
\includegraphics[width=1\linewidth]{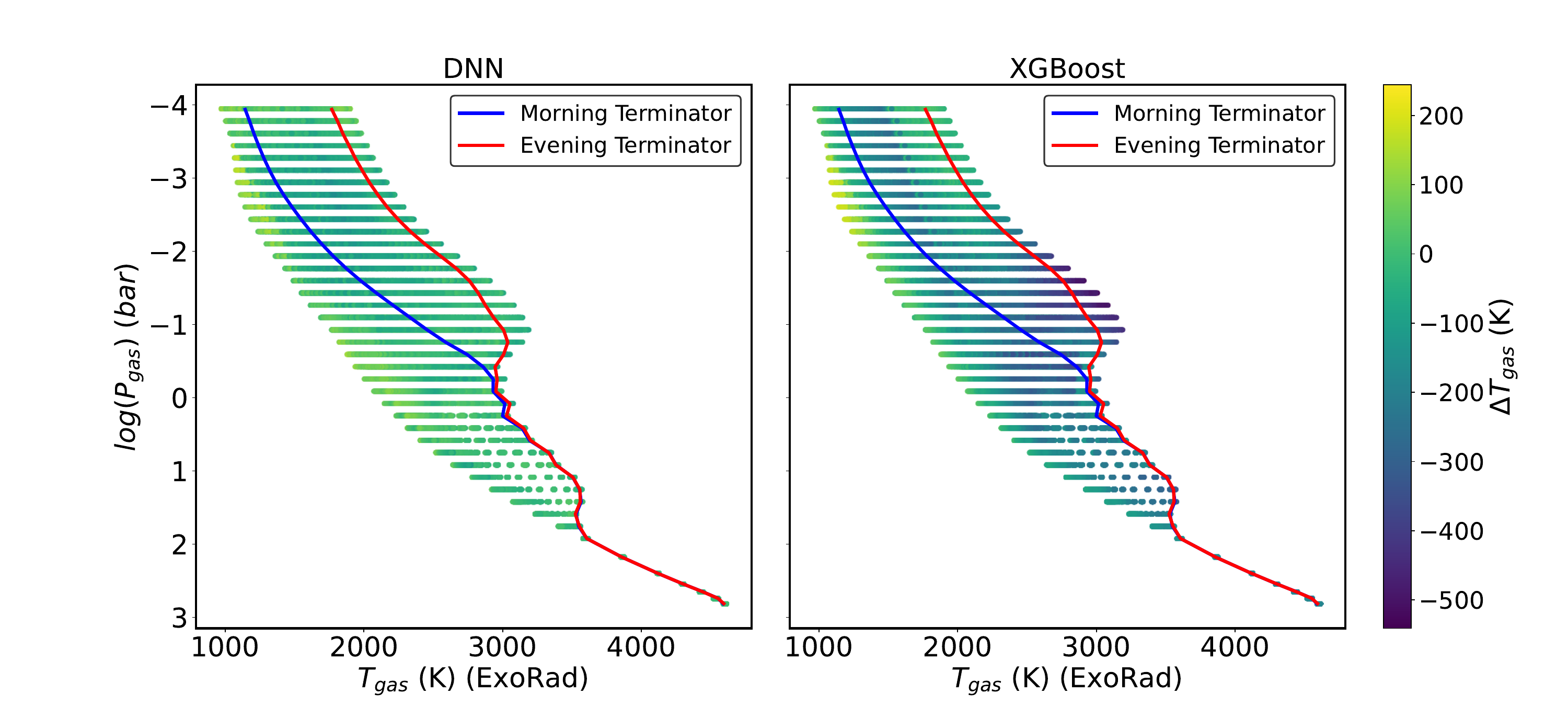}
\caption{WASP-121 b* scatter plot of the absolute gas temperature difference at each point on the planet. Overplotted are the morning (blue) and evening (red) terminator profiles as reference. 
}
\label{fig:WASP121_T_scatter}
\end{figure*}
%-
\begin{figure} 
\centering
\includegraphics[width=1\linewidth]{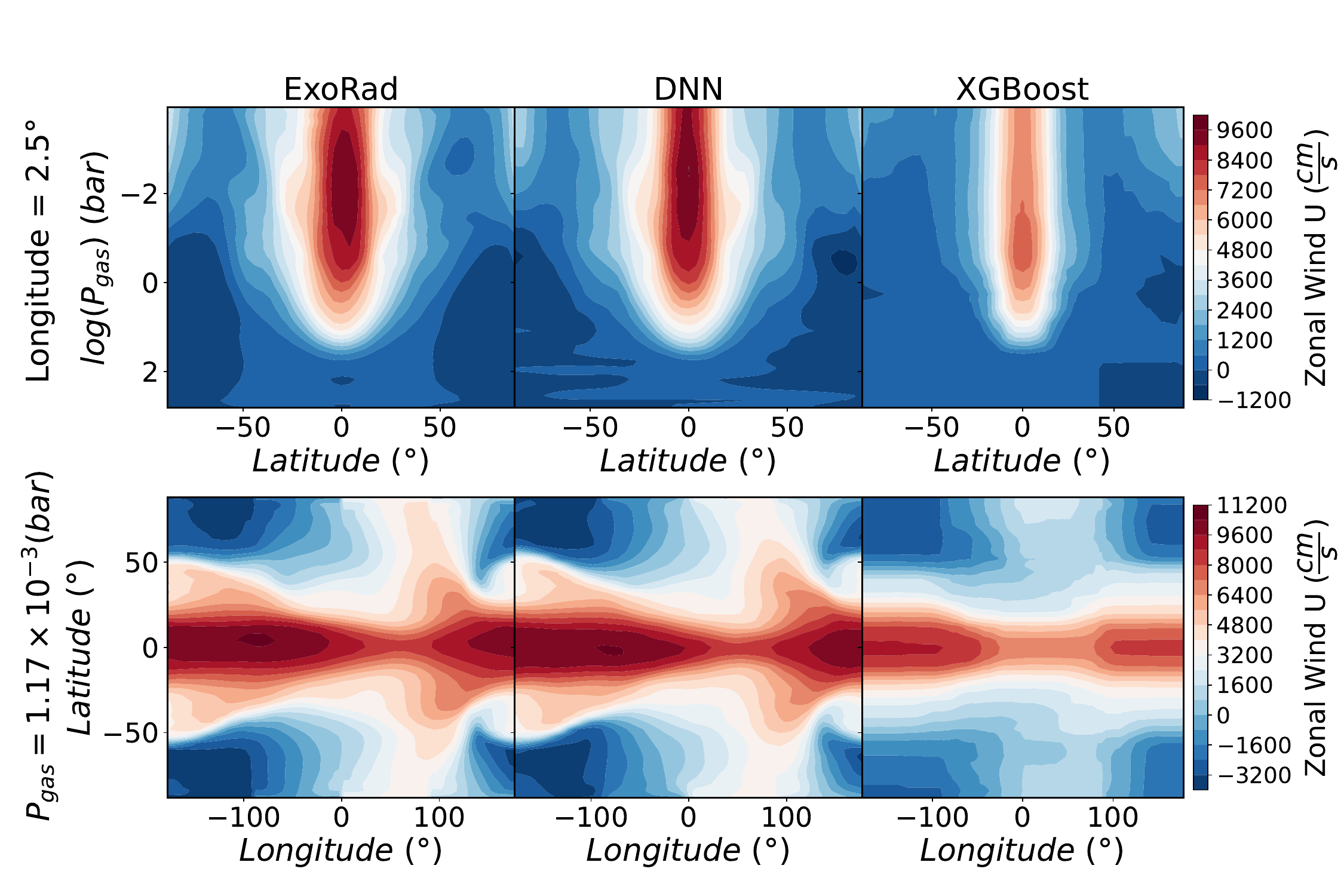}
\caption{WASP-121 b* zonal wind comparison through the atmosphere. The top row shows a latitude-pressure map at a fixed longitude of $2.5^{\circ}$, which is the closest grid point to the substellar point. The bottom row shows a latitude-longitude map at p$_{\rm gas}=1.17\times10^{-3}$ bar.}
\label{fig:Wasp121_U}
\end{figure}
%-
\begin{figure} 
\centering
\includegraphics[width=1\linewidth]{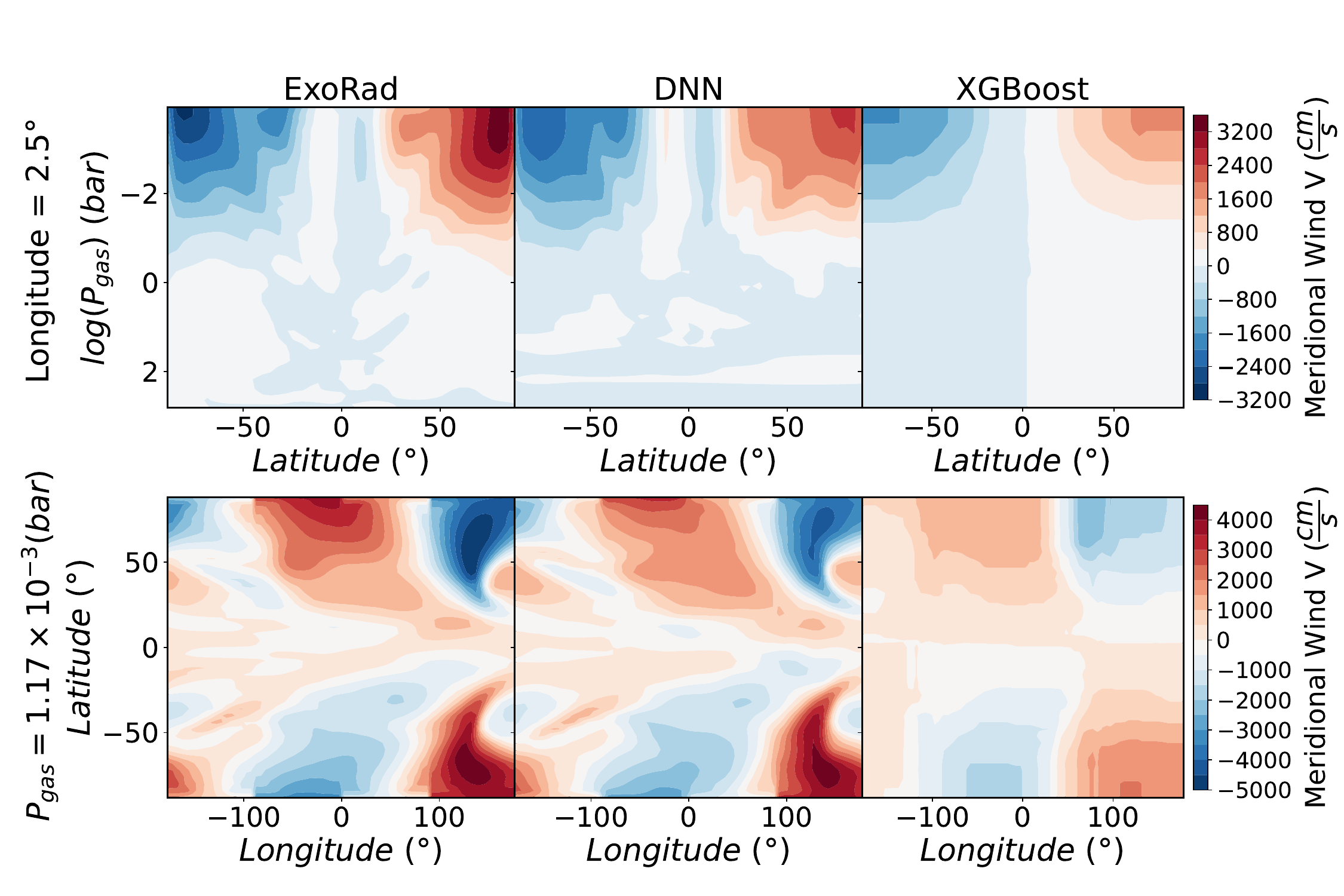}
\caption{WASP-121 b* meridional wind comparison through the atmosphere. With the same setup as in Fig. \ref{fig:Wasp121_U}}
\label{fig:Wasp121_W}
\end{figure}
%-
\begin{figure} 
\centering
\includegraphics[width=1.1\linewidth]{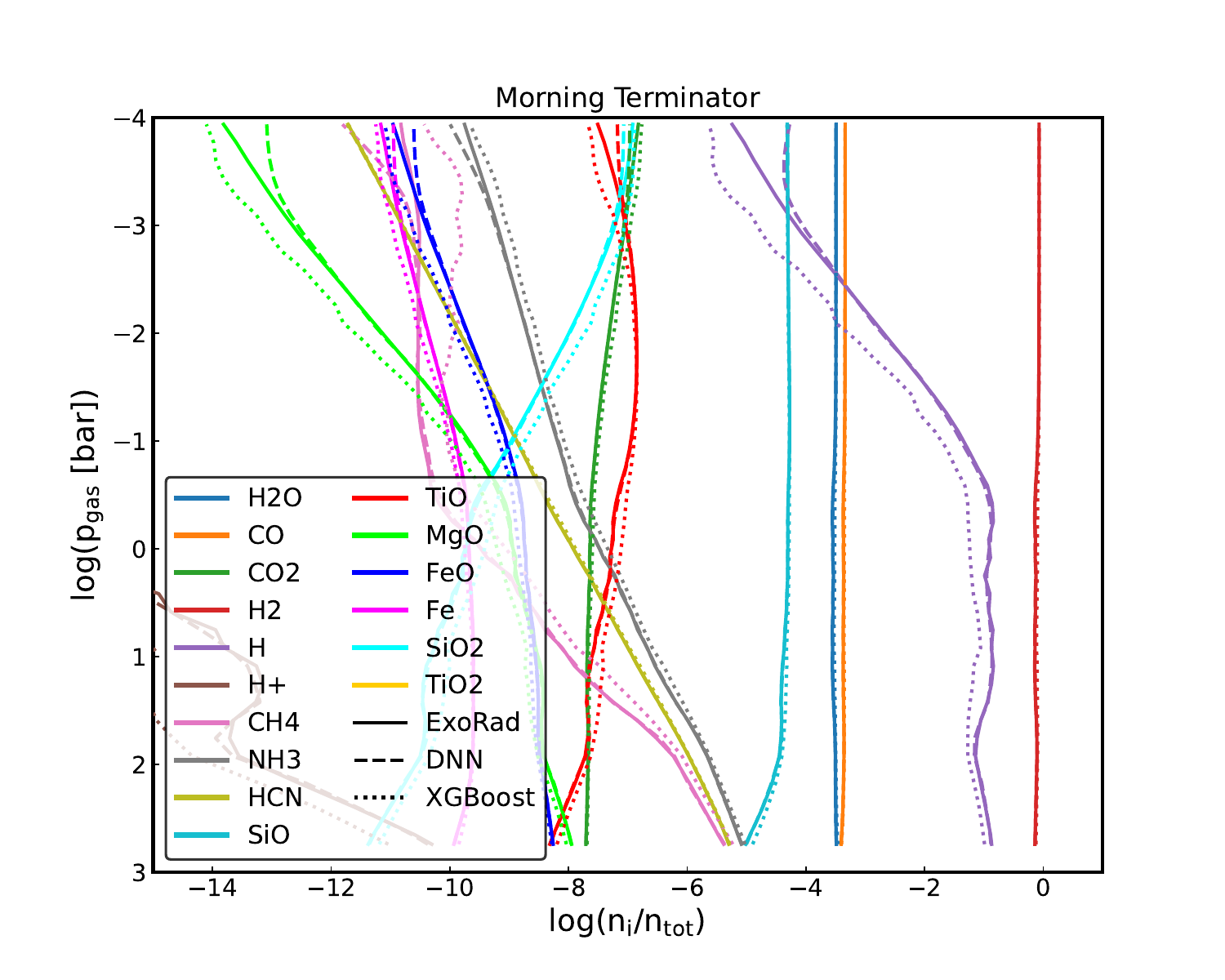}\\[-0.38cm]
\includegraphics[width=1.1\linewidth]{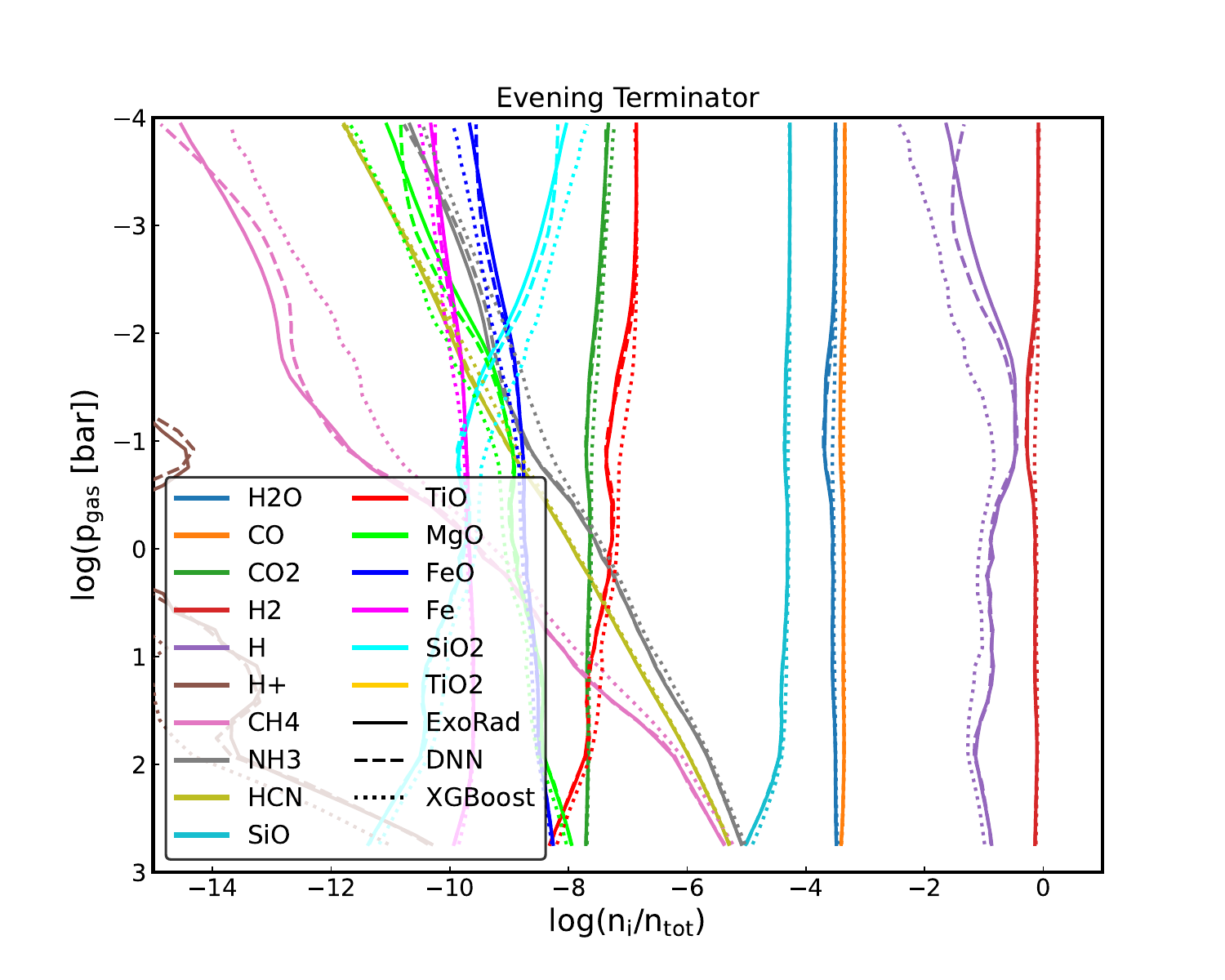}
\caption{WASP-121 b* morning (top) and evening (bottom) terminator gas phase chemistry for selected species. The solid line shows the \texttt{ExoRad} simulated values, the dashed line the DNN prediction and the dotted line the XGBoost prediction.}
\label{fig:Wasp121_Morning}
\end{figure}
%-
\begin{figure} 
\centering
\includegraphics[width=1\linewidth]{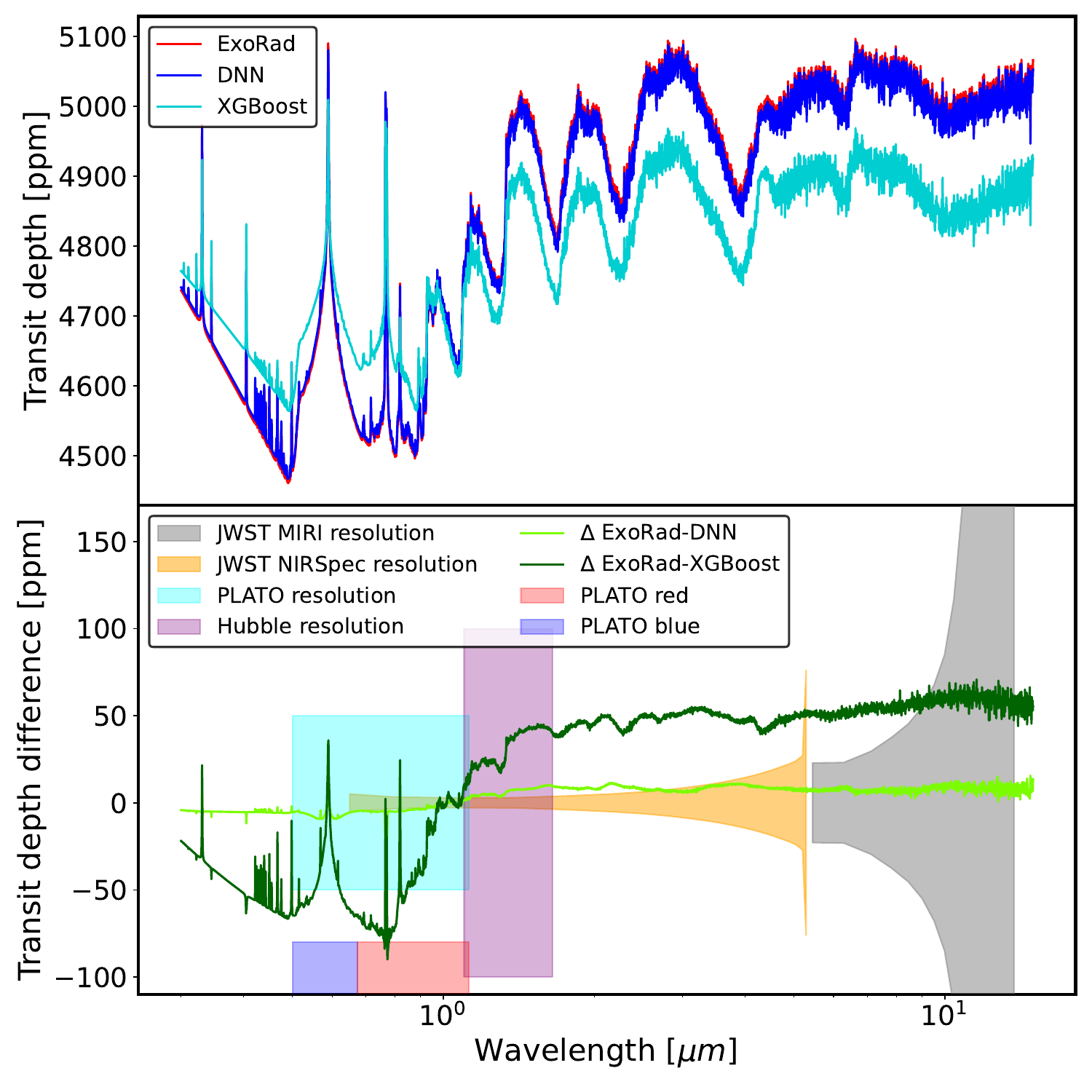}
\caption{WASP-121 b* combined morning and evening terminator transmission spectra. The first row panel shows the \texttt{ExoRad}, DNN and XGBoost results. The second row shows the difference of the ML spectra to \texttt{ExoRad} in comparison to the spectral accuracy of JWST MIRI LRS, JWST NIRSpec PRISM, Hubble and PLATO. }
\label{fig:Wasp121_Spectra}
\end{figure}
%-
Fig.~\ref{fig:Wasp121_T} shows that the DNN predictions closely match the reference model in temperature (i.e., the 3D GCM simulation for WASP-121 b*), with only slight deviations occurring in the uppermost regions of the atmosphere. This close agreement is also reflected in the equilibrium chemistry results (Fig.~\ref{fig:Wasp121_Morning}). The prediction from XGBoost displays a more noticeable, nearly constant offset in the 1D gas temperature profiles. The constant offset and even the larger temperature difference between $10^{1}\,\ldots\,10^{2}$~bar do not significantly affect the chemical composition. The main difference in the gas phase composition is found for the $\text{p}_{\rm gas} = 10^{0}\, \ldots \, 10^{-4}$ bar, and the effects are more prominent at the evening terminator (Fig. \ref{fig:Wasp121_Morning}).
%-

Figs.~\ref{fig:Wasp121_U} and \ref{fig:Wasp121_W} show that the exact values for zonal and meridional wind might differ in certain areas, but the trends are well captured, especially for the DNN. This visual comparison agrees with the numerical findings in the Tab. \ref{tab:ResultsNewPlanets}.  
%-

To ensure that the selected profiles of the substellar point, the anti-stellar point, and the two terminators are not biased selections, the absolute gas temperature difference for each point on the planet is shown in Fig.\ref{fig:WASP121_T_scatter}. Although XGBoost shows different accuracy across the range of temperatures at different pressure layers, the DNN performs more consistently throughout the atmosphere. 
%-

Fig.~\ref{fig:Wasp121_Spectra} presents the transmission spectra computed with \texttt{petitRADTRANS}, a testament to the precision of the prediction from ML models. These transmission spectra are calculated based on equilibrium chemistry data obtained from \texttt{GGchem}. The second row in Fig.~\ref{fig:Wasp121_Spectra}  provides a detailed comparison between the \texttt{ExoRad} and the ML calculated transmission spectra from the row above, with the spectral resolutions of PLATO, Hubble, JWST MIRI LRS, and JWST NIRSpec PRISM. Both predicted spectra align closely with the general shape, and all the absorption features are accurately reproduced. The DNN predictions show a slight offset on the edge of detectability between $\lambda = 1\,\ldots\, 3 \mu$m, but the deviation never exceeds 16 ppm. It is worth noting that at about 1 $\mu$m, we observe a shift from overestimation to underestimation of the features. This trend is more pronounced in the XGBoost prediction, which exhibits the same trend but with a higher offset. In the wavelength range $\lambda> 1 \mu$m, the offset remains nearly constant at around 50 ppm. On the other hand, the error increases to as much as 90 ppm for $\lambda< 1 \mu$m, with distinct differences in spectral features. When we compare the differences in chemistry in Fig. \ref{fig:Wasp121_Morning}, it becomes evident that the almost continuous offset in the spectra is not primarily due to the difference in chemical composition. The key contributing factor is the local gas temperature difference, a significant variable that leads to differences in the atmospheric scale height ($\text{H} = \frac{\text{k}_{\text{b}} \, \text{T}_{\rm gas}}{mg}$).This also explains the larger offset of the XGBoost prediction. As WASP-121\,b is an ultra-hot Jupiter, its atmospheric scale height is so large that even relatively small temperature differences substantially impact the resulting spectrum. The increased column of absorbing material leads to an increase in the line of sight, and therefore, to this type of continuous offset. 
%-
\paragraph{NGTS-1 b*}
%-
\begin{figure*} 
\centering
\includegraphics[width=1\linewidth]{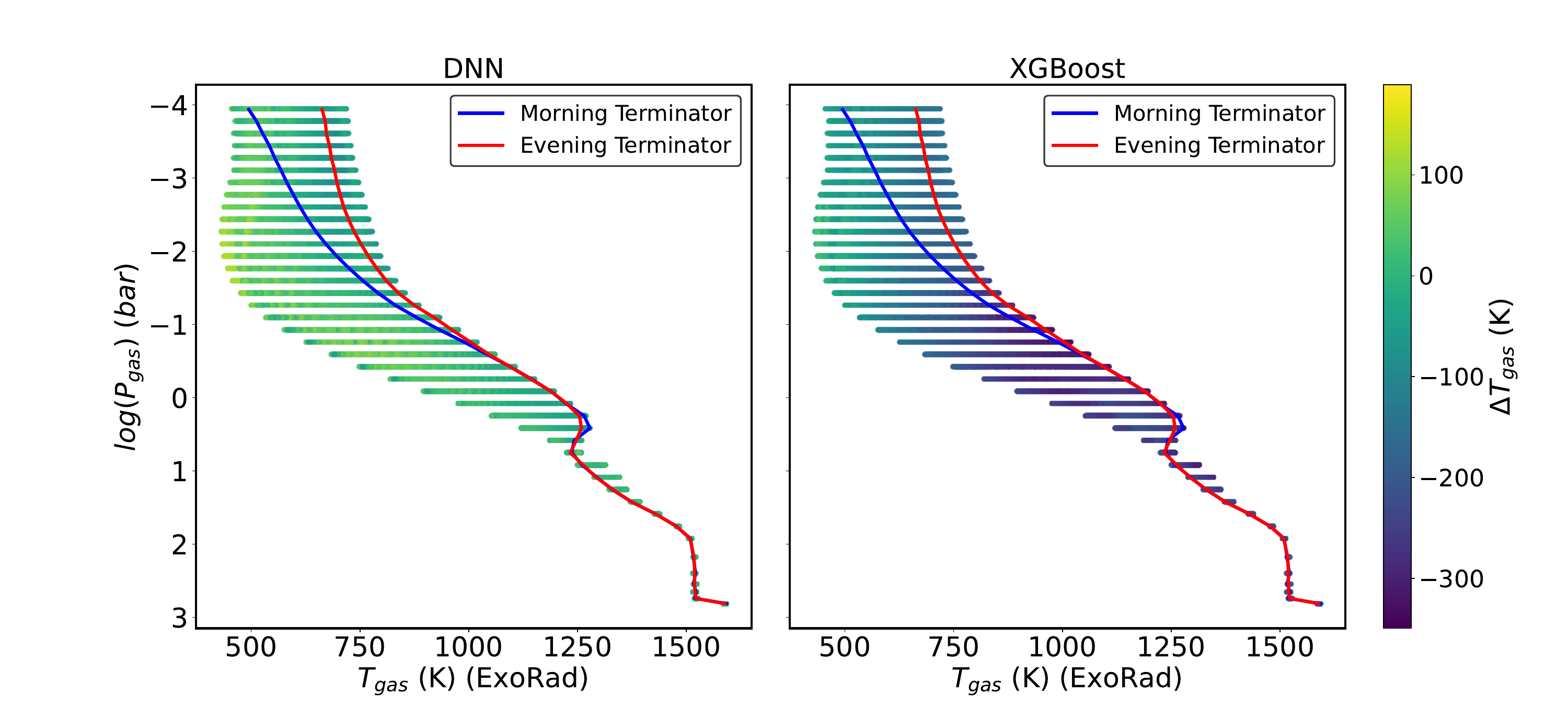}
\caption{NGTS-1 b* scatter plot of the absolute temperature difference at each point on the planet. Overplotted are the morning (blue) and evening (red) terminators as reference.}
\label{fig:NGTS1_T_scatter}
\end{figure*}
\begin{figure} 
\centering
\includegraphics[width=1\linewidth]{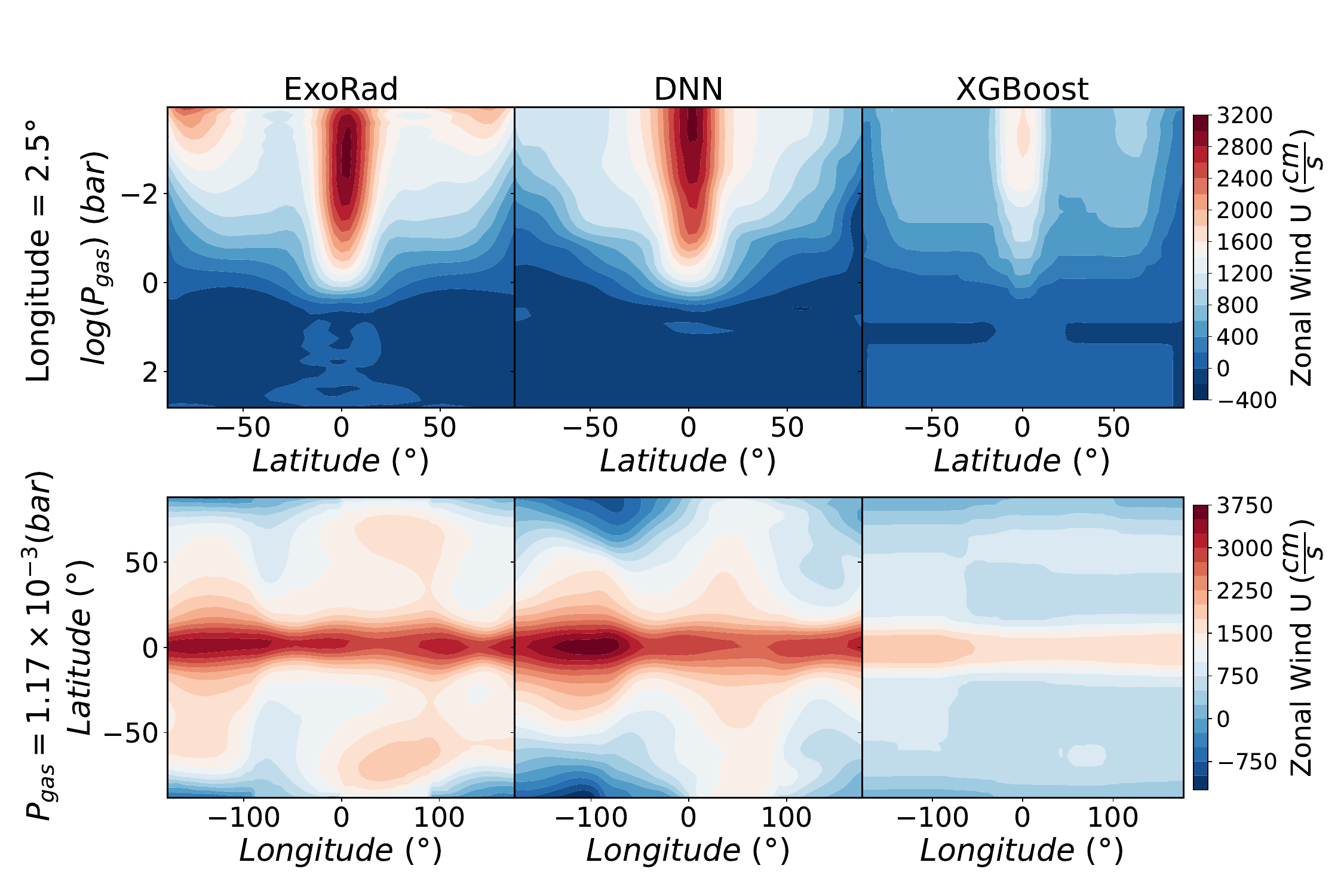}
\caption{NGTS-1 b* zonal wind comparison through the atmosphere. The top row shows a latitude-pressure map at a fixed longitude of $2.5°$, which is the closest grid point to the substellar point. The bottom row shows a latitude-longitude map at a fixed pressure level of $1.17\times10^{-3}$ bar.}
\label{fig:NGTS1_U}
\end{figure}
%-
\begin{figure} 
\centering
\includegraphics[width=1\linewidth]{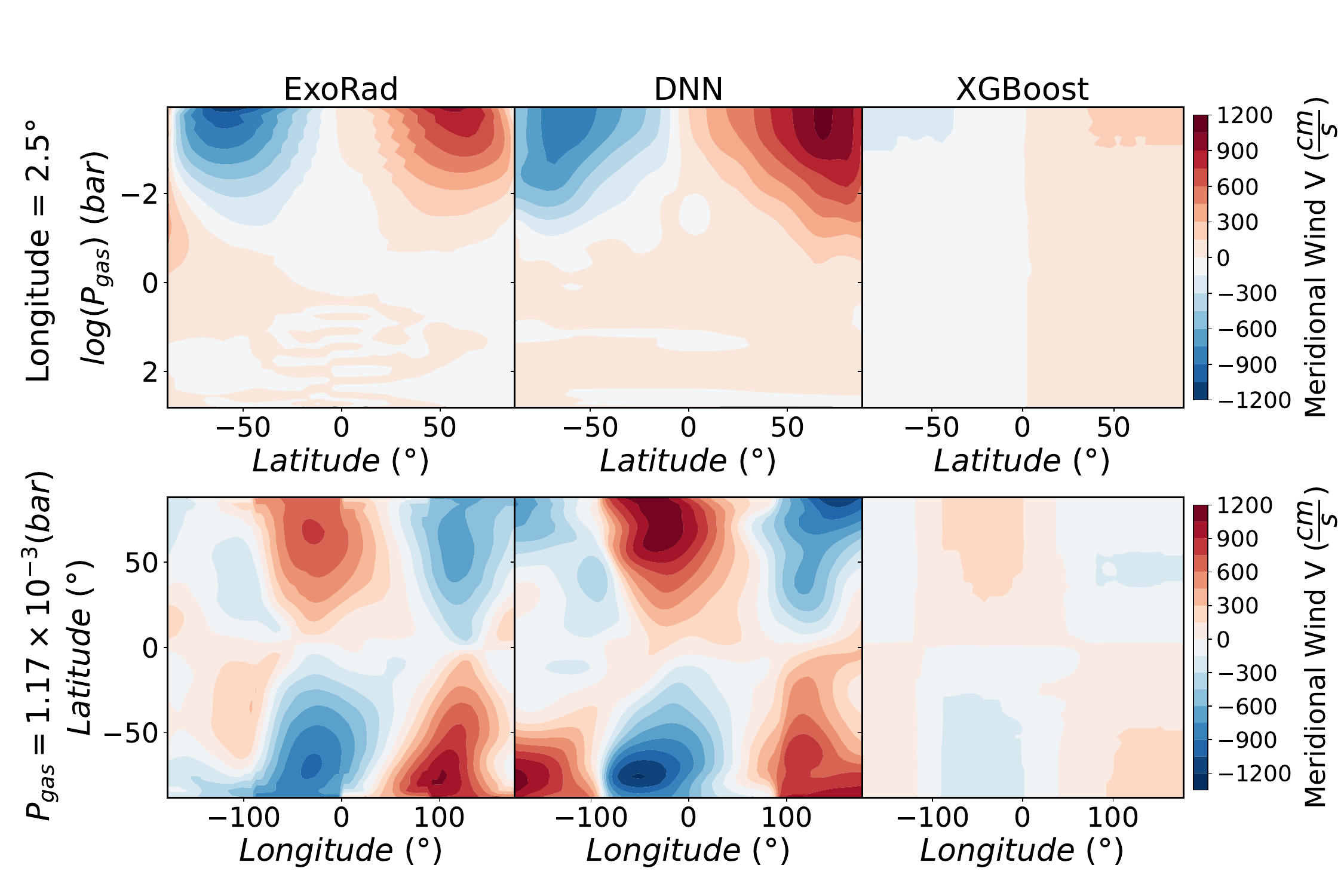}
\caption{NGTS-1 b* meridional wind comparison through the atmosphere. With the same setup as in Fig. \ref{fig:NGTS1_U}.}
\label{fig:NGTS1_W}
\end{figure}
\begin{figure} 
\centering
\includegraphics[width=1.1\linewidth]{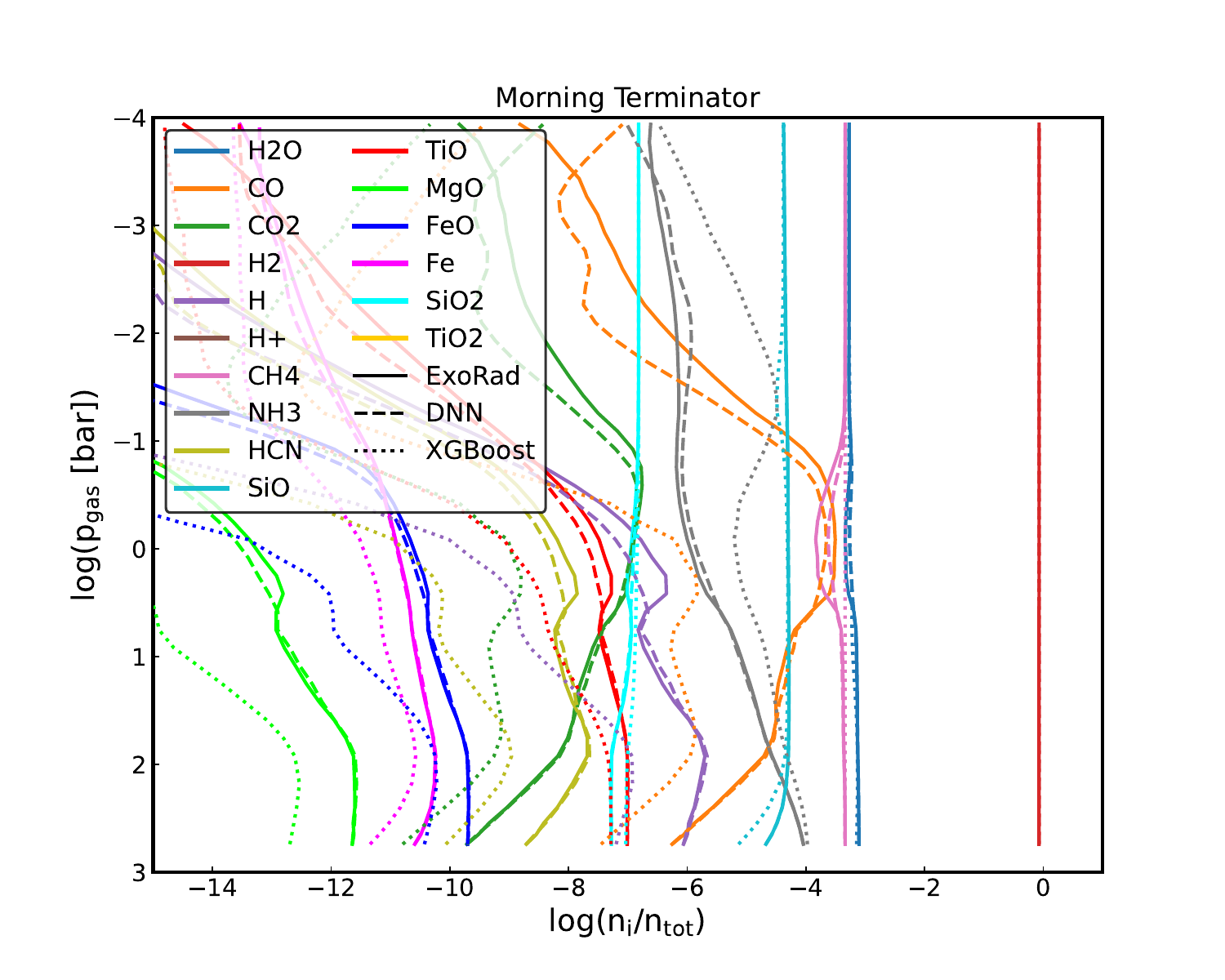}\\[-0.38cm]
\includegraphics[width=1.1\linewidth]{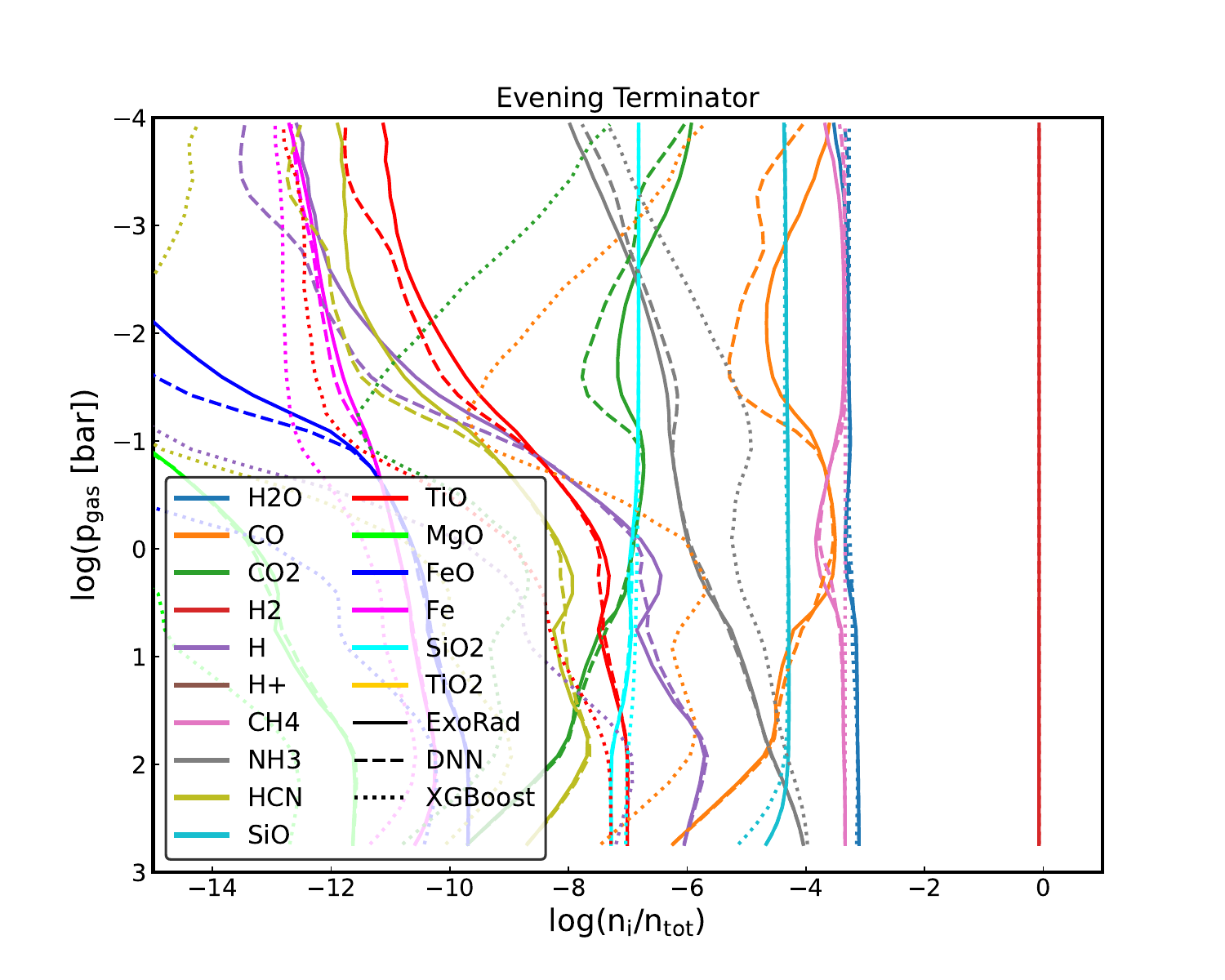}
\caption{NGTS-1 b* morning (top) and evening (bottom) terminator gas phase chemistry for selected species. The solid line shows the \texttt{ExoRad} simulated values, the dashed line the DNN prediction and the dotted line the XGBoost prediction. The differences between these two terminators demonstrate how the terminator temperature asymmetries affect the chemical gas phase composition. }
\label{fig:NGTS1_Morning}
\end{figure}
%-
\begin{figure} 
\centering
\includegraphics[width=1\linewidth]{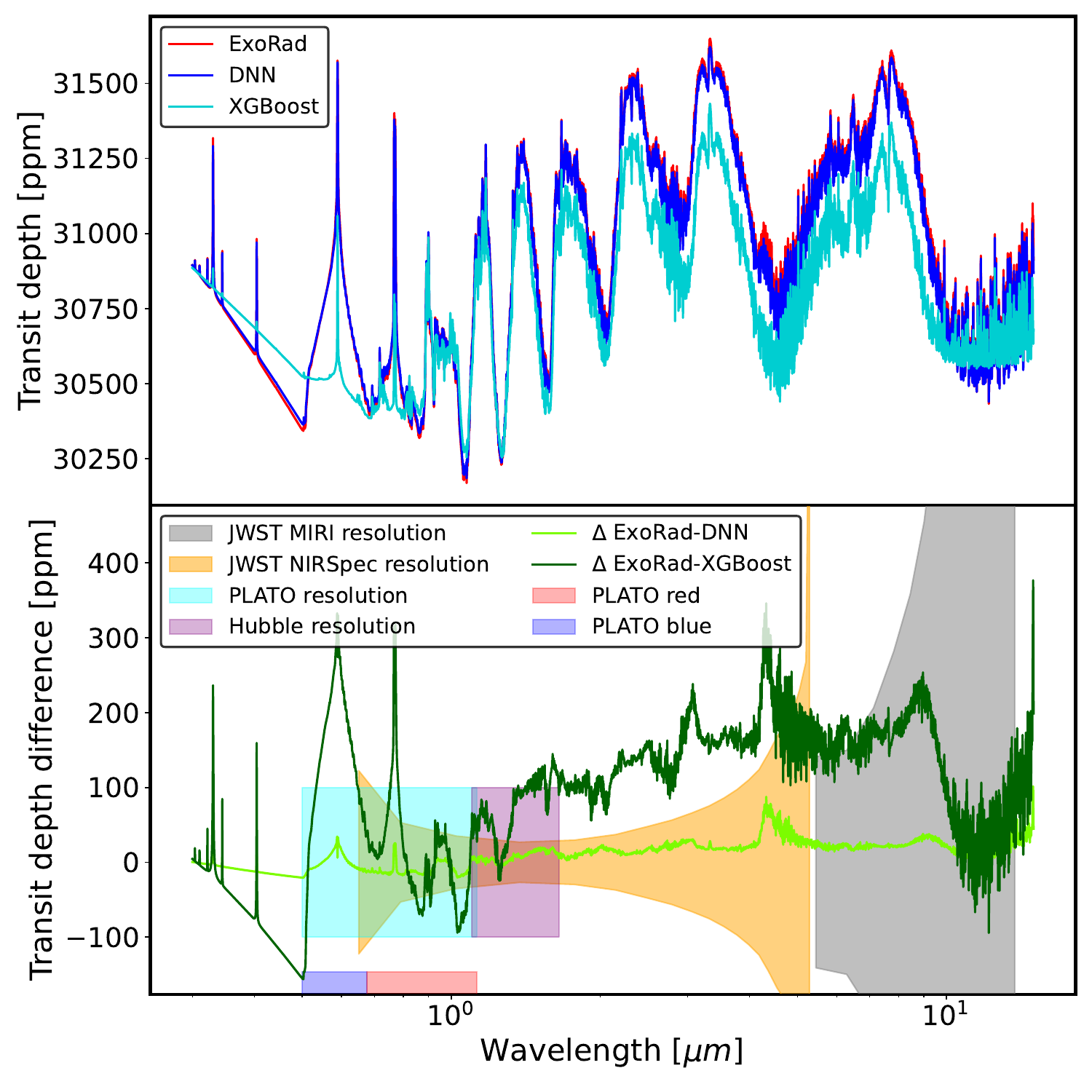}
\caption{NGTS-1 b* combined morning and evening terminator spectra. Same setup as in Fig.~ \ref{fig:Wasp121_Spectra}.}
\label{fig:NGTS1_Spectra}
\end{figure}
%-
\begin{figure} 
\centering
\includegraphics[width=1.1\linewidth]{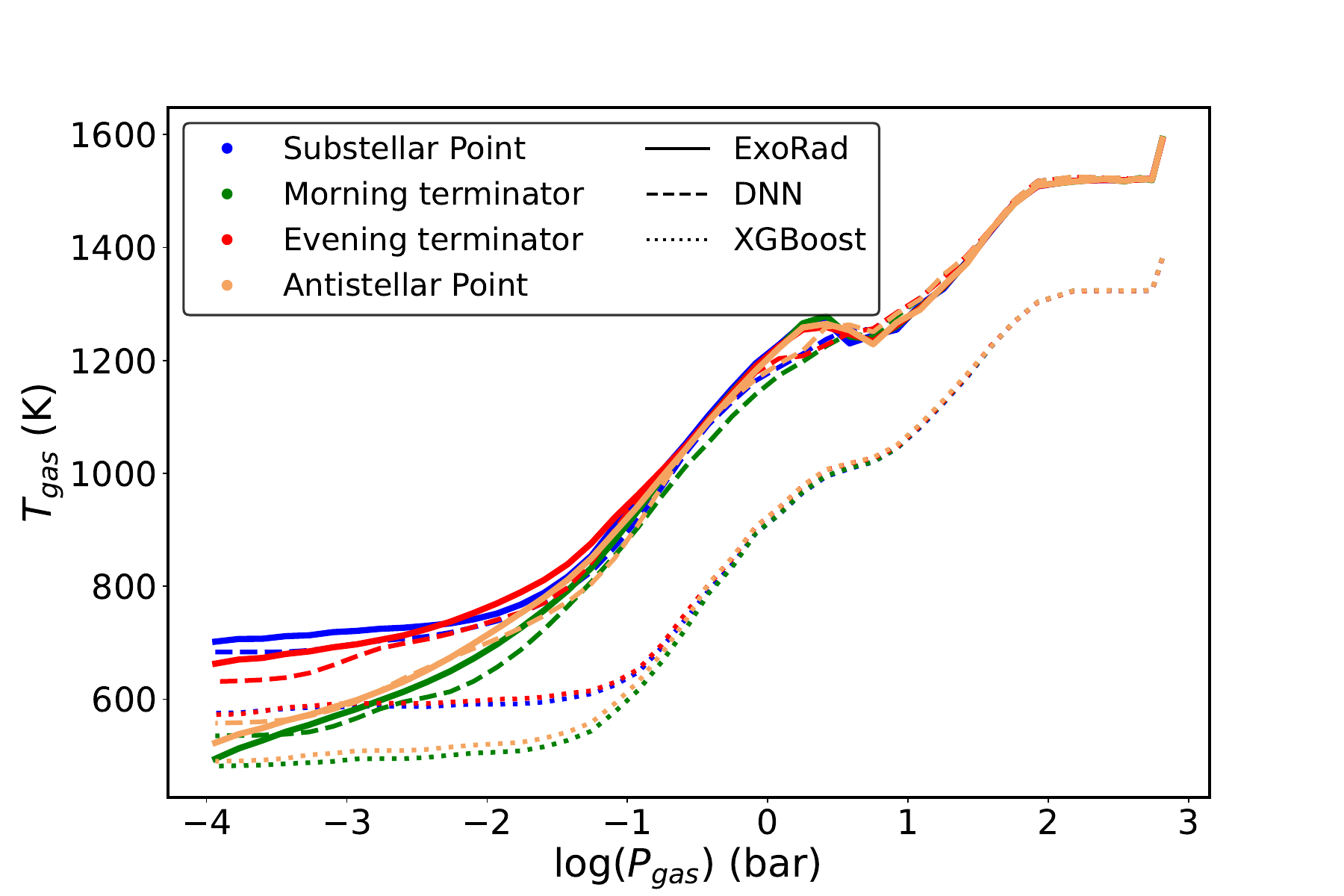}
\caption{NGTS-1 b* 1D (T$_{\rm gas}$, p$_{\rm gas}$)-
profiles for the substellar (blue) and antistellar points (orange), the morning (green) and evening terminators (red). The solid line is the \texttt{ExoRad} simulated values, the dashed line the DNN prediction and the dotted line the XGBoost prediction.}
\label{fig:NGTS1_T}
\end{figure}
%-
After comparing the ML predicted local gas temperature for all five test planets WASP-121 b*, HAST-42b, NGTS-17 b*, WASP-23 b*, and NGTS-1 b*, against \texttt{ExoRad}, we can conclude that the prediction error is highest for NGTS-1 b*. While the DNN prediction shows minor differences in the atmosphere regions  $\text{p}_{\rm gas}<10^{-1}$ bar (Figs. \ref{fig:NGTS1_T}, \ref{fig:NGTS1_U} \ref{fig:NGTS1_W}), the XGBoost shows substantial discrepancies throughout the atmosphere. 
%-

In contrast to WASP-121 b* (see Figs. \ref{fig:Wasp121_T}, \ref{fig:Wasp121_Morning}), where the temperature offset had limited impact on equilibrium chemistry, persistent temperature offset leads to a significant deviation in equilibrium chemistry in the case of NGTS-1 b* (Fig. \ref{fig:NGTS1_Morning}). The inaccurate gas temperature predictions from XGBoost cause various chemical species to differ by multiple orders of magnitude from those derived based on the \texttt{ExoRad} results (considered the ground truth in ML terms; solid vs. dotted lines). The DNN, however, maintains strong predictive performance in most of the atmosphere of NGTS-1 b*. The most noticeable differences between the DNN and the  \texttt{ExoRad} results (solid vs dashed lines in Fig. \ref{fig:NGTS1_Morning} appear in the low-pressure regions ($\text{p}_{\rm gas}<10^{-1}$ bar), high up in the atmosphere. The problematic constituents are CO, CO$_{2}$, TiO, HCN, and H. The carbon-bearing species: CO, CO$_{2}$ and HCN  have the largest deviation at the $\text{p}_{gas} = 10^{-2}$ bar for the DNN results. For $\text{p}_{\rm gas}<10^{-1}$ bar, the DNN fits almost perfectly with the \texttt{ExoRad} result, even for this difficult planet. The strong gas opacity sources H$_2$O and CH$_4$ only appear marginally affected. Potentially interesting species as cloud formation precursors TiO, Fe, and FeO start to differ in the regions $\text{p}_{\rm gas}<10^{-2}$ bar. They are very low in concentration and show an approximately constant offset. SiO and SiO$_2$ appear not to be affected. In agreement with our earlier results in Tab. \ref{tab:ResultsNewPlanets}, NGTS-1 b* shows the strongest differences observed between our ML-accelerated GCM and our \texttt{ExoRad} GCM results. 
%-

The general shape of the resulting transmission spectra in Fig. \ref{fig:NGTS1_Spectra} is well reproduced. However, deviations of up to 100 ppm are seen in the DNN predictions, and up to 400 ppm in those from XGBoost. In the DNN predicted spectrum, all key chemical features are present, though the CO$_2$ band at 4.5 $\mu \text{m}$ and the HCN band at 15 $\mu \text{m}$ are underestimated. This aligns with the reduced concentrations of CO$_2$ and HCN compared to the \texttt{ExoRad} results. However, none of those differences are large enough to be observationally significant. The XGBoost prediction shows detectable errors throughout the whole spectral range. As seen in the case of WASP-121 b*, we can observe an almost constant offset due to the temperature difference. This explains why the predictions are outside the acceptable spectral accuracy for most wavelengths. Despite the substantial discrepancies shown in equilibrium chemistry (see Fig. \ref{fig:NGTS1_Morning}), where most of the abundances differ by several orders of magnitude, the only feature missing is the CO$_{2}$ at 4.5 $\mu \text{m}$. The Na and K lines have also almost completely disappeared in the XGBoost prediction. The temperature predicted by XGBoost at the morning terminator falls to $\sim$500~K, significantly below the expected 600–800 K for pressures $\text{p}_{\rm gas}<10^{-1}$ bar as shown in Fig. \ref{fig:NGTS1_T}. This is the temperature regime where Na and K start to condense out of the gas phase. On the evening terminator side, temperatures hover around $\sim$600~K, which explains why we still get a feature at all.
%-

A direct comparison of the spectral accuracy of WASP-121 b* and NGTS-1 b* for JWST cameras in Figs. \ref{fig:Wasp121_Spectra} and \ref{fig:NGTS1_Spectra} express significantly higher accuracy for the hotter planet orbiting the brighter star. This also explains why the spectra of WASP-121 b* exhibit detectable offsets despite having much smaller absolute errors than those of NGTS-1 b*. The remaining three planets exhibit similar trends, as shown in their spectra in Figs. \ref{fig:HATS42_Spectra}, \ref{fig:NGTS17_Spectra}, and \ref{fig:WASP23_Spectra}. In contrast to the numerical results in Tab. \ref{tab:ResultsNewPlanets}, HATS-42 b* and NGTS-17 b* show such strong agreement across all spectral predictions that no detectable differences arise with any of the selected cameras for both ML methods. 

Local chemical equilibrium is a valid assumption for ultra-hot Jupiters like WASP-121b and is challenging to disentangle from effects of cloud formation (\citealt{2020A&A...635A..31M}). Kinetic chemistry that particular effects CH${}_4$ becomes important for planets like NGTS-1b \citep{Bangera2025,Baeyens1,Baeyens2}. Horizontal advection may also lead to deviations from local chemical equilibrium \citep{Drummond2020,Baeyens1}. However, local chemical equilibrium is the 'ground state' even for kinetic chemistry processes. Furthermore, while the results of the wind prediction are worse, the main effects are still present and especially the strong zonal wind jet is reproduces.
%-
\section{Computational complexity}
\label{sec:time-comp}
%-
Simulating planetary atmospheres using the 3D $\texttt{ExoRad}$ GCM is computationally intensive \citep{H_David_2024}. Both $\texttt{ExoRad}$ and the ML models were run on the same computer system to benchmark performance. The runtime of $\texttt{ExoRad}$ varies with the average global temperature $\text{T}_{\text{global}}$. For hot planets, $\text{T}_{\text{global}} = 1400\,\ldots\,2600$ K, simulations take approximately 40 hours; for intermediate temperatures ($1200\,\ldots\, 1000$ K), around 60 hours; and for colder planets ($800\,\ldots\,400$ K), runtimes can reach up to 240 hours for a 1000-planet-days simulation. These simulations utilise 64 CPU cores on a system with a $\texttt{AMD EPYC 7742}$ processor running at a base clock of 2.25 GHz.
%-
While a single GCM simulation can take anywhere from 40 to 240 hours, depending on the temperature regime, ML models significantly reduce computational time. Training the DNN with \textit{early stopping} takes us approximately 25–28 hours. The XGBoost model trains even faster and only takes 15–20 minutes. For comparison, in each case, even the more time-intensive DNN training is faster than a single GCM run, while XGBoost training is orders of magnitude quicker.
%-

Once trained, both models can generate predictions for one full unknown planet in around 1-2 seconds. This leads to a dramatic reduction in runtime compared to GCM simulations. For instance, a 40-hour (144,000 seconds) simulation for hot planets can be replaced with ML inference, yielding a speed-up of approximately $\mathcal{O}(10^{5})$. For colder planets, where GCM runtimes can reach up to 240 hours (864,000 seconds), the computational gain can reach five to nearly six orders of magnitude.
%-

When evaluating runtime efficiency, it is important to distinguish between the training and inference (testing) phases. While training, particularly for the DNN, can be time-consuming, it is also a one-time cost. This cost may increase with denser training grids, but it does not affect the prediction time for new planets. However, inference time is the correct relevant metric for practical applications, as it determines the cost of predictions for unknown planetary atmospheres. Due to nearly real-time inference, ML models are highly effective for rapid and scalable analysis of exoplanet atmospheres. Instead of running computationally expensive GCM simulations for each new or unknown planet, ML models provide fast and reasonably accurate predictions, enabling efficient exploration of large parameter spaces.
%-

XGBoost significantly reduces training time compared to DNN. However, this comes at a significant trade-off in prediction accuracy. We observed that increasing the density of the training grid (see Appendix~\ref{ss:MLdensityeff}) improves XGBoost's predictive performance, narrowing the accuracy gap between XGBoost and DNN models for equilibrium chemistry calculation. In such cases, XGBoost becomes a more favourable option, particularly when minimising training time is a priority without compromising too much on accuracy.
%-
\section{Summary and conclusion}
\label{s:concl}
%-
We introduce and analyse the 3D AFGKM \texttt{ExoRad} grid for gas giant atmospheres for systematic climate characteristics. Here, we focus on solar metallicity ([Fe/H]=0) and solar C/O ratio (0.55), which is as a valid first assumption for Jupiter mass planets \citep{Sun2024}. Higher metallicity are expected from planetary accretion models typically for planets of Saturn mass or less \citep{Schneider2021} which are not considered here.
From analysing four characteristic properties (day/night and morning/evening temperature differences, maximum zonal wind velocity, and wind jet width), climate states can be identified. It can be concluded that the observation of terminator temperature asymmetries is better suited to diagnose the complexity of wind jet structures on tidally locked gas giants than the canonical day-to-night side temperature differences.
%-

This work demonstrates that of the ML methods we tested, a lightweight DNN is the best at interpolating a grid of exoplanet atmospheres by predicting local gas temperature and horizontal winds. The local gas temperature prediction is so accurate that chemical equilibrium modelling reproduces abundances of most of the key molecules (H$_{2}$O, CO, CO$_{2}$, NH$_{3}$, HCN) within much less than one order of magnitude compared to the \texttt{ExoRad} 'ground truth'.
CO and CO$_{2}$ are the most sensitive to temperature deviations. However, even differences in CO abundances are at most one order of magnitude. With the largest difference in the upper atmosphere ($\text{p}_{\rm gas}< 10^{-1}$) of the coldest (NGTS-1 b*, $\text{T}_{\rm global} = 800$~K) investigated planet. 

The resulting spectra show all of the main absorption features. The differences in spectra are small enough not to be detectable by PLATO, Hubble or JWST in the three planets NGTS-1 b*, HATS-42 b* and NGTS-17 b*. WASP-23 b* ($\text{T}_{\rm global} = 1115$~K) shows single detectable features that are still smaller than 32ppm. WASP-121 b*, the hottest investigated planet, is the only one that shows a general offset of 16~ppm between 1 and 3 $\mu \text{m}$ that could, in principle, be detectable. However, constant offsets in JWST transmission spectra of more than 100~ppm can already be generated by instrumental effects alone \citep{Carter2024}. Therefore, it is generally assumed that the transmission spectra are insensitive to uniform shifts \citep{Rackham2024}. Thus, even this deviation between the spectra generated with the preferred ML method (DNN) and the GCM is inconsequential. We also note that different GCMs can exhibit for one and the same planet differences in local gas temperature of up to 200~K \citep{Bell2024,Coulombe2023,Noti2023}.

While the horizontal wind predictions qualitatively capture global trends, exact values might differ in specific regions, and thus, wind predictions should not be considered accurate at individual points. The wind speed deviations are in many cases, however, smaller than 20\% and in case smaller than the deviations in wind speed from different GCMs. Here, deviations of up to 50\% in maximum wind velocity and in some cases basic wind jet structure due to different numerical set-ups and the choice of Navier-Stokes equations were found \citep{Heng2011,Mayne2019,Noti2023}. The vast majority of GCMs, including this one, are in qualitative agreement in general wind stricture for tidally locked irradiated gas giants. The GCM used here also shows little time variability (by few 10s of K in temperature and less than 10\% for wind speed) during longer run times once the simulations have converged, when accounting for numerical sources of instabilities \citep[e.g.][]{Komacek2025,Carone2020,Schneider2022b}. The stability of the GCM is also elucidated by Fig.~\ref{fig:Grid_Temp_Windjet_Diag} top right, for example, where the efficacy of superrotation shaping the evening and morning terminator temperatures can be clearly linked to orbital periods and global temperatures. In this context, the ML predictions are very well within the variations given by the underlying model framework, both for the local gas temperature and the horizontal wind velocities.

This work presents the first proof-of-concept study in which ML models are explored to directly predict the three major atmospheric variables ($\text{T}_{\rm gas}$, U, and V). The developed ML models are lightweight and hence reduce memory usage during training and inference. As a result, they can predict the complete 3D solution of an unknown planet within the grid space in nearly real-time. In conclusion, ML models offer a computationally efficient alternative and can serve as fast and efficient climate emulators after training them on a traditional GCM grid. 
%-
\begin{acknowledgements}
Vienna Science Cluster (VSC) HPC facility CPU time project 3D GCMs for Exoplanets (72245).
%-
%-
\end{acknowledgements}
%-
\bibliographystyle{aa} % style aa.bst
\bibliography{references} % your references Yourfile.bib
%-
\newpage
\begin{appendix}
%-
\section{The complete 3D AFGKM ExoRad model grid}
\label{apx:completgris}
\begin{figure*} 
\centering
\includegraphics[width=1\linewidth]{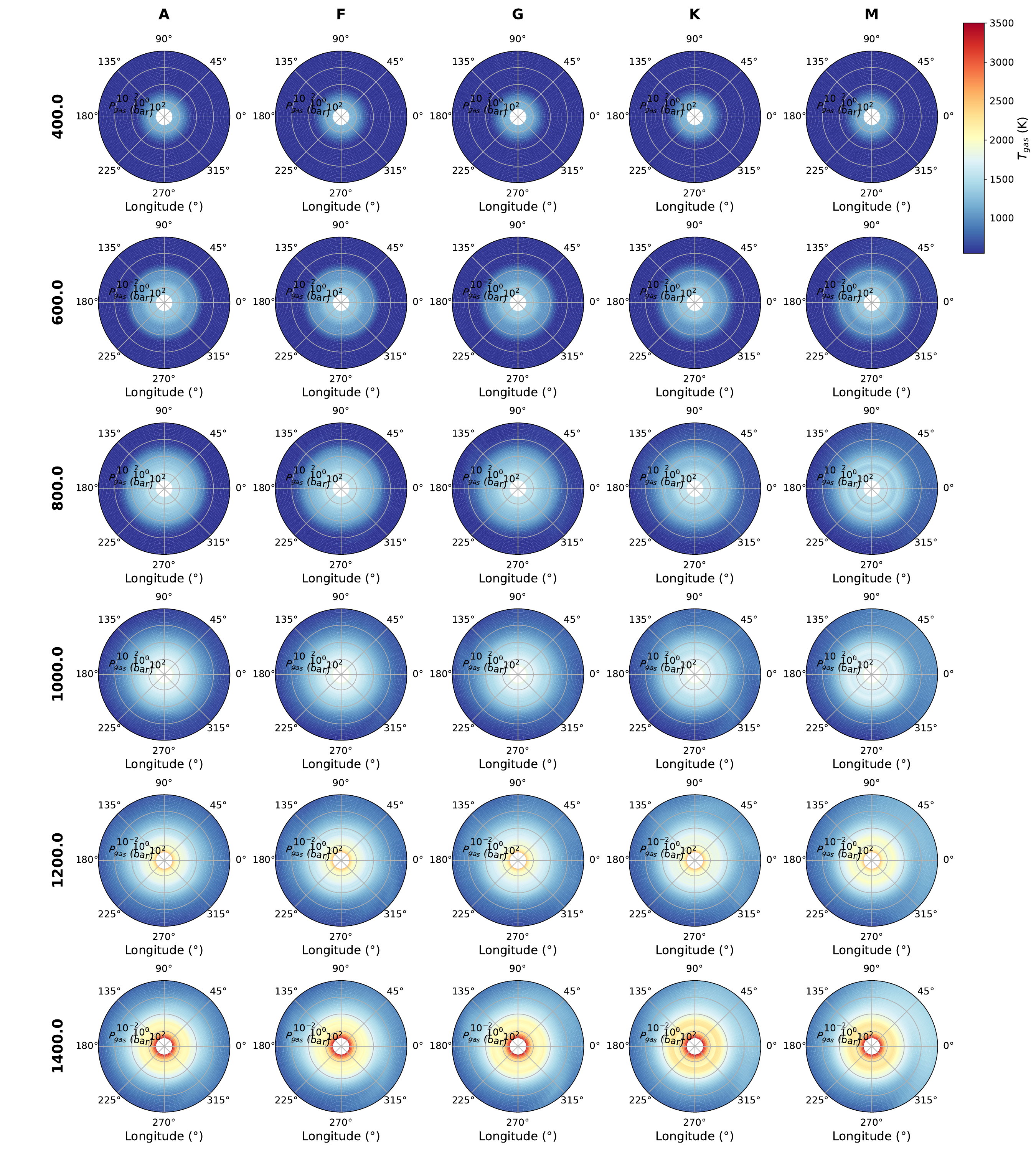}
\caption{The radial gas temperature maps of planets from the 3D AFGKM \texttt{ExoRad} with a global temperatures, T$_{\rm global}$ of 400K - 1400K (top to bottom) that orbit different host stars (left to right: A, F, G, K, M). The (T$_{\rm gas}, \text{p}_{\rm gas}$)-maps gas are shown as equatorial slice plots. 
}
\label{fig:Slice_full}
\end{figure*}

\begin{figure*} 
\centering
\includegraphics[width=1\linewidth]{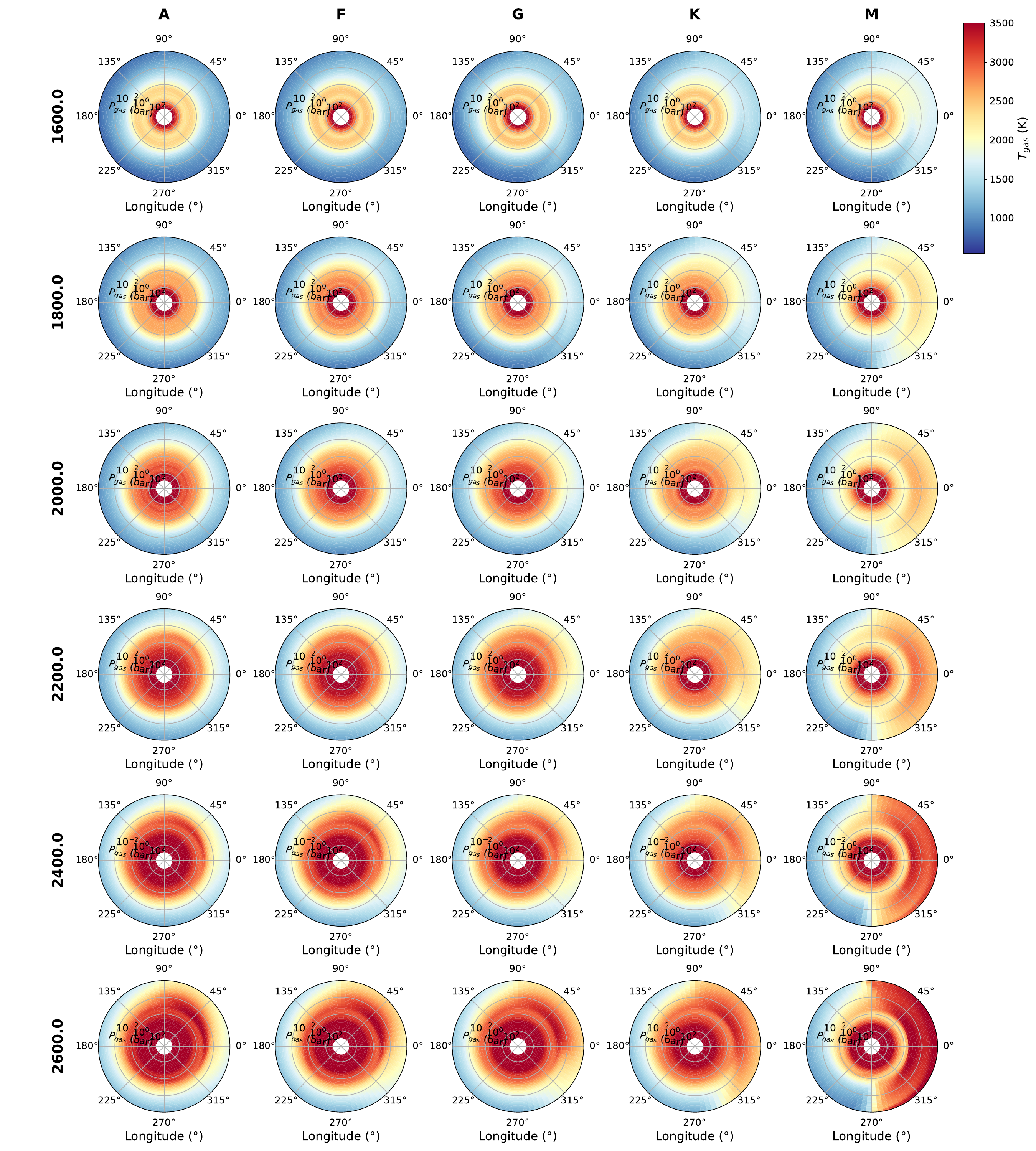}
\caption{The radial gas temperature maps of planets from the 3D AFGKM \texttt{ExoRad} with a global temperatures, T$_{\rm global}$ of 1600K - 2600K (top to bottom) that orbit different host stars (left to right: A, F, G, K, M). The (T$_{\rm gas}, \text{p}_{\rm gas}$)-maps gas are shown as equatorial slice plots. 
}
\label{fig:Slice_full2}
\end{figure*}

%-
The gas temperature maps for the complete set of 60  3D \texttt{ExoRad} GCM  models are provided for all host stars (AFGKM) included in this gas giant atmosphere grid. The host star parameters are shown in Tab.~\ref{tab: stellar_params}. Tabs.~\ref{tab: A_params} - ~\ref{tab: M_params} summarises the planetary global parameters $\text{T}_{\rm global}$ [K], $\text{T}_{\rm int}$ [K], the semi-major axis $\text{a}$ [au], and the orbital period $\text{P}_{\rm orb}$ [d].  Figs.~\ref{fig:Slice_full} and ~\ref{fig:Slice_full2} provide the equatorial gas temperature maps. The 1D stellar and anti-stellar profiles 
are shown in Sec.~\ref{ss:3dAFGKM} for all planets around A, G and M type stars.

%-
\begin{table}[ht]
\caption{Stellar parameters for the 3D \texttt{ExoRad} GCM  grid.}
\centering
\begin{tabular}{l c c c}
\hline\hline
 {Spectral type} &  {$\text{T}_{\mathrm eff}$ [K]} &  {${\text{R}_*}$ [$  {\rm R_{\mathrm Sun}}$]} &  {$  {\text{M}_*}$ [$  {\rm M_{\mathrm Sun}}$]}\\
\hline
A5V & 8100 & 1.79 & 1.88 \\
F5V & 6550 & 1.47 & 1.33 \\
G5V & 5660 & 0.98 & 0.98 \\
K5V & 4400 & 0.70 & 0.70 \\
M5V & 3060 & 0.20 & 0.16 \\
\hline
\end{tabular}
\label{tab: stellar_params}
\end{table}
%-
\begin{table}[ht]
\caption{A main sequence host star planet properties}
\centering
\begin{tabular}{l c c c}
\hline\hline
$  {\text{T}_{\rm global}}$  {[K]}& $  {\text{T}_{\rm int}}$  {[K]} & { \text{a}  [AU]}&  {$  {\text{P}_{\rm orb}}$ [days]} \\
\hline
400 & 200 & 1.758 & 620.68\\
600 & 200 & 0.761 & 176.85 \\
800 & 200 & 0.426 & 74.14 \\
1000 & 222 & 0.273 & 37.89 \\
1200 & 383 & 0.189 & 21.91 \\
1400 & 526 & 0.139 & 13.80 \\
1600 & 623 & 0.106 & 9.24 \\
1800 & 667 & 0.084 & 6.49 \\
2000 & 664 & 0.068 & 4.73 \\
2200 & 628 & 0.056 & 3.55 \\
2400 & 571 & 0.047 & 2.74 \\
2600 & 503 & 0.040 & 2.15 \\
\hline
\end{tabular}
\label{tab: A_params}
\end{table}
%-
\begin{table}[ht]
\caption{F main sequence host star planet properties}
\centering
\begin{tabular}{l c c c}
\hline\hline
${\text{T}_{\rm global}}$  {[K]}& $  {\text{T}_{\rm int}}$  {[K]} & { \text{a}  [AU]}&  {$  {\text{P}_{orb}}$ [days]} \\
\hline
400 & 200 & 0.949 & 292.48 \\
600 & 200 & 0.411 & 83.34 \\
800 & 200 & 0.230 & 34.93 \\
1000 & 222 & 0.147 & 17.86 \\
1200 & 384 & 0.102 & 10.33 \\
1400 & 526 & 0.075 & 6.50 \\
1600 & 625 & 0.057 & 4.35 \\
1800 & 668 & 0.045 & 3.06 \\
2000 & 665 & 0.037 & 2.23 \\
2200 & 624 & 0.030 & 1.67 \\
2400 & 578 & 0.026 & 1.29 \\
2600 & 509 & 0.022 & 1.01 \\
\hline
\end{tabular}
\label{tab: F_params}
\end{table}
%-
\begin{table}[ht]
\caption{G main sequence host star planet properties}
\centering
\begin{tabular}{l c c c}
\hline\hline
$  {\text{T}_{\rm global}}$  {[K]}& $  {\text{T}_{\rm int}}$  {[K]} & {\text{a}  [AU]}&  {$  {\text{P}_{orb}}$ [days]} \\
\hline
400 & 200 & 0.470 & 118.74 \\
600 & 200 & 0.203 & 33.83 \\
800 & 200 & 0.114 & 14.18 \\
1000 & 221 & 0.073 & 7.25 \\
1200 & 379 & 0.051 & 4.19 \\
1400 & 528 & 0.037 & 2.64 \\
1600 & 628 & 0.028 & 1.77 \\
1800 & 669 & 0.022 & 1.24 \\
2000 & 663 & 0.018 &  0.91\\
2200 & 627 & 0.015 & 0.68 \\
2400 & 581 & 0.013 & 0.52 \\
2600 & 513 & 0.011 & 0.41 \\
\hline
\end{tabular}
\label{tab: G_params}
\end{table}
%-
\begin{table}[ht]
\caption{K main sequence host star planet properties}
\centering
\begin{tabular}{l c c c}
\hline\hline
$  {\text{T}_{\rm global}}$  {[K]}& $  {\text{T}_{\rm int}}$  {[K]} & { \text{a}  [AU]}&  {$  {\text{P}_{orb}}$ [days]} \\
\hline
400 & 200 & 0.207 & 41.21 \\
600 & 200 & 0.090 & 11.74 \\
800 & 200 & 0.050 & 4.92 \\
1000 & 224 & 0.032 & 2.52 \\
1200 & 390 & 0.022 & 1.46 \\
1400 & 537 & 0.016 & 0.92 \\
1600 & 613 & 0.013 & 0.61 \\
1800 & 666  & 0.010 & 0.43 \\
2000 & 664 & 0.008 & 0.31 \\
2200 & 641 & 0.007 & 0.24 \\
2400 & 597 & 0.006 & 0.18 \\
2600 & 526 & 0.005 & 0.14 \\
\hline
\end{tabular}
\label{tab: K_params}
\end{table}
%-
\begin{table}
\caption{M main sequence host star planet properties}
\centering
\begin{tabular}{l c c c}
\hline\hline
$  {\text{T}_{\rm global}}$  {[K]}& $  {\text{T}_{\rm int}}$  {[K]} & { \text{a}  [AU]}&  {$  {\text{P}_{orb}}$ [days]} \\
\hline
400 & 200 & 0.0275 & 4.14\\
600 & 200 & 0.0119 & 1.18 \\
800 & 200 & 0.067 & 0.49 \\
1000 & 219 & 0.0043 & 0.25 \\
1200 & 378 & 0.0030 & 0.15 \\
1400 & 522 & 0.0022 & 0.09 \\
1600 & 618 & 0.0017 & 0.06 \\
1800 & 668 & 0.0013 & 0.04 \\
2000 & 667 & 0.0011 & 0.03 \\
2200 & 633 & 0.0009 & 0.02 \\
2400 & 548 & 0.0007 & 0.02 \\
2600 & 480 & 0.0006 & 0.01 \\
\hline
\end{tabular}
\label{tab: M_params}
\end{table}
%-
%-
\clearpage
\section{Supplementary results from ML testing}
%-
This section presents additional results to support the performance of ML models. Sec.~\ref{s:App_train} provides a study of the training accuracy of our ML models. Sec.~\ref{s:App_val} presents results from the sanity check discussed in Sec.~\ref{s:on-grid val} related to other four pairs of planets: (F1800, K800), (G1800, K1800), (F2000, G2200) and (G2000, K600). Sec. \ref{ss:MLdensityeff} delves into improving the prediction quality for XGBoost with a higher density of data points.
%-
\subsection{Learning efficiency of ML models}
\label{s:App_train}
\begin{figure}[ht!]
\includegraphics[width=1\linewidth]{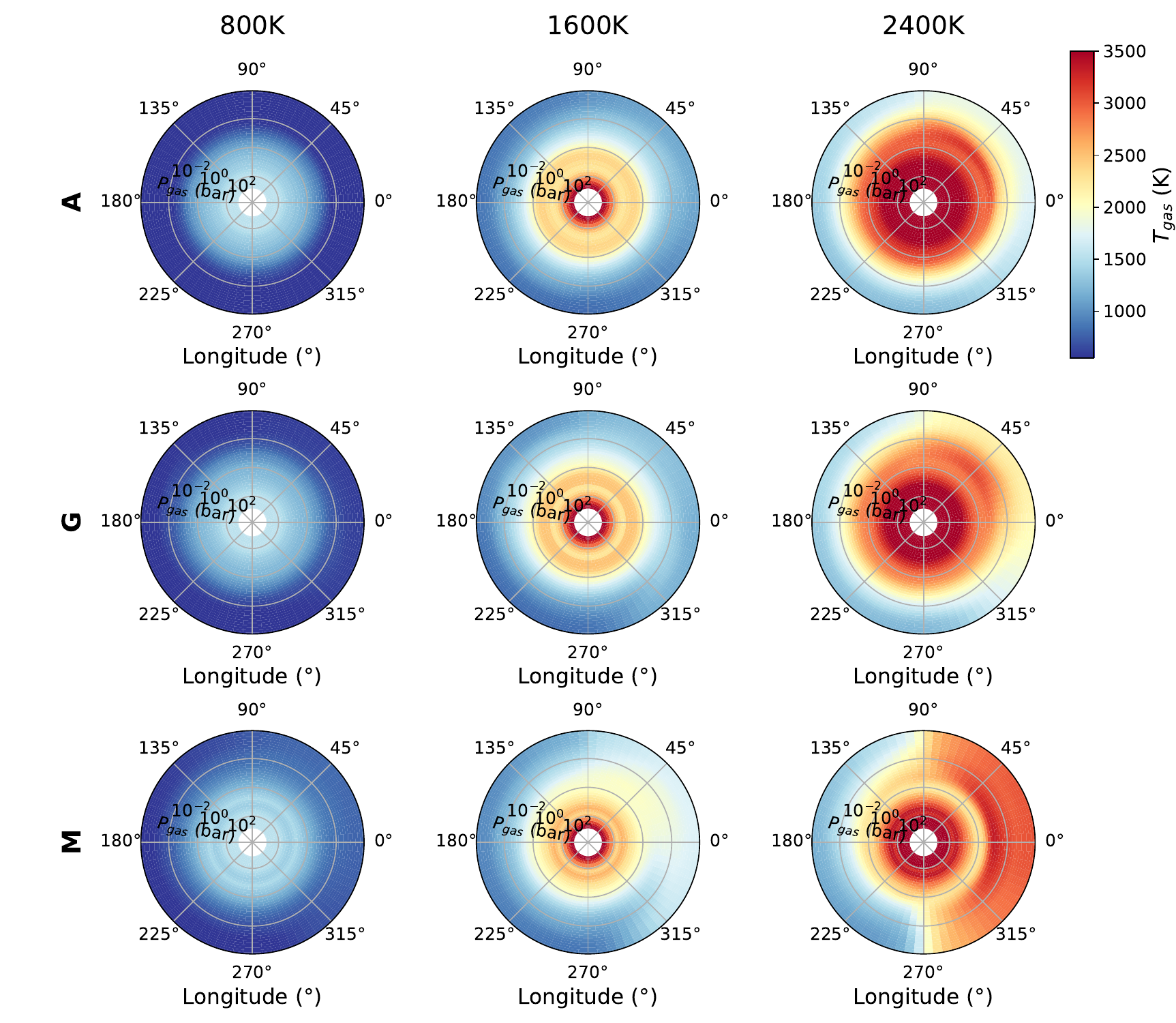}
\caption{
The recreated equatorial slice plots as in Fig.~\ref{fig:EQ_slice}. The DNN has been trained on this data and is subsequently used to reproduce the same slices.}
\label{fig:EQ_sliceDNN}
\end{figure}

\begin{figure}[ht!]
\includegraphics[width=1\linewidth]{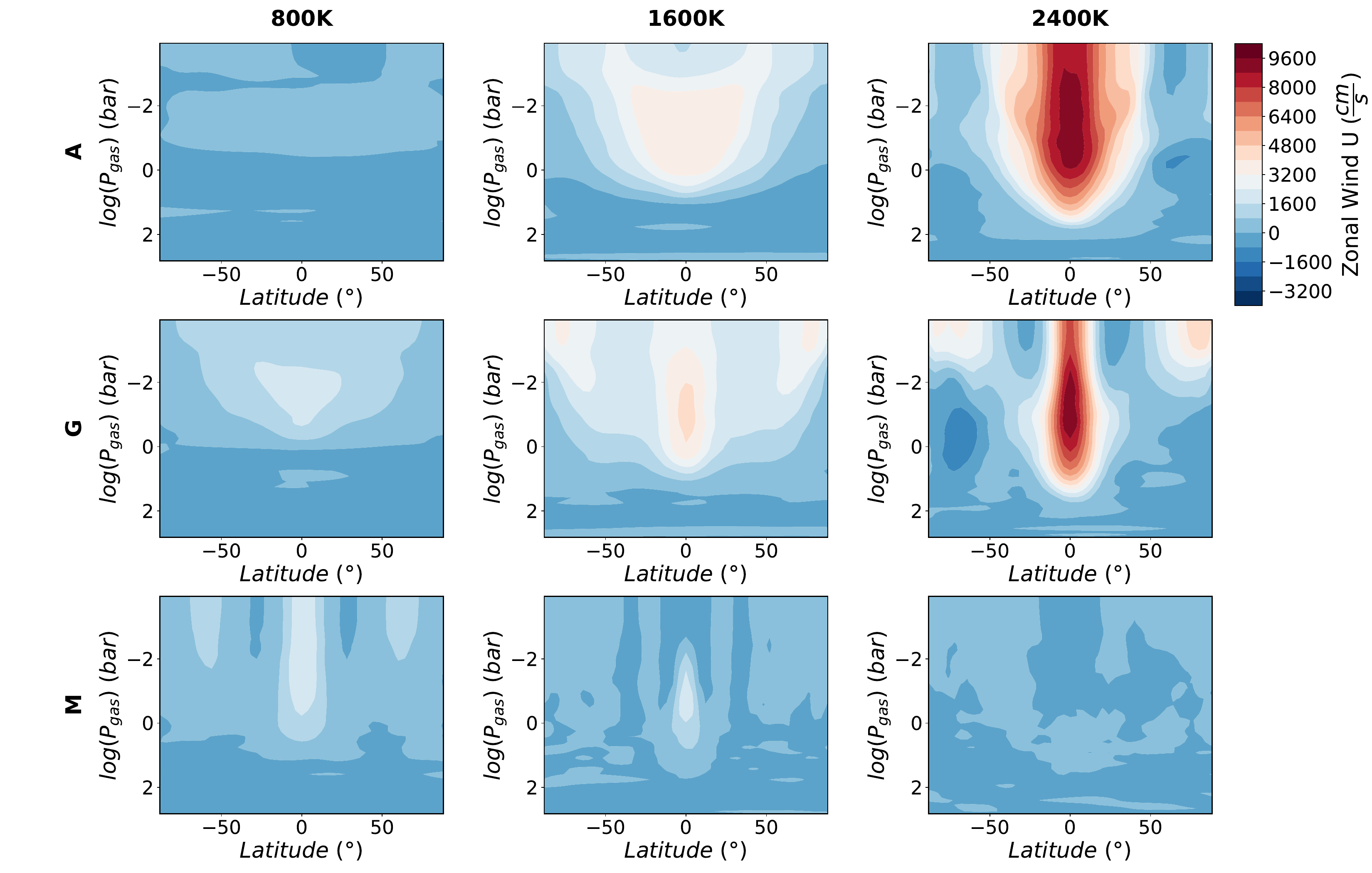}
\caption{
The zonal wind speeds from Fig.~\ref{fig:zonal3x3} reproduced by the DNN, which was trained and evaluated using the same data.
}
\label{fig:zonal3x3DNN}
\end{figure}

\begin{figure}
\includegraphics[width=1\linewidth]{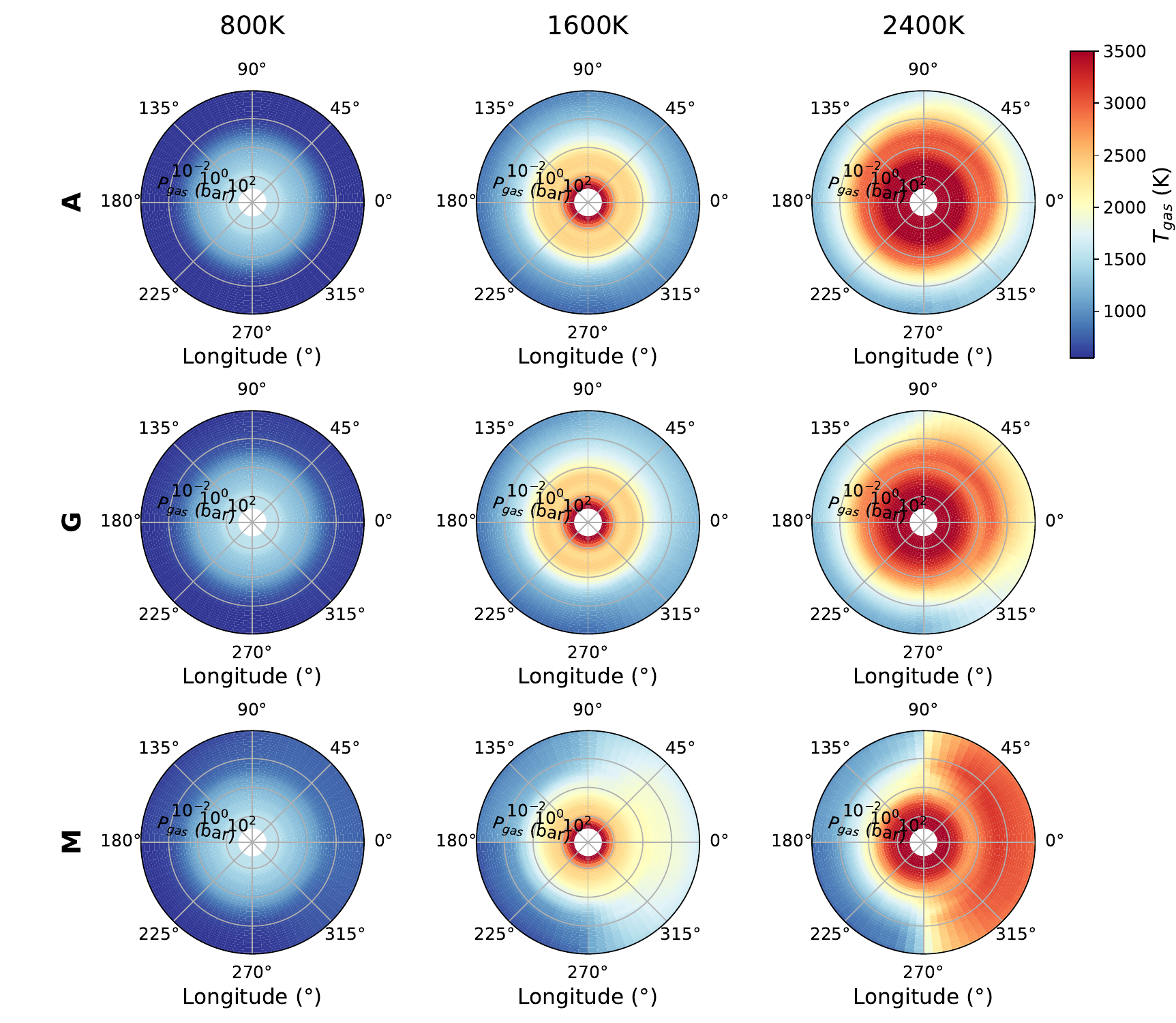}
\caption{
The recreated equatorial slice plots as in Fig.~\ref{fig:EQ_slice}. The XGBoost has been trained on this data and is subsequently used to reproduce the same slices.
}
\label{fig:EQ_sliceXG}
\end{figure}

\begin{figure}
\includegraphics[width=1\linewidth]{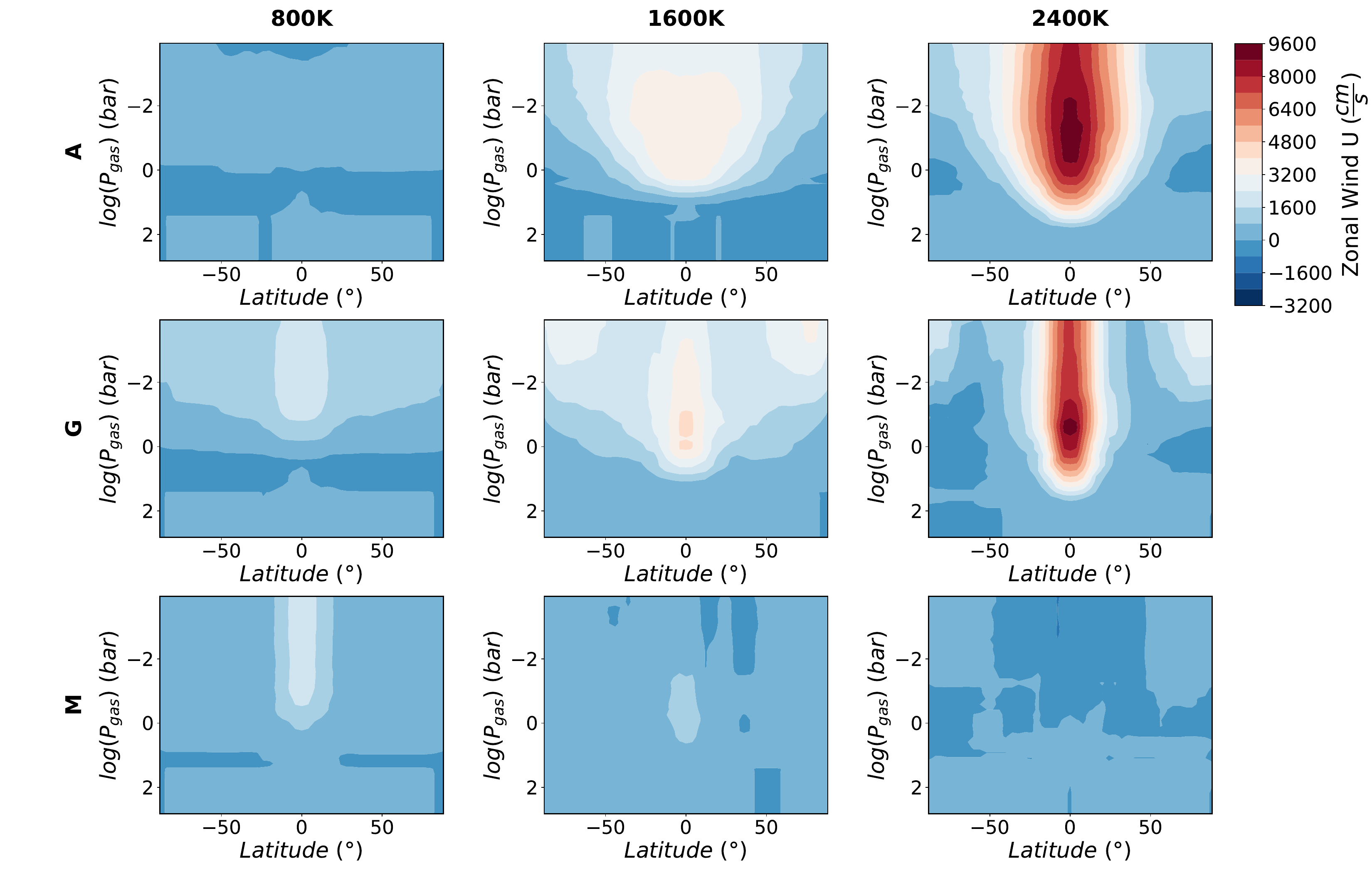}
\caption{
The zonal wind speeds from Fig.~\ref{fig:zonal3x3} reproduced by XGBoost, which was trained and evaluated using the same data.
}
\label{fig:zonal3x3XG}
\end{figure}

%-
Investigating the learning capacity and efficiency of ML models is a key part of the training phase. The simple and effective way to understand the learning capacity is to estimate the quality of the predictions in the training data itself. The prediction error indicates the issue of overfitting or underfitting, and this form of verification can support the selection of appropriate model architectures. It also helps to assess the quality of training data, revealing potential issues such as sparsity (i.e., regions with low data density). Figs. \ref{fig:EQ_sliceDNN} and \ref{fig:EQ_sliceXG} illustrate the reconstruction of the temperature using the DNN and XGBoost models, respectively, based on training data presented in Fig. \ref{fig:EQ_slice}. That means that Figs. \ref{fig:EQ_sliceDNN} and \ref{fig:EQ_sliceXG} are the reconstructed versions of Fig.~\ref{fig:EQ_slice}. Similarly, Figs.~\ref{fig:zonal3x3DNN} and \ref{fig:zonal3x3XG} present the reconstruction results for zonal wind slices corresponding to the input data in Fig.~\ref{fig:zonal3x3}.

%-
%-
%-
%-
%-
\subsection{1D profiles for pairs of test planets}
\label{s:App_val}
\begin{figure}[ht!] 
\centering
\includegraphics[width=0.85\linewidth]{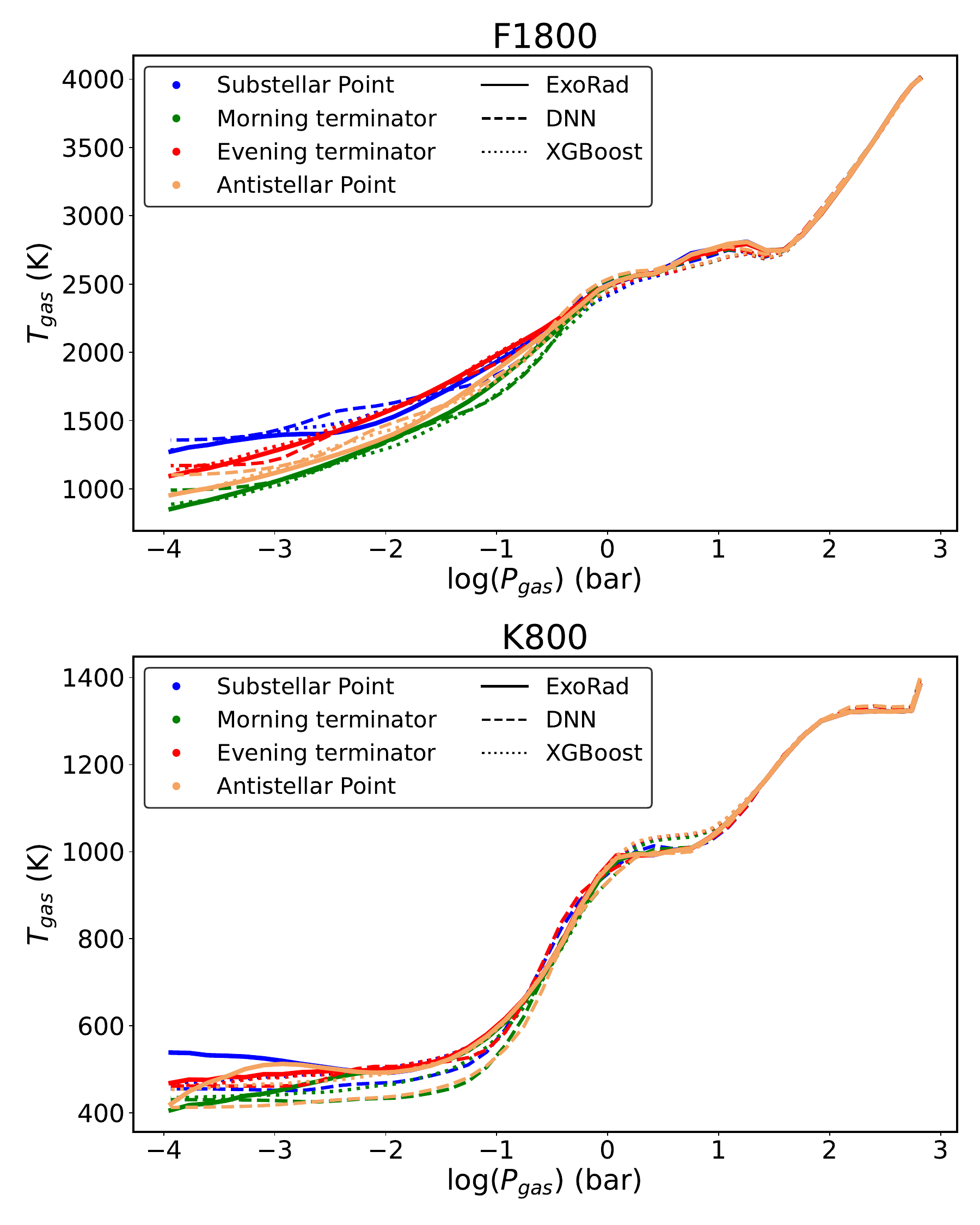}
\caption{1D (T$_{\rm gas}$, p$_{\rm gas}$)-profiles of the two planets chosen for test (F1800, K800). Same setup as in Fig. \ref{fig:1DTeststackF400F1600}.}
\label{fig:1DTeststackF1800K800}
\end{figure}

\begin{figure}[ht!]
\centering
\includegraphics[width=0.85\linewidth]{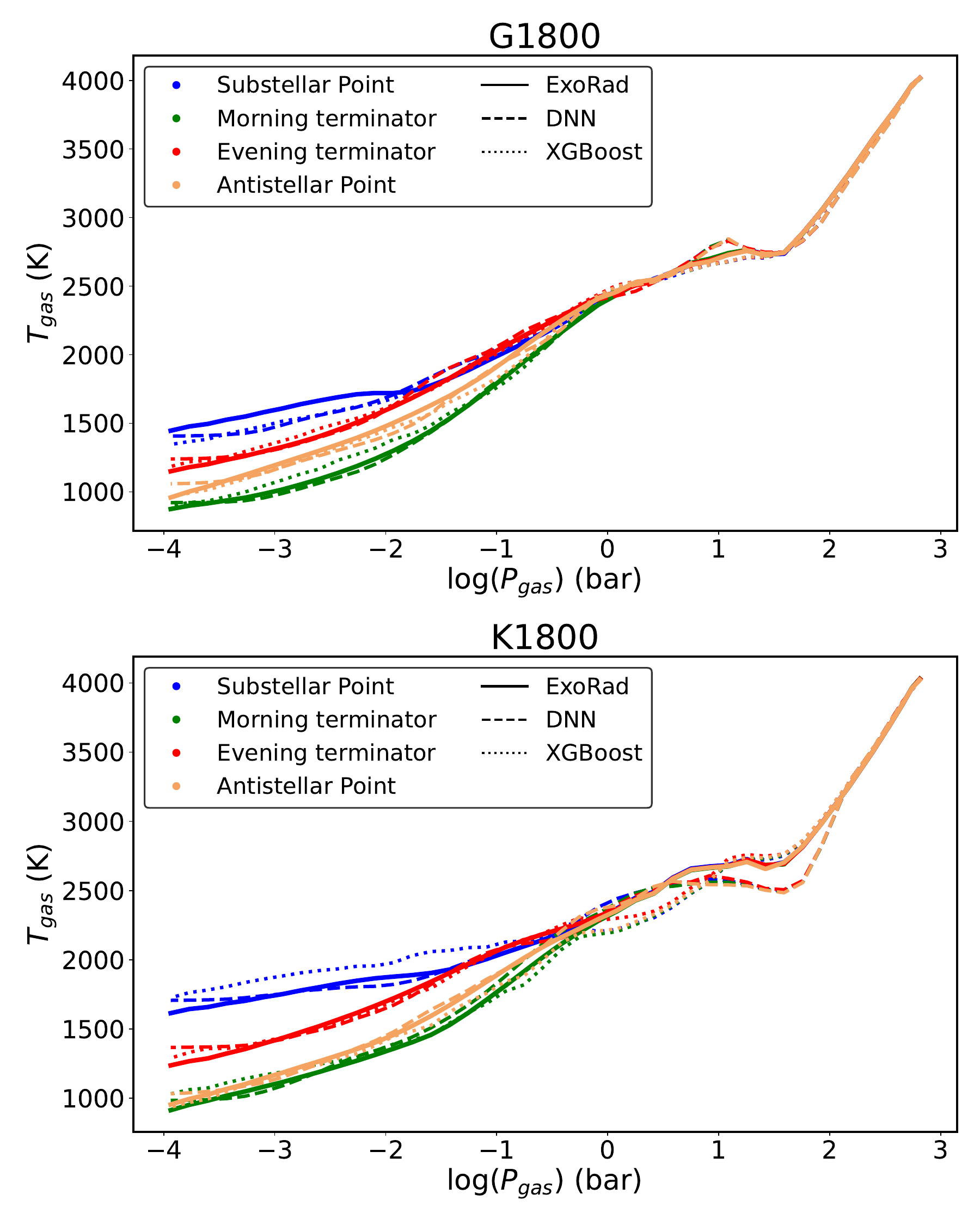}
\caption{1D (T$_{\rm gas}$, p$_{\rm gas}$)-profiles of the two planets chosen for test (G1800, K1800). Same setup as in Fig. \ref{fig:1DTeststackF400F1600}.}
\label{fig:1DTeststacG1800K1800}
\end{figure}

\begin{figure}[ht!]
\centering
\includegraphics[width=0.85\linewidth]{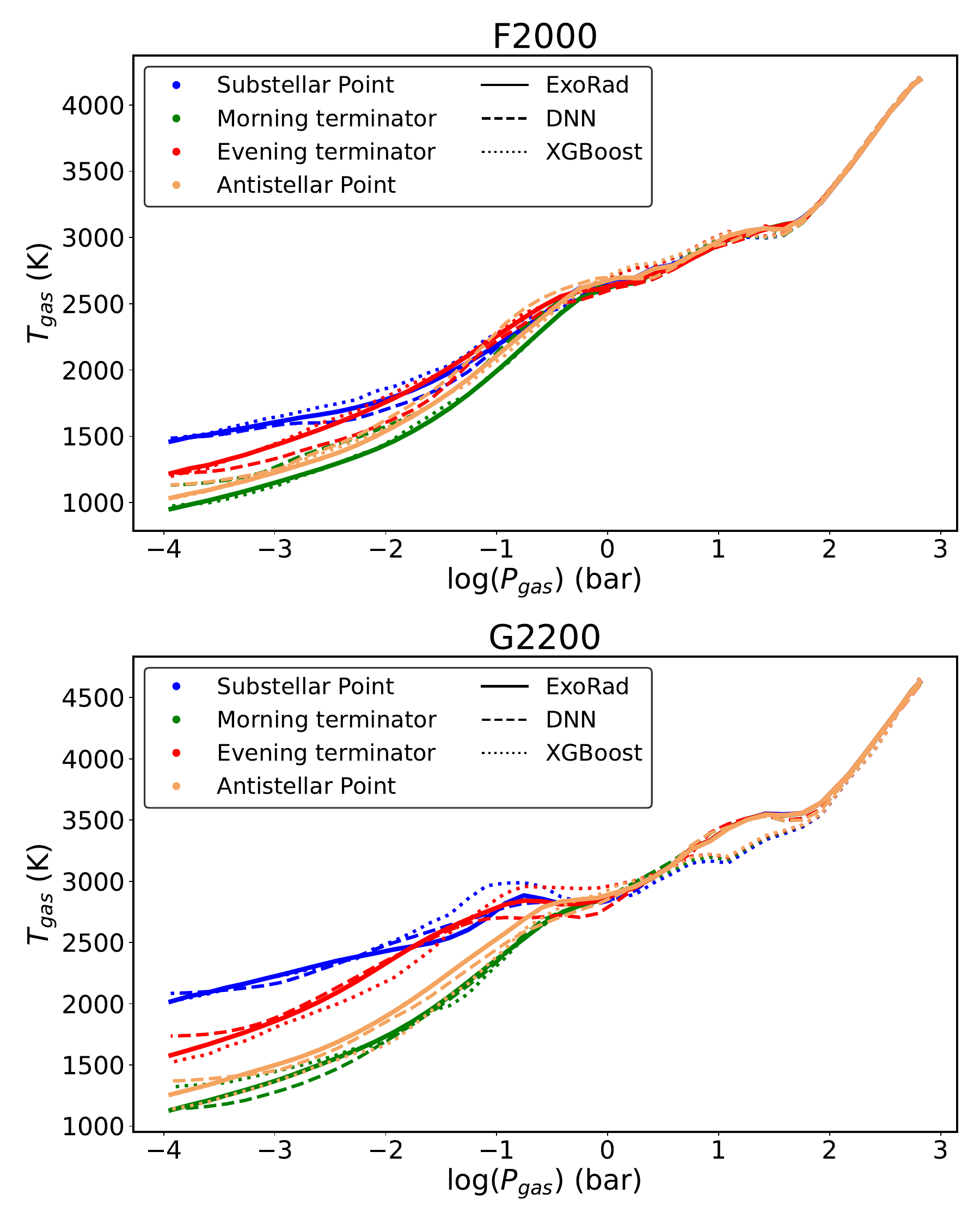}
\caption{1D (T$_{\rm gas}$, p$_{\rm gas}$)-profiles of the two planets chosen for test (F2000, G2200). Same setup as in Fig. \ref{fig:1DTeststackF400F1600}.}
\label{fig:1DTeststackF2000G2200}
\end{figure}

\begin{figure}[ht!]
\centering
\includegraphics[width=0.85\linewidth]{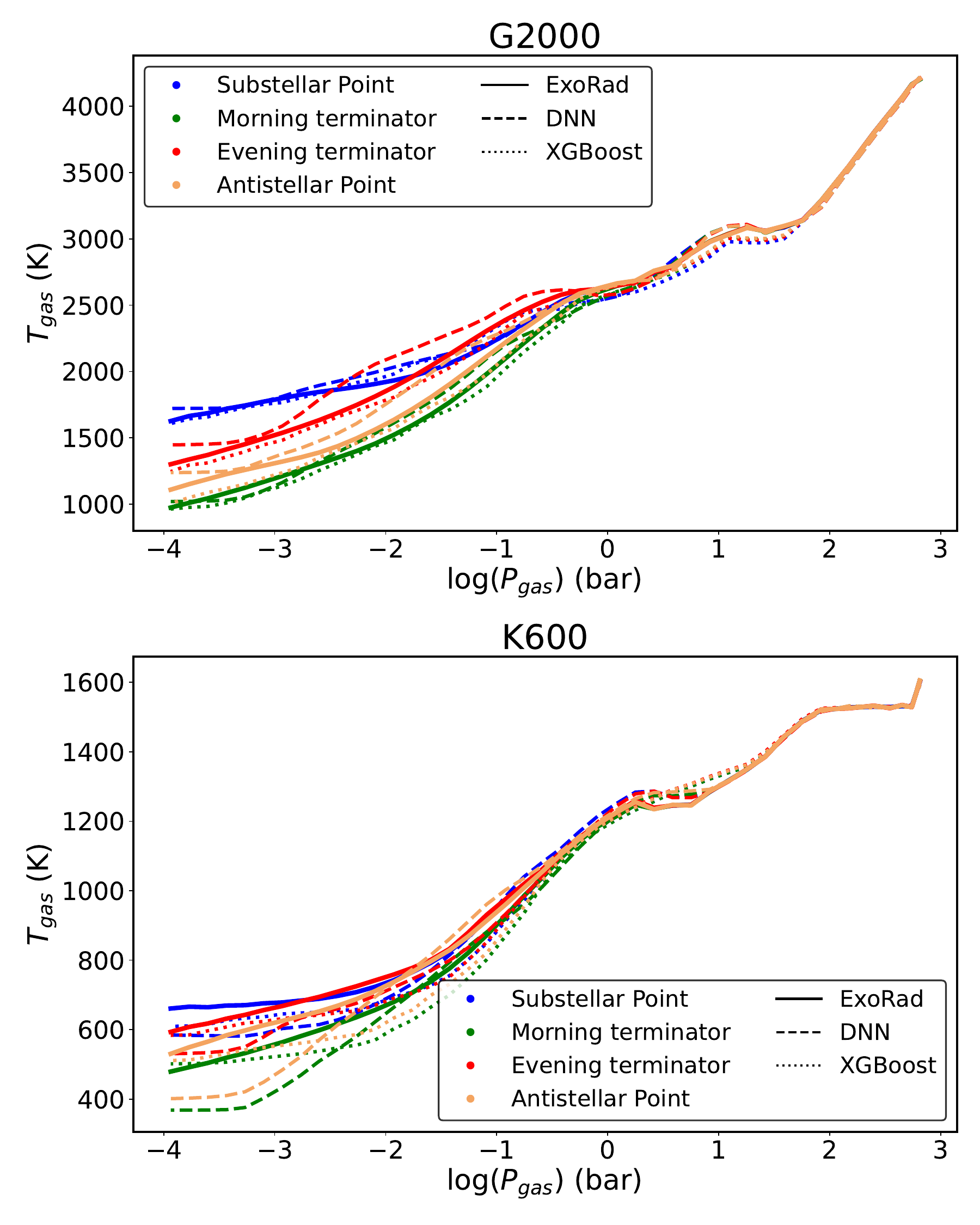}
\caption{1D (T$_{\rm gas}$, p$_{\rm gas}$)-profiles of the two planets chosen for test (G2000, K600). Same setup as in Fig. \ref{fig:1DTeststackF400F1600}.}
\label{fig:1DTeststackG2000K600}
\end{figure}

%-
This section extends the results presented in Sec.~\ref{s:on-grid val}, which focused on a specific pair of planets (F400, F1600). Here, we present the prediction quality of the ML models for all sets of test planet pairs: (F1800, K800), (G1800, K1800), (F2000, G2200) and (G2000, K600). Figs.~\ref{fig:1DTeststackF1800K800} - \ref{fig:1DTeststackG2000K600} show the predicted 1D (T$_{\rm gas}$, p$_{\rm gas}$) profiles at the substellar point, the morning terminator, the evening terminator and the antistellar point for each of these pairs of planets.
%-
%-
%-
%-
%-
\clearpage
\subsection{Testing data density effects on XGBoost}
\label{ss:MLdensityeff}
%-
We hypothesized that ML models, particularly XGBoost, require more data points (i.e., $\{\text{T}_{\rm eff}, \text{T}_{\rm global}, \text{p}_{\rm gas}, \theta, \phi\}$) for improved performance. Ideally, this hypothesis should be validated using additional data points generated from the \texttt{ExoRad} 3D GCM model. However, producing such a dense set of points is computationally expensive, making a direct validation infeasible. To address this, we conducted a sub-optimal yet informative test by using DNN predictions as a proxy ground truth. While this approach does not provide a direct measure of accuracy against physical simulations, it effectively evaluates the sensitivity of XGBoost to increased training data point density.
%-
\begin{figure}[ht!]
\centering
\includegraphics[width=1\linewidth]{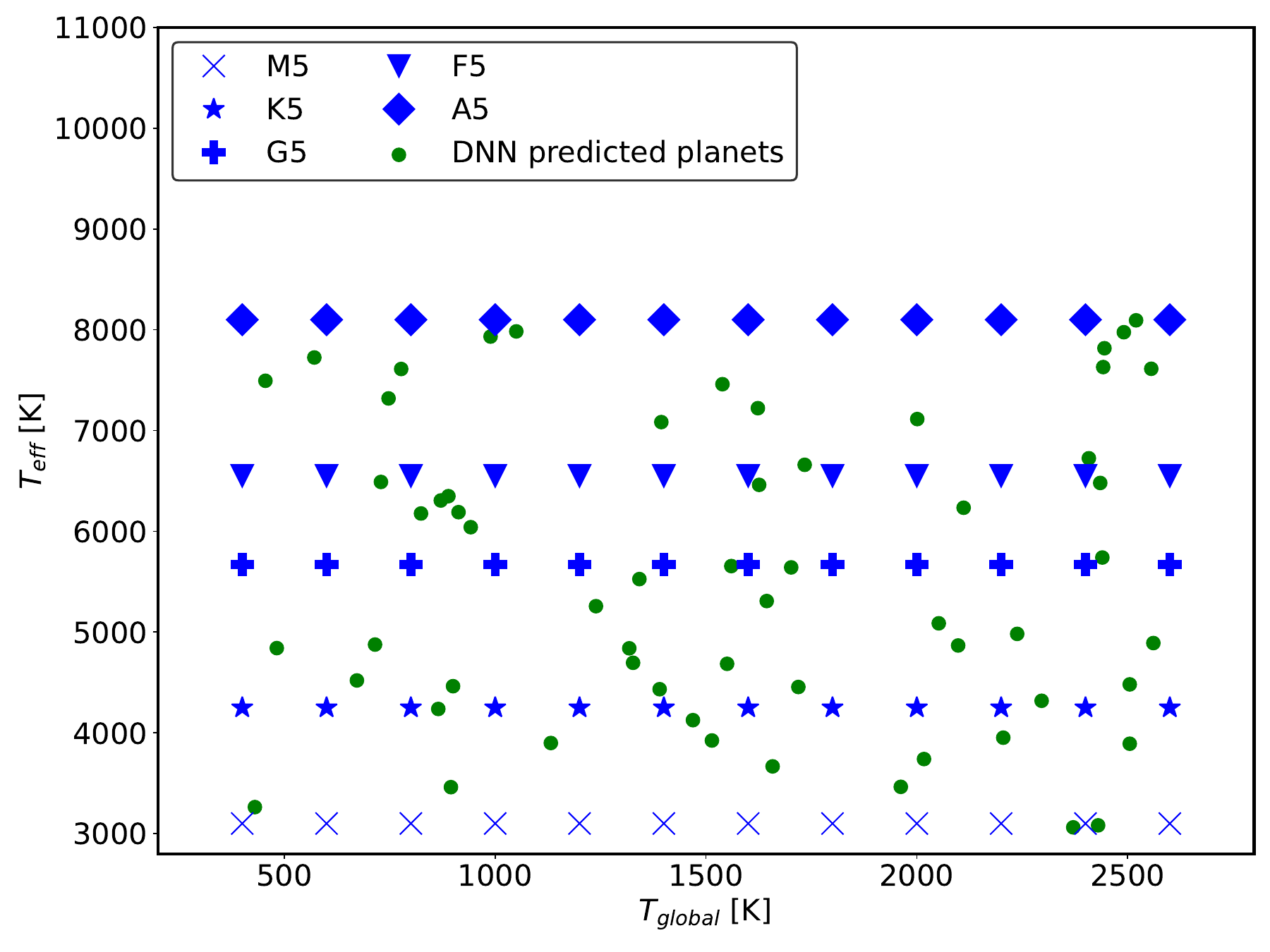}
\caption{
The scatter plot highlights the 60 randomly selected planets, shown in green. These data points and their corresponding prediction of four atmospheric variables by the DNN were integrated into the original training set. This expanded data set (120 planets in total) is utilised to evaluate the impact of increased data density on the prediction quality of the XGBoost model.
}
\label{fig:Motivation_Generative}
\end{figure}
%-
\begin{figure} 
\centering
\includegraphics[width=1.1\linewidth]{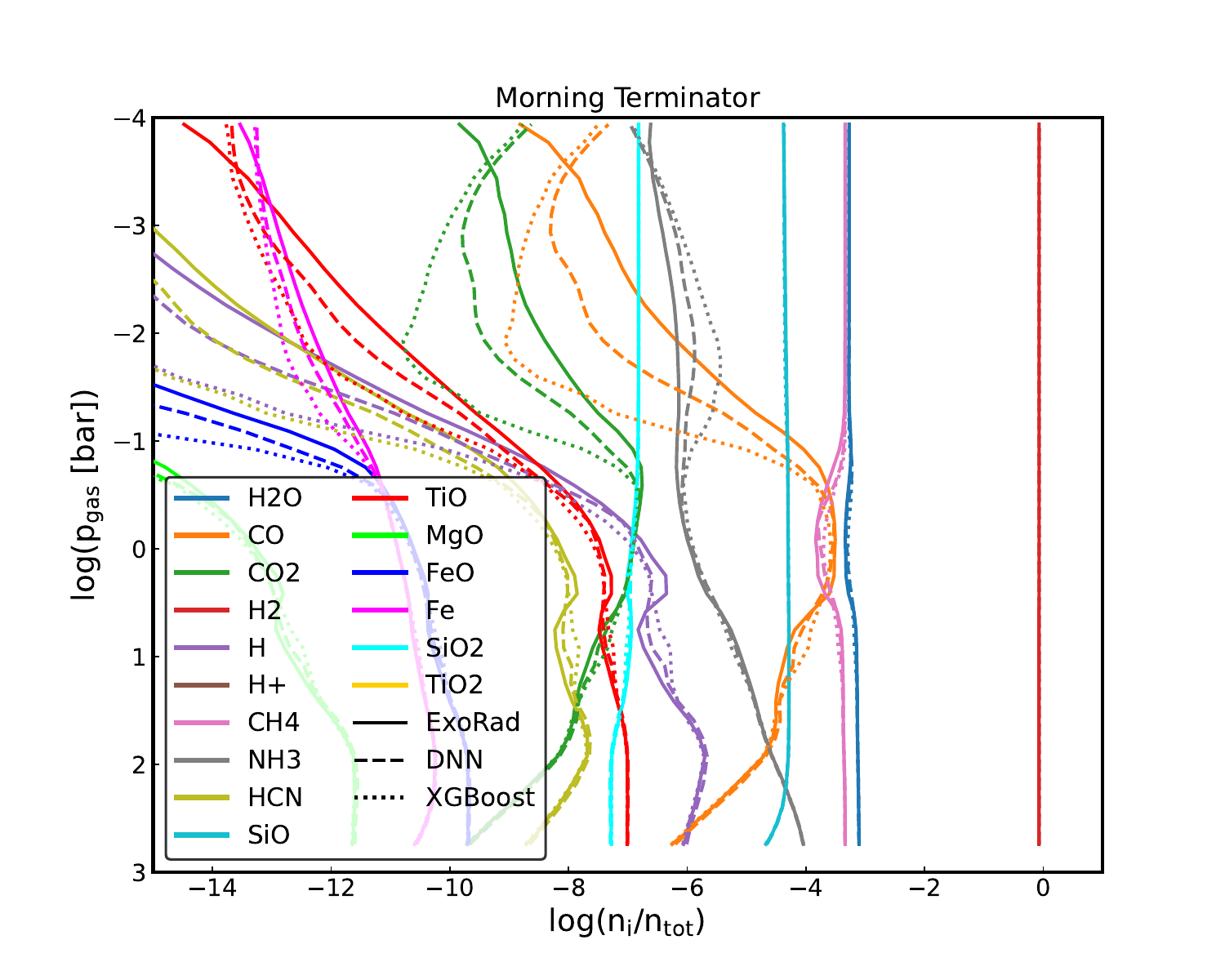}\\[-0.38cm]
\includegraphics[width=1.1\linewidth]{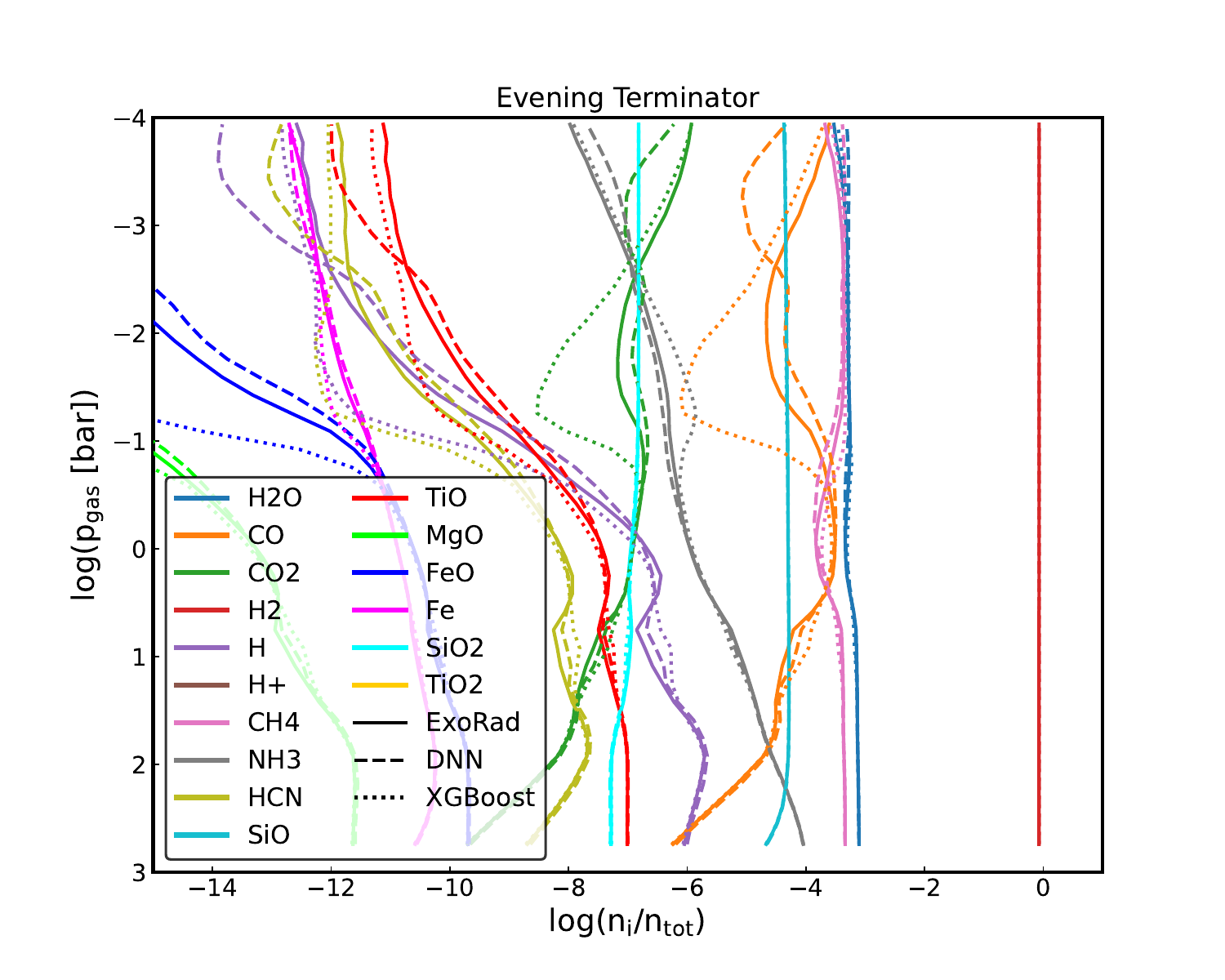}
\caption{
NGTS-1 b* morning (top) and evening (bottom) terminator gas phase chemistry for selected species. The solid line shows the \texttt{ExoRad} simulated values, the dashed line the DNN prediction, and the dotted line the XGBoost prediction. The ML models are trained with 120 planets as shown in Fig.~\ref{fig:Motivation_Generative}. With denser data points, the prediction accuracy of the XGBoost model has drastically improved (See Fig.~\ref{fig:NGTS1_Morning}) compared to those previously trained on 60 planets. 
}
\label{fig:NGTS1_Morning_Generative}
\end{figure}
%-
We used the DNN model to predict atmospheric variables (i.e., $\{\text{T}_{\rm gas}, \text{U}, \text{V}, \text{W} \}$) on 60 additional randomly selected planets within the parameter space (see green points in Fig.\ref{fig:Motivation_Generative}). These new points were combined with the original training set, resulting in a denser dataset comprising 120 samples. Both the XGBoost and DNN models were re-trained using this expanded dataset. Upon retraining, we observed a significant improvement in the equilibrium chemistry predictions from XGBoost (Fig.\ref{fig:NGTS1_Morning_Generative}). As expected, the DNN's performance remained unchanged, as it has been retrained on data derived from its own earlier predictions, thus failing to access new features which would improve the performance. 
%-
It is important to reiterate that the sole goal of this experiment is to explicitly validate the hypothesis that XGBoost could benefit from a denser sampling of the training data points, even when the additional data is obtained from another ML model rather than a physical simulation.
%-
\section{Supplementary results for the WASP-23, NGTS-17 and HATS-42 spectra }
\label{sec: Spectra_all}
%-
In Sec.~\ref{ss:dtdm}, transmission spectra for the hottest and coldest investigated test planets are shown and discussed. The other three planets lie between those two extremes, and their simulated transmission spectra are shown in Figs. \ref{fig:WASP23_Spectra},\ref{fig:NGTS17_Spectra} and \ref{fig:HATS42_Spectra}. One noticeable outlier is the spectra for WASP-23 b* that we discuss further here.

\begin{figure} 
\centering
\includegraphics[width=1\linewidth]{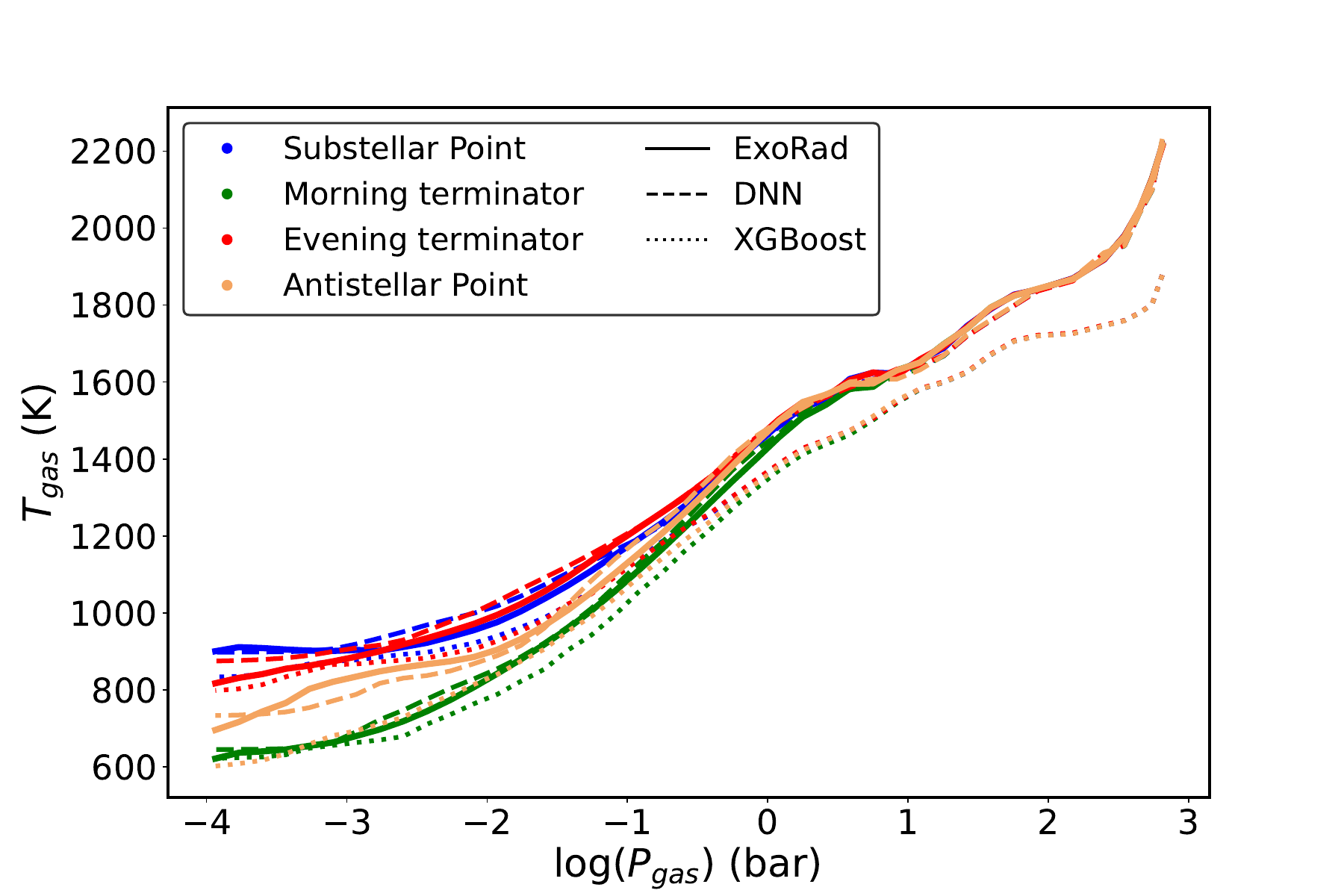}
\caption{WASP-23 b* 1D (T$_{\rm gas}$, p$_{\rm gas}$)-
profiles for the substellar (blue) and antistellar points (orange), the morning (green) and evening terminators (red). The solid line is the \texttt{ExoRad} simulated values, the dashed line the DNN prediction and the dotted line the XGBoost prediction.}
\label{fig:WASP23_T}
\end{figure}

\begin{figure} 
\centering
\includegraphics[width=1\linewidth]{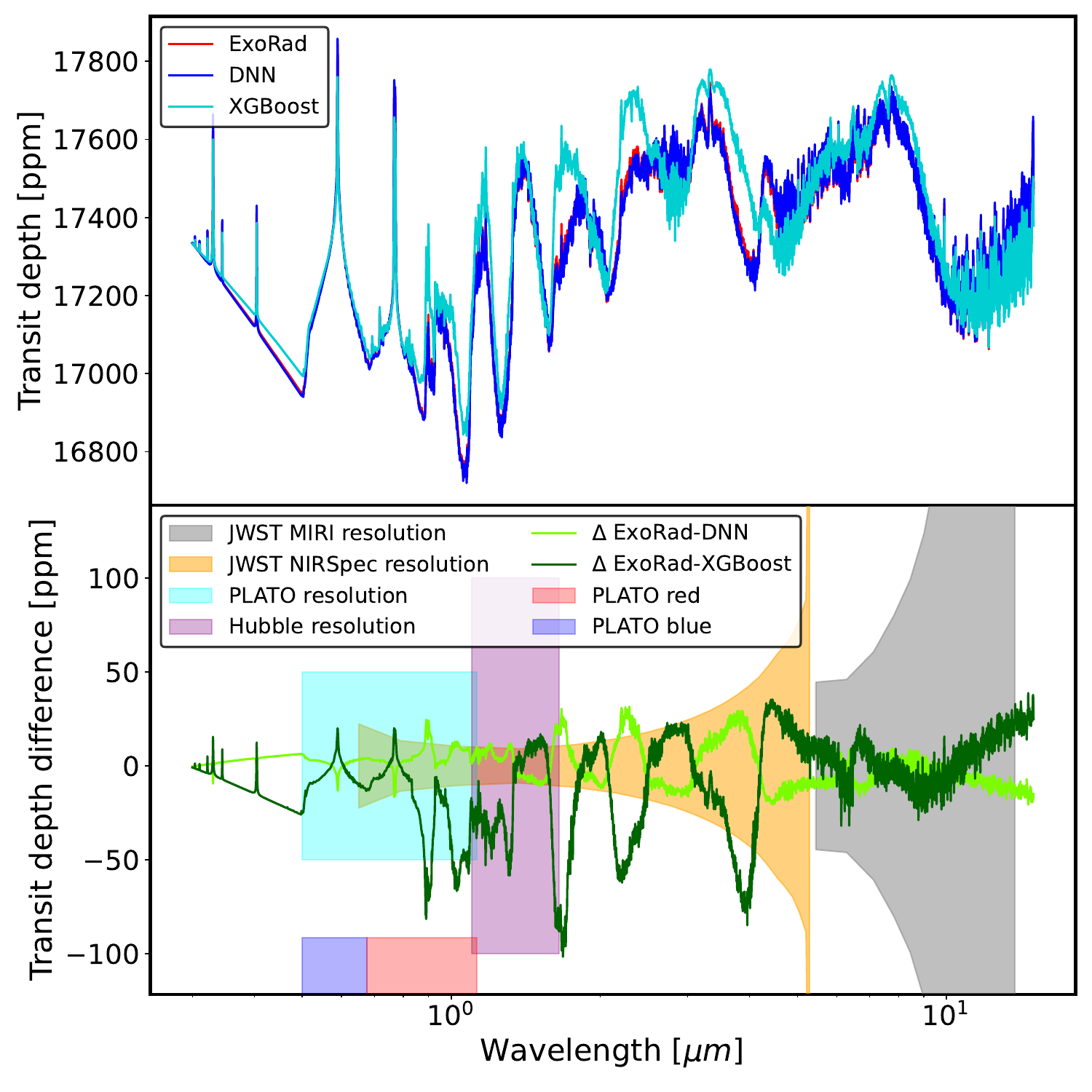}
\caption{Combined morning and evening terminator spectra of WASP-23 b*. Same setup as in Fig.~ \ref{fig:Wasp121_Spectra}.}
\label{fig:WASP23_Spectra}
\end{figure}

\begin{figure} 
\centering
\includegraphics[width=1.1\linewidth]{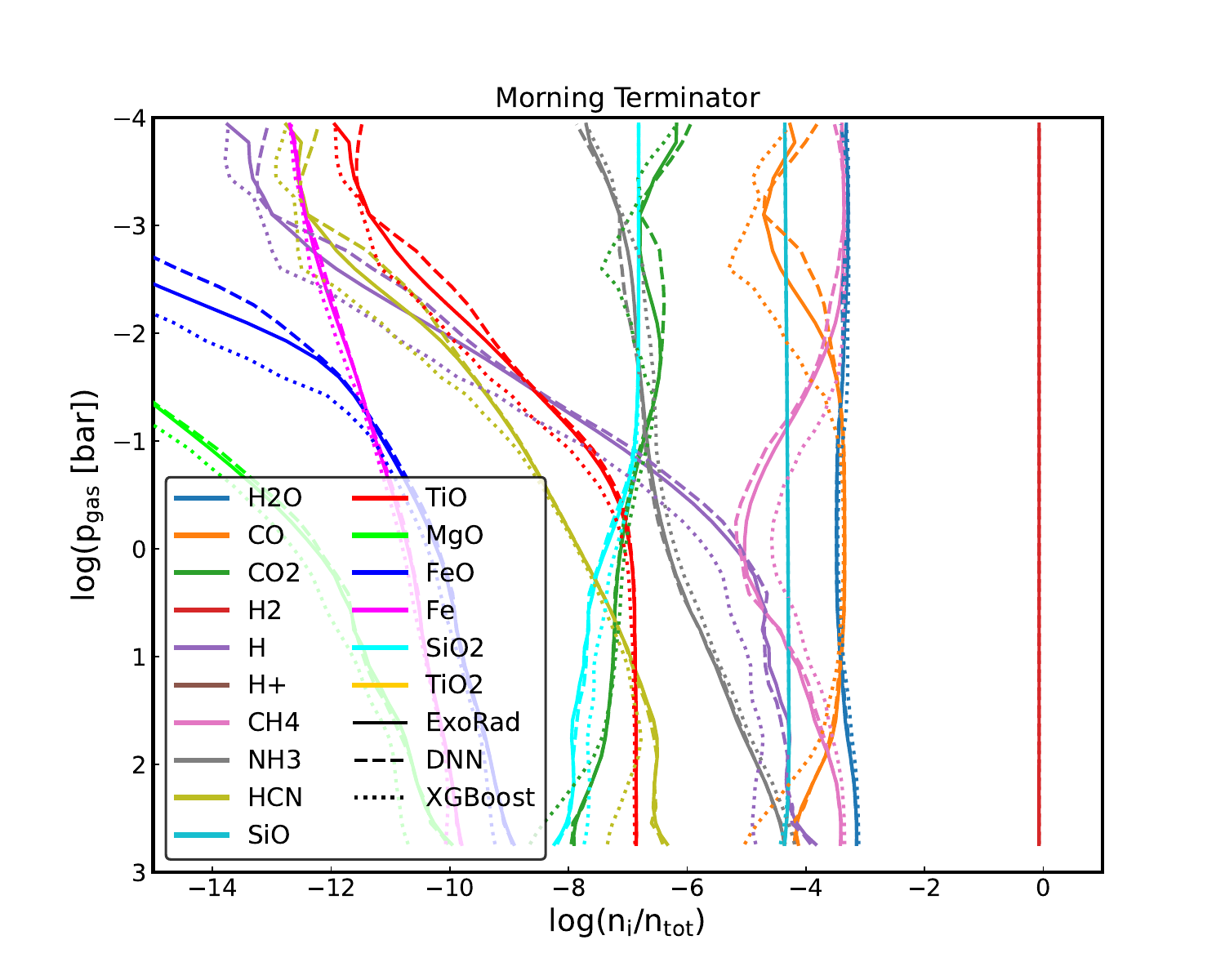}\\[-0.38cm]
\includegraphics[width=1.1\linewidth]{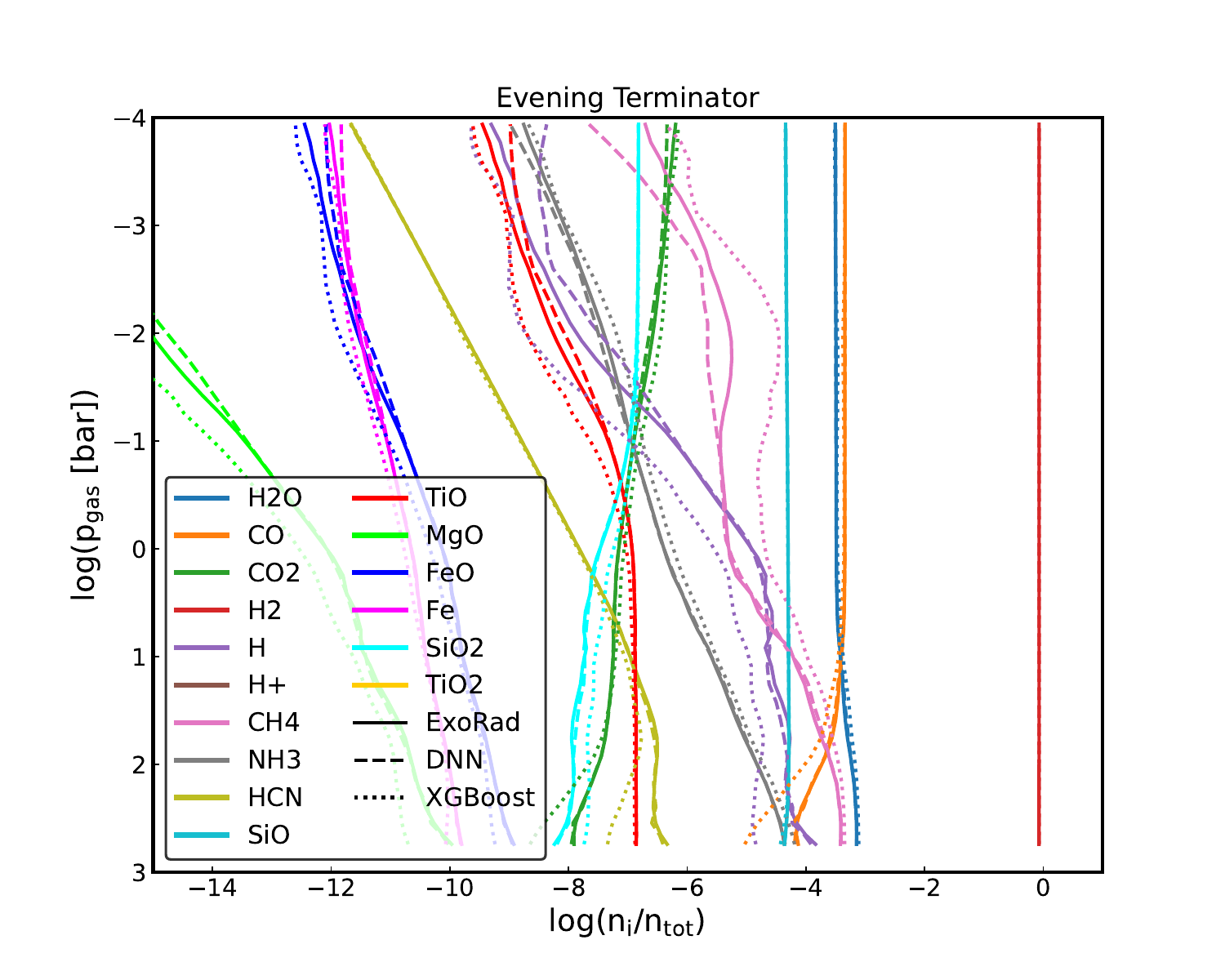}
\caption{WASP-23 b* morning (top) and evening (bottom) terminator gas phase chemistry for selected species. The solid line shows the \texttt{ExoRad} simulated values, the dashed line the DNN prediction and the dotted line the XGBoost prediction. The differences between these two terminators demonstrate how the terminator temperature asymmetries affect the chemical gas phase composition. }
\label{fig:WASP23_Morning}
\end{figure}

\begin{figure} 
\centering
\includegraphics[width=1\linewidth]{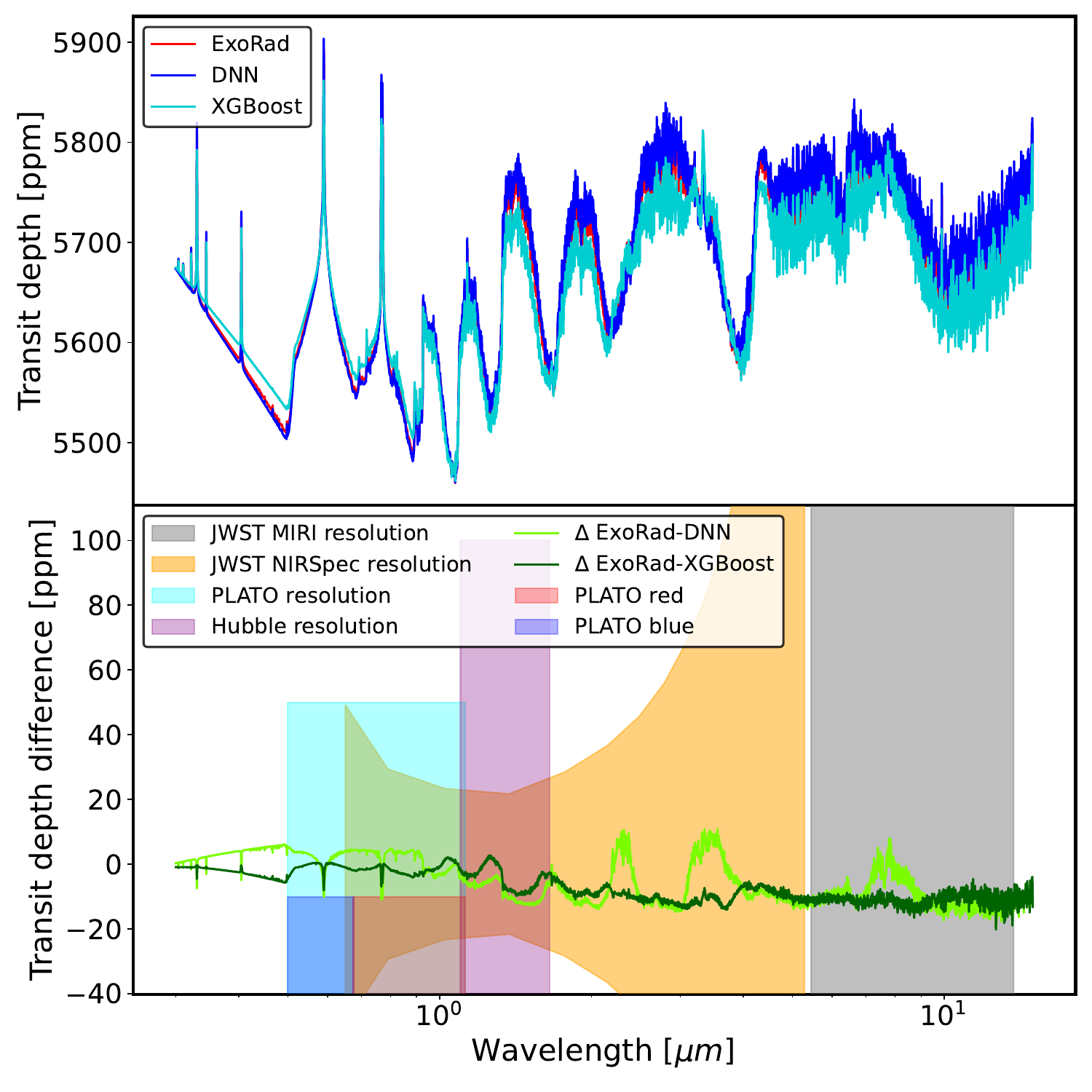}
\caption{Combined morning and evening terminator spectra of NGTS-17 b*. Same setup as in Fig.~ \ref{fig:Wasp121_Spectra}.}
\label{fig:NGTS17_Spectra}
\end{figure}

\begin{figure} 
\centering
\includegraphics[width=1\linewidth]{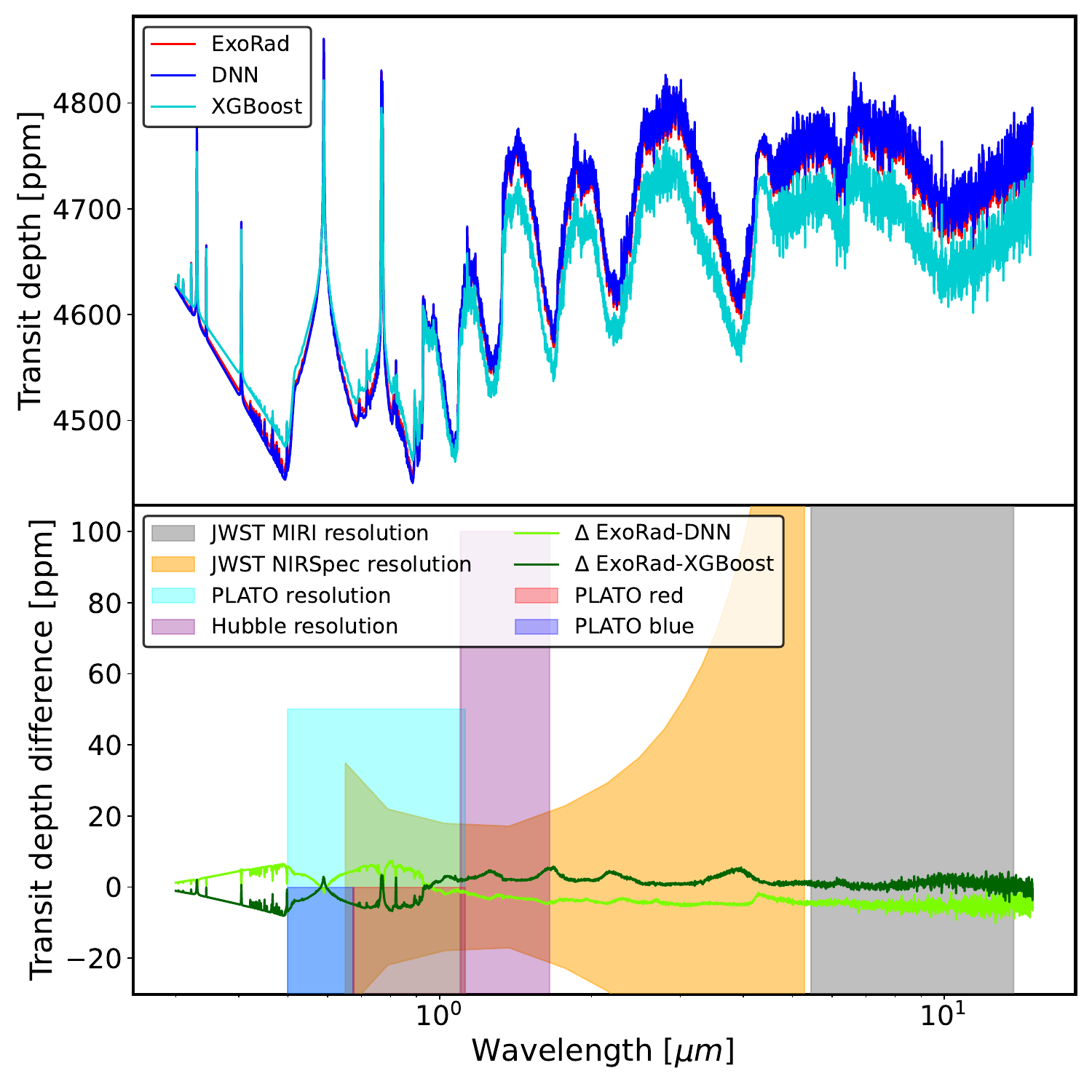}
\caption{Combined morning and evening terminator spectra of HATS-42 b*. Same setup as in Fig.~ \ref{fig:Wasp121_Spectra}.}
\label{fig:HATS42_Spectra}
\end{figure}

\noindent
\paragraph{Transition in carbon chemistry:} The global temperature of WASP-23 b* was assumed to be 1115~K. The local gas temperature thus crosses the CO - CH$_4$ chemical conversion threshold. For hotter planets, CO is more abundant than  CH$_4$, whereas for cooler planets, CH$_4$ is more abundant than CO, assuming equilibrium chemistry.
%-
The tendency of the XGBoost prediction to underestimate the local gas temperature (Fig. \ref{fig:WASP23_T}), thus results in a situation where more CH$_4$, less CO and CO$_2$ and more H$_2$O is produced compared to \texttt{ExoRad} and the DNN prediction. This effect is visible in Fig. \ref{fig:WASP23_Morning}. Looking at the differences between the hot evening and the cold morning terminator shows how these temperature differences affect the carbon chemistry.
%-
Thus, we identified in this work two cases, where the comparatively large temperature deviations of the XG Boost prediction can result in different spectral features due to critical chemical regimes. For WASP-23 b*, it is the carbon chemistry transition. For NGTS-1 b* (Fig.~\ref{fig:NGTS1_Spectra}), it is condensation that suppresses the amplitude of N and K spectral features in the XG Boost prediction.
%-
%-
%-
%-
%-
%-
\end{appendix}
%-
\end{document}